\documentclass[12pt,oneside]{article}

\usepackage{graphicx}

\usepackage[T1]{fontenc}
\usepackage{fancyhdr}           
\usepackage{setspace}           
\usepackage{titlesec}           

\usepackage[numbers,sort&compress]{natbib}
\usepackage[nottoc]{tocbibind}

\usepackage[hidelinks,linktoc=all]{hyperref}

\usepackage[margin=1.25in]{geometry}

\usepackage{caption}
\usepackage[titletoc]{appendix}

\titleformat{\section}[hang]%
{\bfseries\singlespacing\Large}%
{\thesection}%
{0.25in}%
{}

\titleformat{\subsection}[hang]%
{\bfseries\singlespacing\large}%
{\thesubsection}%
{0.25in}%
{}

\usepackage{amsmath,amsfonts,mathtools}
\usepackage[cbgreek]{textgreek}
\usepackage{bm}

\renewcommand{\sp}{\hspace{1pt}}
\newcommand{\nsp}{\hspace{-1pt}}

\newcommand{\ud}{\mathrm{d}}

\newcommand{\bx}{{\bf x}}
\newcommand{\bp}{{\bf p}}
\newcommand{\lrangle}[1]{\langle #1 \rangle}
\newcommand{\pbar}{\bar{p}}
\newcommand{\tI}{\tilde{I}}
\newcommand{\tbI}{\tilde{\bf I}}

\newcommand{\shortminus}{\scalebox{0.85}[1.0]{$-$}}

\usepackage{scalefnt}
\makeatletter
\newcommand\mfrac[2]{%
	\dfrac{\text{\raisebox{-0.5ex}{\scalefont{0.85}{$\m@th#1$}}}}%
	{\text{\raisebox{0.35ex}{\scalefont{0.85}{$\m@th#2$}}}}%
}
\makeatother

\usepackage{placeins}

\makeatletter
\newcommand\new@setfontsize[3]{%
	\ifx \protect \@typeset@protect \let \@currsize #1\fi \fontsize {#2}{#3}\selectfont
}
\let\orig@setfontsize\@setfontsize
\let\orig@cases\cases
\let\endorig@cases\endcases

\renewenvironment{cases}{%
	\let\@setfontsize\new@setfontsize
	\setstretch{\setspace@singlespace}%
	\let\setfontsize\orig@setfontsize
	\orig@cases
}{%
	\endorig@cases
}
\makeatother

\usepackage{array,booktabs,multirow} 
\newcolumntype{M}[1]{>{\centering\arraybackslash}m{#1}}
\usepackage{longtable,tabularx}

\usepackage{mmacells}
\newcommand{\mmasize}{\footnotesize\singlespacing}
\definecolor{MmaCommentColor}{gray}{.6}
\newcommand{\refellipse}{\color{MmaCommentColor}{\ref{fig:MLE_ellipses_Ex9}}}

\usepackage{xcolor}

\usepackage{authblk}

\title{Tutorial: Maximum likelihood estimation in the context of an optical measurement}
\author{Anthony Vella}
\affil{\small The Institute of Optics, University of Rochester\\Rochester, NY 14627, USA}
\affil{avella{\fontfamily{ptm}\selectfont @}optics.rochester.edu}
\date{\vspace{-.6\baselineskip}\small\today}

\begin{document}
\maketitle
\thispagestyle{empty}


\begin{abstract}
The method of maximum likelihood estimation (MLE) is a widely used statistical approach for estimating the values of one or more unknown parameters of a probabilistic model based on observed data. In this tutorial, I briefly review the mathematical foundations of MLE, then reformulate the problem for the measurement of a spatially-varying optical intensity distribution. In this context, the detection of each individual photon is treated as a random event, the outcome being the photon's location. A typical measurement consists of many detected photons, which accumulate to form a spatial intensity profile. Here, I show a straightforward derivation for the likelihood function and Fisher information matrix (FIM) associated with a measurement of multiple photons incident on a detector comprised of a discrete array of pixels. An estimate for the parameter(s) of interest may then be obtained by maximizing the likelihood function, while the FIM determines the uncertainty of the estimate. To illustrate these concepts, several simple examples are presented for the one- and two-parameter cases, revealing many interesting properties of the MLE formalism, as well as some practical considerations for optical experiments. Throughout these examples, connections are also drawn to optical applications of quantum weak measurements, including off-null ellipsometry and scatterometry.
\end{abstract}

\onehalfspacing

\newpage
\thispagestyle{empty}

\renewcommand*\contentsname{Table of Contents}
\tableofcontents

\newpage
\section{Introduction}
The method of maximum likelihood estimation (MLE) was introduced by R.A. Fisher in the early 20$^\text{th}$ century as a way to estimate the parameters associated with an observed quantity based on some statistical model \cite{Fisher_1922,Fisher_1925,Fisher_1935}. Since then, it has been used in wide-ranging applications in the physical and social sciences \cite{Refregier_2003,Gailmard_2014,King_1998,Ly_2017}. This tutorial concentrates on its application to the measurement of an optical intensity distribution $I(\bx;\bp)$ that depends on some vector of unknown physical parameters $\bp=(p_1,\ldots,p_N)$, for example, the physical dimensions or refractive index of an unknown substrate. These parameters can take a continuous range of values, and in general they might each have different units.  In this context, the goal of MLE is to determine the most likely value of $\bp$ from a measurement of $I$. The spatial variable $\bx$ is typically a two-dimensional coordinate in the plane perpendicular to the direction of light propagation, although in some instances it may be replaced by a one-dimensional (1D) coordinate $x$. The treatment shown in this discussion emphasizes the information gained from the shape of $I$ (i.e., its dependence on $\bx$) without regard for the overall intensity (i.e., the total power incident on the detector). One advantage of this approach is that the accuracy of the parameter estimate is not influenced by power fluctuations of the light source, which would otherwise be especially problematic when operating under low-light conditions, as discussed further in Section \ref{sect:MLE_example3}.


Useful in-depth tutorials on MLE and the related topic of Fisher information can be found in Refs.~\cite{Ly_2017,Myung_2003}. The key concepts are summarized in Section \ref{sect:MLE_overview} for the case of a discrete random variable that depends on one or more parameters $p_n$. This situation applies directly to most real-world optical measurements, in which the detector is divided into a discrete pixel array, implying that a measurement consisting of a finite number of photon detections has a finite number of possible outcomes. A mathematical description of this scenario is derived explicitly in Section \ref{sect:MLE_optics}. For context and further insight, the results are then compared in Section \ref{sect:Bayesian_statistics} to the Bayesian statistical approach employed in Ref.~\cite{Ramkhalawon_2013}. Lastly, Sections \ref{sect:MLE_examples_1param} and \ref{sect:MLE_examples_2param} contain a number of simple one- and two-parameter examples illustrating the procedure of MLE for optical measurements, as well as the role of Fisher information in evaluating and optimizing the accuracy of an experiment. The Mathematica code for these calculations is provided in the appendix. 

\begin{sloppypar}
The theory developed in Sections \ref{sect:MLE_overview} through \ref{sect:Bayesian_statistics} is presented for the multiple-parameter case (vector-valued $\bp$), which can trivially be reduced to the single-parameter case when needed (as in Section \ref{sect:MLE_examples_1param}). The key results of this tutorial are those established in Section \ref{sect:MLE_optics}.
\end{sloppypar}


\section{Overview of MLE: likelihood, Fisher information, and the Cram\'er-Rao bound}\label{sect:MLE_overview}
Before discussing its application to an optical measurement, in this section the basic concepts of MLE are reviewed in a general context. Consider a discrete random variable $Y$, and let $P(y|\bp)$ denote the probability mass function (PMF) specifying the conditional probability of the outcome $Y=y$ given some vector of parameters $\bp$. The PMF is normalized such that
\begin{equation}
\sum_{y\in\mathcal{Y}}P(y|\bp) = 1,\label{eq:MLE_pmf_normalization}
\end{equation}
where $\mathcal{Y}$ is the set of all possible outcomes of $Y$. It should be emphasized that the PMF is interpreted as a function of $y$. That is, given a fixed value of $\bp$, the function $P(y|\bp)$ provides the probability of each possible outcome $y$. In a typical measurement, however, we require just the opposite: given an observed value of $y$, we wish to determine the value of $\bp$ that is most likely to have produced the measured outcome. This inverse problem is solved by introducing the likelihood function, defined as\footnote{Often, the likelihood is used to describe of a set of measurements $\mathcal{S}=(y_1,y_2,\ldots)$, in which case it could be denoted as $L(\mathcal{S}|\bp)$. In this discussion, the notation $L(\bp|y)$ is used with the understanding that $y$ could represent either a single measurement or an ensemble of measurements, e.g., an optical intensity distribution, which is a collection of many individual photon detection events.}\linebreak[3] $L(\bp|y)=P(y|\bp)$. Although the likelihood function and the PMF appear to be mathematically identical (and indeed they are in their unevaluated symbolic forms), they actually have quite different meanings. In contrast to the PMF, the likelihood function is regarded as a continuous function of $\bp$ for some fixed value of $y$. It is not subject to any normalization condition over $\bp$. Given an observation $Y\nsp=\nsp y$,\linebreak[4] $L(\bp|y)$ represents the likelihood (relative probability) of a vector $\bp$ of candidate parameter values. Accordingly, the maximum likelihood estimate (also abbreviated as MLE) for the unknown parameter values is obtained by determining the value of $\bp$ that maximizes $L(\bp|y)$. For computational convenience, the log-likelihood function $\ell(\bp|y)=\ln L(\bp|y)$ is often equivalently maximized instead.

Next, consider the related problems of (1) evaluating the uncertainty of a maximum likelihood estimate and (2) designing an experiment for optimal sensitivity. These problems both pertain to the Fisher information, which quantifies the amount of information about $\bp$ that is contained within a measurement of $Y$. For the case of $N$ parameters, the Fisher information matrix (FIM) $\mathbb{J}(\bp)$ is defined as the $N\times N$ symmetric, positive semi-definite matrix with elements%
\begin{subequations}\label{eq:Fisher_1stders}
	\begin{align}
	[\mathbb{J}(\bp)]_{mn} &= \mathrm{E}\left[
	\left(\mfrac{\partial}{\partial p_m}\ell(\bp|y)\right)\!\nsp
	\left(\mfrac{\partial}{\partial p_n}\ell(\bp|y)\right)\right] \label{eq:Fisher_1stders_E}\\[2pt]
	&= \sum_{y\in\mathcal{Y}}
	\left(\mfrac{\partial}{\partial p_m}\ell(\bp|y)\right)\!\nsp
	\left(\mfrac{\partial}{\partial p_n}\ell(\bp|y)\right)\!
	L(\bp|y),\label{eq:Fisher_1stders_sum}
	\end{align}
\end{subequations}
where $\mathrm{E}$ denotes the expectation value over $\mathcal{Y}$. Under mild regularity conditions \cite{Rao_2017}, the FIM is equivalently defined as\footnote{To prove this result, one can expand the derivatives in Eq.~(\ref{eq:Fisher_2ndder_sum}) using the chain rule and product rule. This produces the RHS of Eq.~(\ref{eq:Fisher_1stders_sum}) plus an additional term $-\sum_{y\in\mathcal{Y}}\frac{\partial^2}{\partial p_m\partial p_n}L(\bp|y)=-\frac{\partial^2}{\partial p_m\partial p_n}\sum_{y\in\mathcal{Y}}L(\bp|y)$. By Eq.~(\ref{eq:MLE_pmf_normalization}), the sum over $L(\bp|y)$ is equal to 1, so its derivative is zero. The ``regularity conditions'' for this proof essentially require that $L(\bp|y)$ is twice differentiable and that the order of summation and differentiation can be swapped. In practice, these conditions are met in all but the most pathological cases.}%
\begin{subequations}\label{eq:Fisher_2ndder}
	\begin{align}
	[\mathbb{J}(\bp)]_{mn} &= -\mathrm{E}\left[\mfrac{\partial^2}{\partial p_m\partial p_n}\ell(\bp|y)\right] \label{eq:Fisher_2ndder_E}\\[2pt]
	&= -\sum_{y\in\mathcal{Y}}
	\left(\textstyle\mfrac{\partial^2}{\partial p_m\partial p_n}\ell(\bp|y)\right)\!
	L(\bp|y).\label{eq:Fisher_2ndder_sum}
	\end{align}
\end{subequations}
Since $\mathbb{J}(\bp)$ represents the information contained in a single observation of the random variable $Y$, it is sometimes called the \emph{unit} Fisher information. If the measurement is repeated for $T$ independent trials, it can be shown that the total information obtained is $T\,\mathbb{J}(\bp)$. Note that while the Fisher information is a function of the true parameter values $\bp$, it is independent of $y$. This indicates that $\mathbb{J}(\bp)$ is not a property of an individual measurement, but rather of the measurement scheme (and its expected outcome). 
For this reason, $\mathbb{J}(\bp)$ is often referred to as the \emph{expected} Fisher information. Some texts also define the \emph{observed} Fisher information $\mathbb{J}^\text{(obs)}(\bp;y)$ associated with a particular measured outcome $y$ by dropping the expectation values from Eqs.~(\ref{eq:Fisher_1stders_E}) and (\ref{eq:Fisher_2ndder_E}) and evaluating at the maximum likelihood estimate for $\bp$. There has been debate regarding the conditions under which it is more appropriate to use the observed or expected Fisher information \cite{Efron_1978,Cao_2013}. In the asymptotic limit of a large number of observations, it can be shown that the two definitions are equivalent \cite{Newey_1994}.

The statistical significance of the FIM is that its inverse $\mathbb{J}^{-1}(\bp)$ places a lower limit on the covariance matrix $\mathbb{C}(\bp)$ for a maximum likelihood estimate of $\bp$. More precisely, for any unbiased estimator\footnote{In general, the MLE can be biased. However, it is asymptotically unbiased for a sufficiently large sample size \cite{Naftali_2001}. The form of the Cram\'er-Rao bound given in Eq.~(\ref{eq:Cramer-Rao}) only applies when the MLE is unbiased.}, the Cram\'er-Rao bound \cite{Refregier_2003} states that the matrix $\mathbb{C}-\mathbb{J}^{-1}$ must be positive semi-definite, i.e., for any vector $\bp$,
\begin{equation}
\bp^{\rm T}\mathbb{C}\,\bp \geq \bp^{\rm T}\mathbb{J}^{-1}\bp. \label{eq:Cramer-Rao}
\end{equation}
The diagonal elements $[\mathbb{J}^{-1}]_{nn}$ provide the minimum variance of each parameter $p_n$, while the off-diagonal elements $[\mathbb{J}^{-1}]_{mn}$ (where $m\neq n$) represent the expected covariances between parameters $p_m$ and $p_n$. The uncertainty of the measurement can be visualized as an ellipsoid in $N$-dimensional parameter space (centered at the MLE) representing the standard deviation confidence interval. The principal axis orientations of the ellipsoid are given by the eigenvectors of $\mathbb{J}^{-1}$, and the semi-axis lengths are the square roots of the corresponding eigenvalues \cite{Friendly_2013}. Four examples are illustrated in Table~\ref{tbl:ellipse_examples} for the case of a two-parameter measurement in which the true parameter values for $p_1$ and $p_2$ are both zero. Since $\mathbb{J}^{-1}$ is a function of $\bp$, in general the size and shape of the error ellipsoid also varies over the parameter space. This dependence can be visualized for the two-parameter case (or a 2D slice of a higher-dimensional parameter space) by plotting a grid of ellipses over a selection of parameter values, as seen in Section \ref{sect:MLE_examples_2param} and in Ref.~\cite{Vella_2018_fbs_arxiv}.
	
\begin{table}
	\begin{center} 
		\begin{tabular}{M{1in}M{1.1in}M{1.5in}M{1.7in}}
			\toprule
			$\mathbb{J}^{-1}$ & Eigenvalues & Eigenvectors & Error ellipse \\
			\midrule
			$\left[\!\begin{array}{cc}  1 & 0 \\ 0 & 1  \end{array}\!\right]$ 
			& 1, 1
			& $\left[\!\begin{array}{r}  1 \\ 0  \end{array}\!\right]$\nsp,
			$\left[\!\begin{array}{r}  0 \\ 1  \end{array}\!\right]$
			& \includegraphics[width=1.7in]{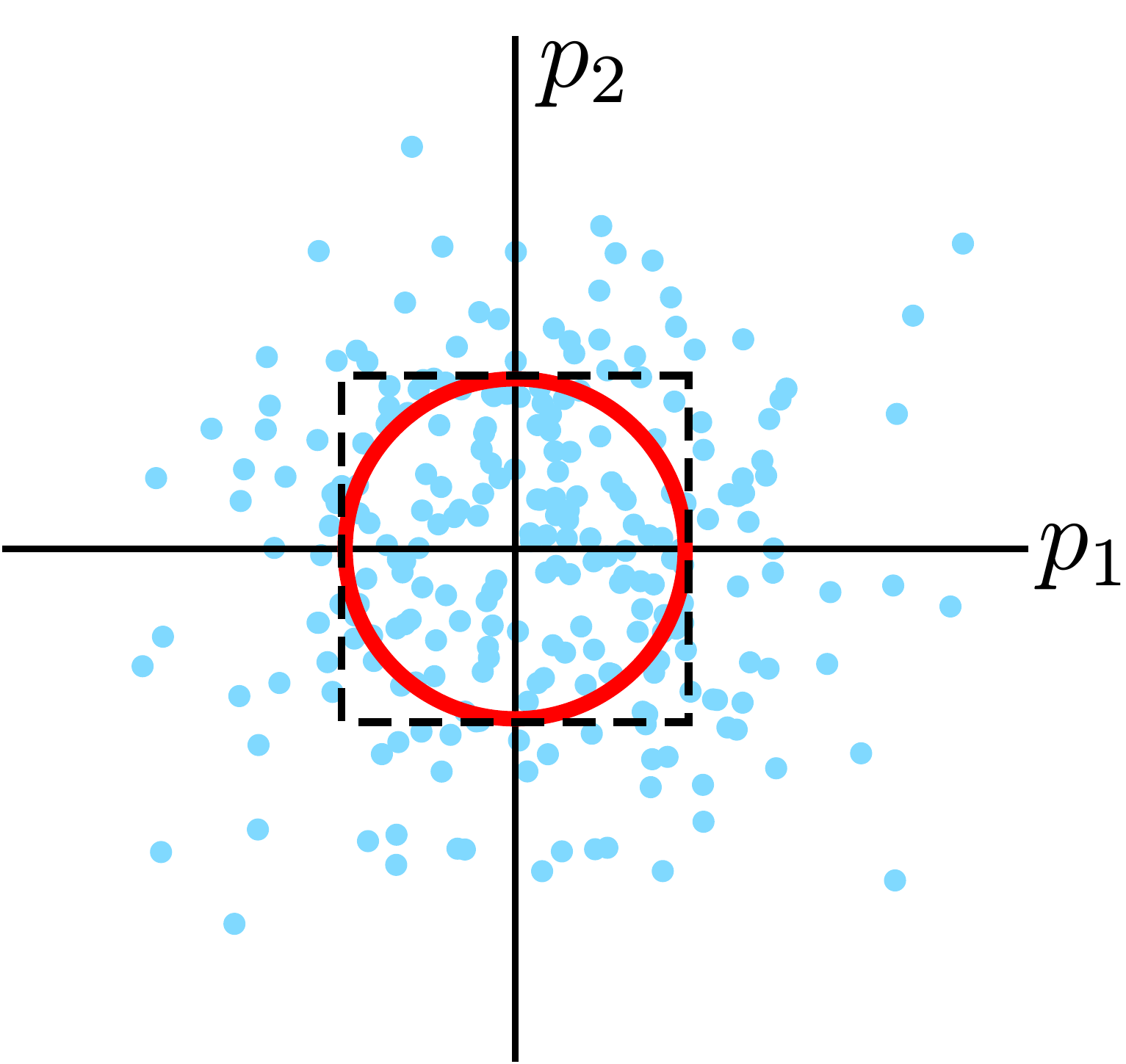} \\
			
			$\left[\!\begin{array}{cc}  1 & 0 \\ 0 & 0.2  \end{array}\!\right]$ 
			& 1, 0.2
			& $\left[\!\begin{array}{r}  1 \\ 0  \end{array}\!\right]$\nsp,
			$\left[\!\begin{array}{r}  0 \\ 1  \end{array}\!\right]$
			& \includegraphics[width=1.7in]{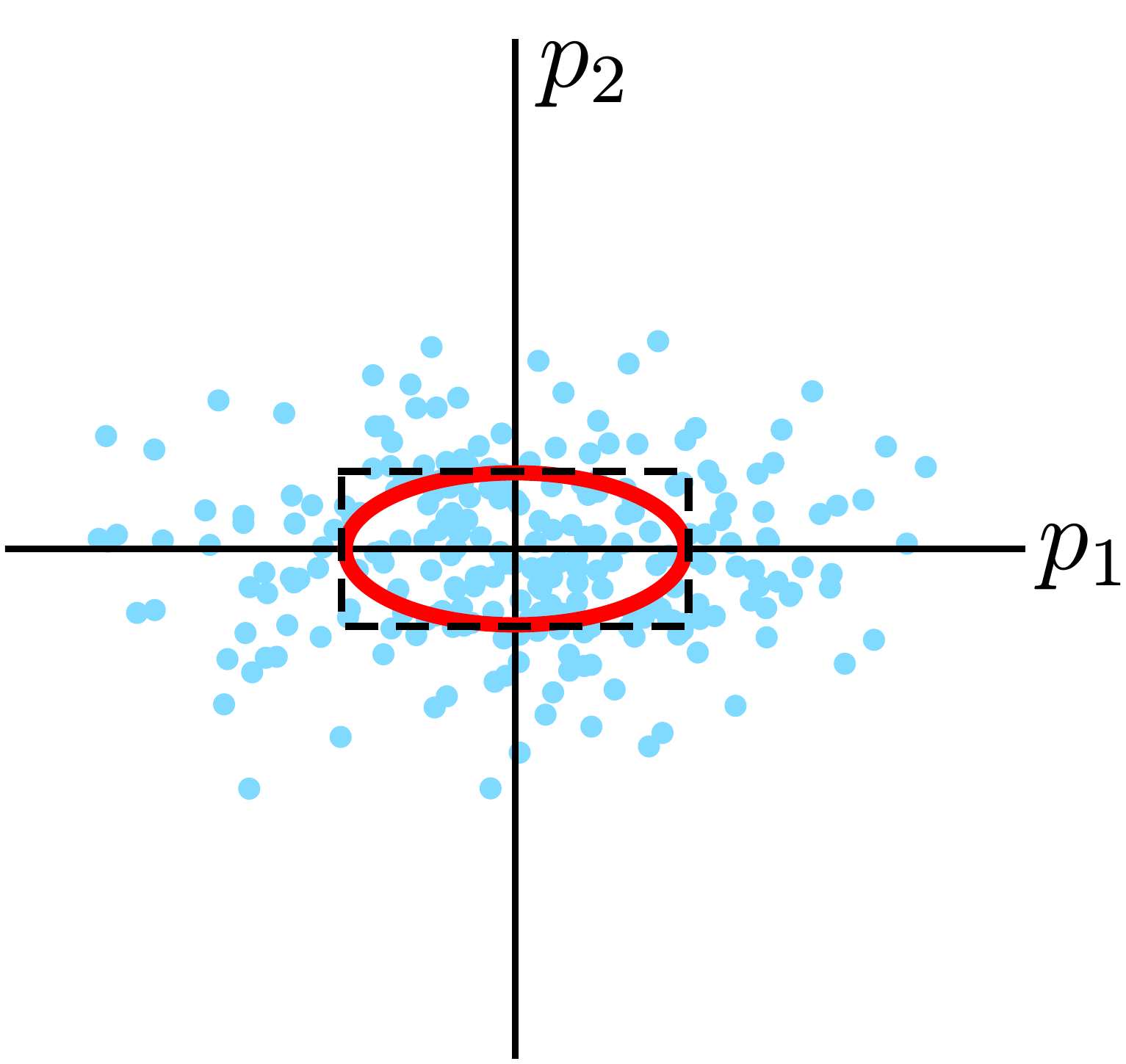} \\
			
			$\left[\!\begin{array}{cc}  1 & 0.5 \\ 0.5 & 1  \end{array}\!\right]$ 
			& 1.5, 0.5
			& $\left[\!\begin{array}{r}  0.71 \\ 0.71  \end{array}\!\right]$\nsp,
			$\left[\!\!\begin{array}{r}  0.71 \\ \shortminus 0.71  \end{array}\!\right]$
			& \includegraphics[width=1.7in]{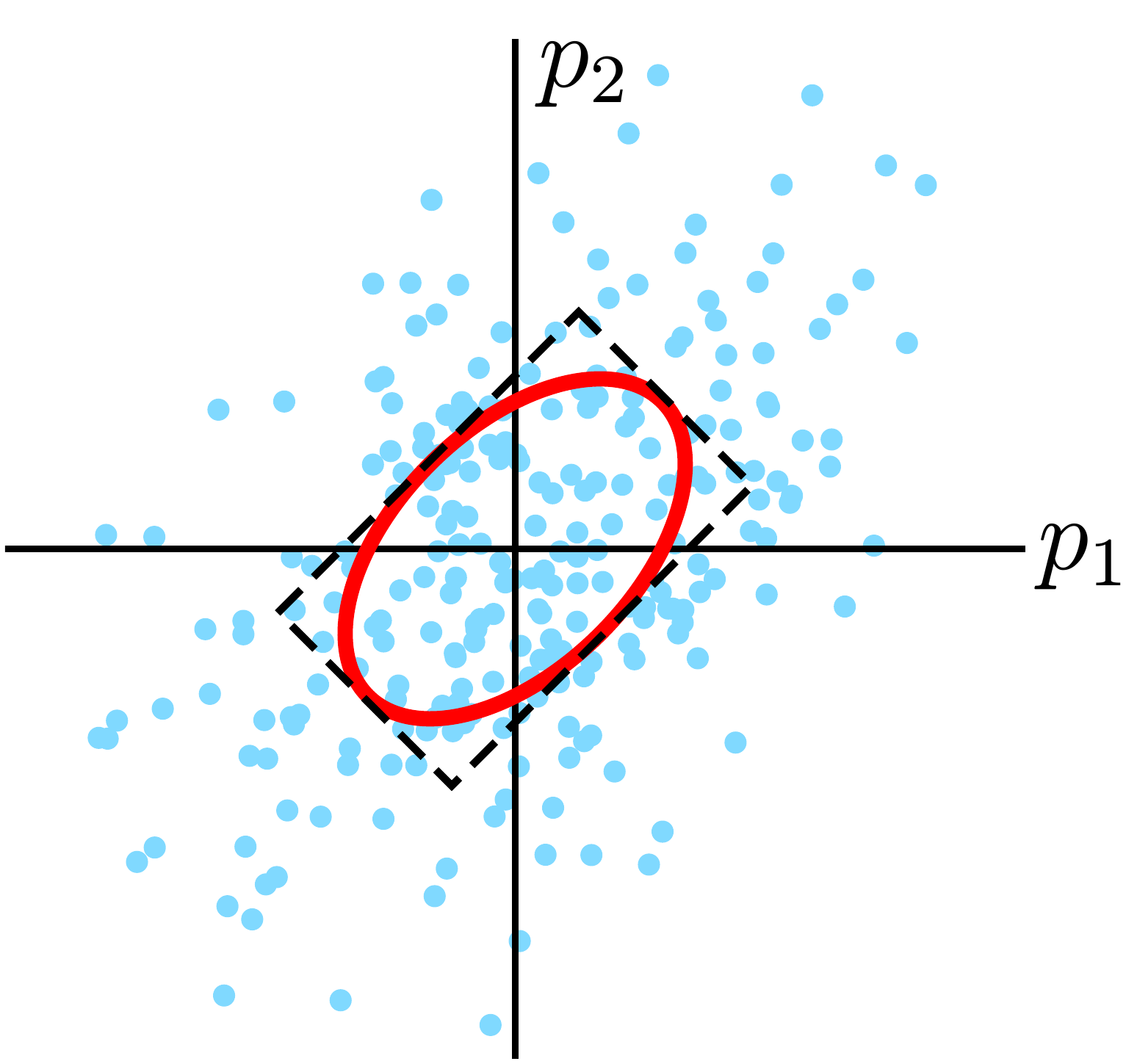} \\
			
			$\left[\!\!\begin{array}{cc}  \,0.2 & \shortminus 0.5 \\ \shortminus 0.5 & \,2  \end{array}\!\right]$ 
			& 2.13, 0.07
			& $\left[\!\!\begin{array}{r}  \shortminus 0.25 \\ 0.97  \end{array}\!\right]$\nsp,
			$\left[\!\begin{array}{r}  0.97 \\ 0.25  \end{array}\!\right]$
			& \includegraphics[width=1.7in]{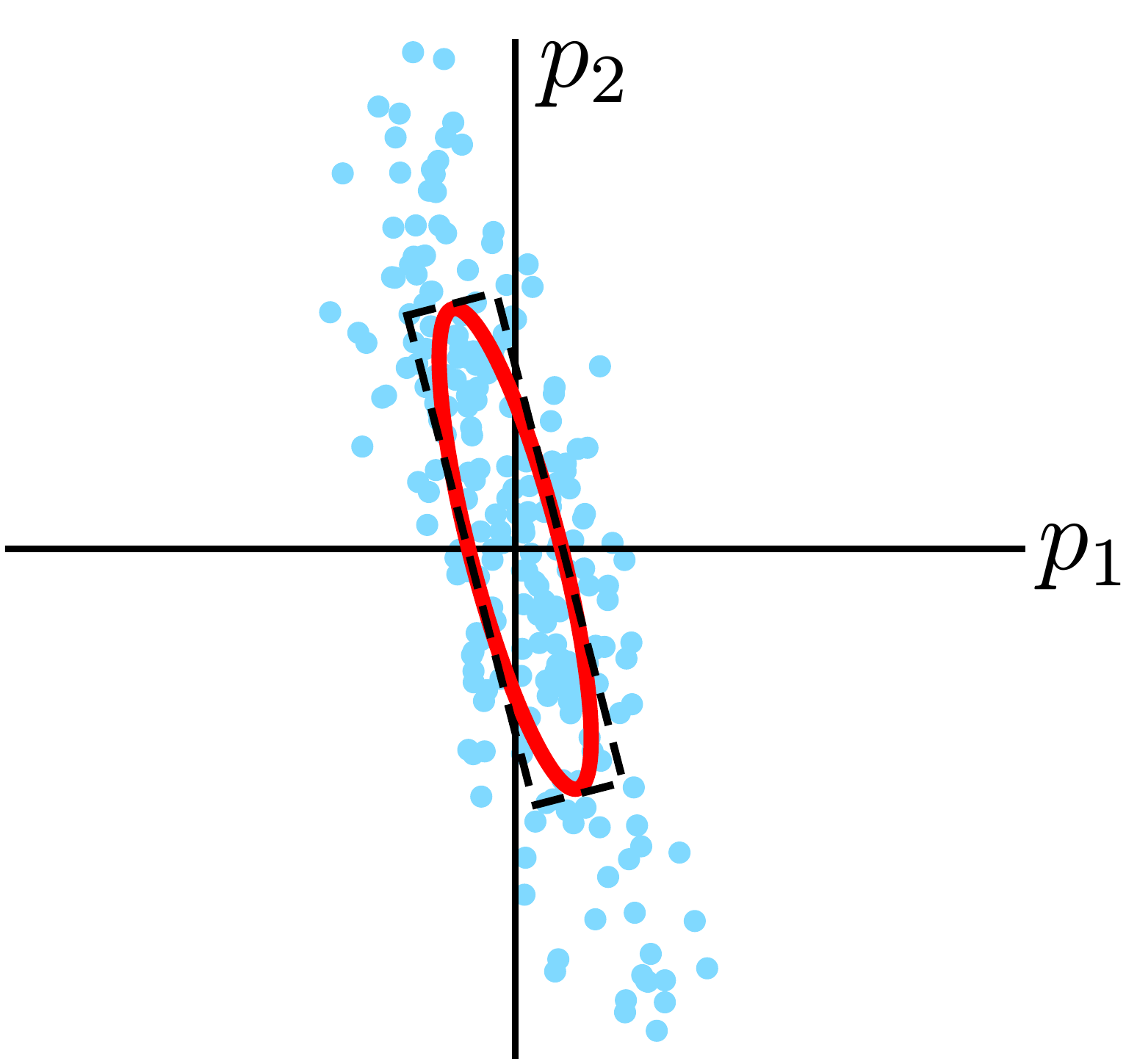}\\
			\bottomrule
		\end{tabular}
		\caption{Plots of the error ellipses associated with four different $2\times 2$ Fisher information matrices. The square roots of the eigenvalues of $\mathbb{J}^{-1}$ determine the semi-axis lengths of the ellipse, i.e., the dimensions of the bounding rectangle, while the eigenvectors determine the orientation. The blue points in each plot represent the estimated parameters from 250 observations of the random variable $Y$ (assuming a bivariate normal distribution) given true parameter values $p_1=p_2=0$. In these examples, $p_1$ and $p_2$ are taken to be unitless, and they are plotted over the range $-3\leq p_1,p_2\leq 3$.}
		\label{tbl:ellipse_examples}
	\end{center}
\end{table}

In summary, the Cram\'er-Rao lower bound can be used to assess the minimum expected error of a maximum likelihood estimate based on the inverse of the expected Fisher information matrix for the measurement. In a similar manner, the FIM can be used to predict and optimize the accuracy of an experiment before any measurements are taken. This is done by minimizing a suitable merit function (chosen based on the desired relative accuracies of each parameter) over the range of interest of $\bp$. It is often convenient to reparametrize $\bp$ to be unitless, such that the intervals $-1\leq p_n\leq 1$ (for $n=1,\ldots,N$) correspond to each physical parameter's range of interest.\footnote{One of the advantages of MLE is that it is invariant to the choice of parametrization \cite{Refregier_2003}.\label{footnote:MLE_parametrization}
} Then one reasonable choice for the merit function would be the product of the eigenvalues of $\mathbb{J}$, which is inversely proportional to the square root of the area (for two parameters) or volume/hypervolume (for three or more parameters) of the error ellipsoid. Another option is the root mean square (RMS) of the eigenvalues of $\mathbb{J}^{-1}$, which is half of the diagonal length of the rectangle/box containing the ellipse/ellipsoid. This second merit function is used in Ref.~\cite{Vella_2018_fbs_arxiv} since it has a lower tendency to heavily prioritize the accuracy of one parameter at the expense of another.

\section{MLE formalism for an optical measurement}\label{sect:MLE_optics}
The MLE formalism is now applied to the optical measurement described previously, in which one or more parameters $\bp$ are to be estimated from a measurement of an intensity distribution $I(\bx;\bp)$. The functional form of $I(\bx;\bp)$ (not to be confused with the measured intensity $\tbI$ defined below) is generally obtained from either a theoretical model, simulated data, experimental calibration data, or some combination thereof. Suppose that the detector is discretized into a finite number of pixels $i=1,2,\ldots$ centered at coordinates $\bx_i$, and assume the pixels are sufficiently small so that $I(\bx;\bp)$ is nearly constant over the area of one pixel. Then, given some vector of true parameter values $\bp$, the probability that a single incident photon will hit the detector at pixel $i$ is prescribed by the normalized intensity distribution:
\begin{equation}
P(i|\bp) = \frac{I(\bx_i;\bp)}{\sum_i I(\bx_i;\bp)},\label{eq:photon_pmf}
\end{equation}
where the sum is taken over all pixels.\footnote{This approximation for small pixels is acceptable for most applications involving sensors with dense pixel arrays. For large pixels, however, one should instead use the exact expression $P(i|\bp)=\lrangle{I}_i/\sum_i\lrangle{I}_i$, where $\lrangle{I}_i$ is the integral of $I(\bx;\bp)$ over the area of pixel $i$. For experiments in which the expected intensity distribution is obtained from a set of calibration images (which themselves are discretized), Eq.~(\ref{eq:photon_pmf}) is an exact result.\label{footnote:P(i|p)_exact}} 
This equation represents the PMF for a single detected photon. Notice that in this context, the outcome of a measurement (denoted as $y$ in the previous section) is the pixel $i$ where a photon is detected. For a classical measurement, each photon detection can be considered as an independent event, so the probability of $M$ photons hitting pixels $i_1,\ldots,i_M$ is given by the product
\begin{equation}
P(i_1\cap\cdots\cap i_M|\bp) = \prod_{m=1}^M P(i_m|\bp).
\end{equation}

Now consider a measured intensity $\tbI=(\tI_1,\tI_2,\ldots)$, where $\tI_i$ is the number of photons detected at pixel $i$. Since the detector is indifferent to the order in which photons arrive (i.e., photons are indistinguishable), the probability of obtaining this distribution is
\begin{equation}
P(\tbI|\bp) = P_0\, \prod_i P(i|\bp)^{\tI_i},\label{eq:P(tbI|p)}
\end{equation}
where the leading factor $P_0=(\sum_i \tI_i)!\, / \prod_i \tI_i!$ accounts for all possible permutations. When regarded as a function of $\bp$, the right-hand side of Eq.~(\ref{eq:P(tbI|p)}) represents the likelihood function $L(\bp|\tbI)$. The log-likelihood is therefore given by
\begin{equation}
\ell(\bp|\tbI) = \ln P_0 + \sum_i \tI_i \ln P(i|\bp).\label{eq:ell(p|tbI)}
\end{equation}
Since $P_0$ is a constant, the maximum likelihood estimate for $\bp$ is obtained by maximizing the sum in the second term of this expression. As described in Section \ref{sect:MLE_overview}, the inverse of the Fisher information matrix places a lower bound on the covariance matrix for this estimate. The expected FIM for a single photon can be calculated using Eq.~(\ref{eq:Fisher_1stders}) or (\ref{eq:Fisher_2ndder}), with $y$ replaced by the pixel index $i$ specifying the photon's location. For a measurement of $\mathcal{N}$ photons, the total information is\footnote{Here the FIM is written in terms of the PMF $P(i|\bp)$ to emphasize the dependence on the normalized intensity distribution. However, the likelihood function $L(\bp|i)$ associated with pixel $i$, which has the same functional form, could also be used. Also, note that in this analysis $\mathcal{N}$ is taken as an integer representing the actual number of measured photons (i.e., the number of photoelectrons registered by the detector), as opposed to the mean or expected number of photons over a particular time interval.}
\begin{subequations}\label{eq:MLE_Fisher}
\begin{align}
[\sp\mathcal{N}\sp\mathbb{J}(\bp)]_{mn} &= \mathcal{N} \sum_i P(i|\bp)
	\left(\mfrac{\partial}{\partial p_m} \ln P(i|\bp)\right)\!
	\left(\mfrac{\partial}{\partial p_n} \ln P(i|\bp)\right)\label{eq:MLE_Fisher_1der}\\
&= -\mathcal{N} \sum_i P(i|\bp) \left(\mfrac{\partial^2}{\partial p_n\partial p_m} \ln P(i|\bp)\right)\nsp.
\end{align}
\end{subequations}
On the other hand, the observed FIM associated with a particular measurement $\tbI$ is obtained by summing the derivatives of $\ell(\bp|\tbI)$ over all detected photons:
\begin{subequations}\label{eq:MLE_Fisher_obs}
	\begin{align}
	\sp[\mathbb{J}^\text{(obs)}(\bp;\tbI)]_{mn} &= \sum_i \tI_i 
	\left(\mfrac{\partial}{\partial p_m} \ln P(i|\bp)\right)\!
	\left(\mfrac{\partial}{\partial p_n} \ln P(i|\bp)\right)\label{eq:MLE_Fisher_obs_1der}\\
	&= - \sum_i \tI_i \left(\mfrac{\partial^2}{\partial p_n\partial p_m} \ln P(i|\bp)\right)\nsp.
	\end{align}
\end{subequations}
Since $\tI_i\approx\mathcal{N}P(i|\bp)$ when a large number of photons are measured, the expected and observed information converge in the limit as $\mathcal{N}\to\infty$, in agreement with the claim made in the previous section. In practice, they should yield nearly identical results in most applications, with the exception of extreme low-light measurements using single-photon detectors.

In the above analysis, it has been implicitly assumed that the detector is capable of measuring any arbitrary number of photons incident on a pixel, i.e., that it can resolve individual photons. However, most real detectors have a finite bit depth, meaning that they can only resolve some finite number of distinct intensity levels. For example, in an 8-bit sensor, each pixel has an integer readout value between 0 and 255. This discretization of pixel values is analogous to the discreteness of photons; therefore, in this situation, Eqs.~(\ref{eq:P(tbI|p)}) through (\ref{eq:MLE_Fisher_obs}) can be used with $\tI_i$ interpreted as the readout value of pixel $i$. In the absence of thermal noise or other sources of error, the equivalent ``photon count'' of the signal from a sensor with finite bit depth must be less than or equal to $\mathcal{N}$, the actual number of photons incident on the detector. As needed, the effective bit depth of the sensor can be increased by averaging the output signal over multiple exposures. This time-averaging has the added benefit of reducing the impact of electronic shot noise.

\section{Comparison to Bayesian statistics}\label{sect:Bayesian_statistics}
The method of MLE is considered a ``frequentist'' approach in the sense that it does not assign a probability distribution to the unknown parameter $\bp$, but rather it estimates the value of $\bp$ that is most consistent with the observed data. A popular alternative is the Bayesian approach, which is predicated on the calculation of a posterior probability density function (PDF) $P(\bp|\tbI)$ describing the probability of every possible value of $\bp$ given an observed intensity $\tbI$. In general, $P(\bp|\tbI)$ depends on a prior distribution $P(\bp)$ as well as the observed intensity. The prior distribution $P(\bp)$ may be uniformly distributed (i.e., constant), or it may be used to introduce known (or assumed) information about $\bp$ before the measurement takes place. For example, in the polarimetry experiment discussed in Ref.~\cite{Ramkhalawon_2013} (with $\bp=(p_1,p_2,p_3)$ representing the normalized Stokes parameters), $P(\bp)$ could be used to incorporate prior knowledge about the source's polarization. Another example is the focused beam scatterometry experiment discussed in Ref.~\cite{Vella_2018_fbs_arxiv}, in which it might be possible in some cases to assign a prior distribution $P(\bp)$ based on the fabrication process of the sample under test.

Using Bayes' theorem, the posterior PDF can be written as
\begin{equation}
P(\bp|\tbI) = \frac{P(\bp)}{P(\tbI)} P(\tbI|\bp),\label{eq:P(p|tbI)}
\end{equation}
where the constant term in the denominator, given by
\begin{equation}
P(\tbI) = \int\nsp P(\bp) P(\tbI|\bp)\sp \ud^Np,
\end{equation}
ensures the normalization condition $\int\nsp P(\bp|\tbI)\sp\ud^Np=1$. Substituting Eq.~(\ref{eq:P(tbI|p)}) into Eq.~(\ref{eq:P(p|tbI)}), one obtains
\begin{subequations}
\begin{align}
P(\bp|\tbI) &= \frac{P(\bp)}{P(\tbI)}P_0 \prod_i P(i|\bp)^{\tI_i} \\[2pt]
&= \frac{P(\bp)}{P(\tbI)}P_0 \exp\!\left(\nsp\sum_i\, \tI_i \ln P(i|\bp) \nsp\right)\nsp.
\end{align}
\end{subequations}
Notice that $P(\bp|\tbI)$ is proportional to the prior distribution times the likelihood. If no prior information is assumed about $\bp$ (as is the case for all examples discussed throughout this tutorial), then $P(\bp)$ is constant and the peak of $P(\bp|\tbI)$ coincides with the maximum likelihood estimate for $\bp$. More generally, if $P(\bp)$ is nonuniform, the two values converge in the limit as $\mathcal{N}\to\infty$, assuming that $P(\bp)$ is smooth and nonzero near the true value of $\bp$.

As discussed in Ref.~\cite{Ramkhalawon_2013}, if the measurement is limited by photon noise (as opposed to other noise mechanisms or systematic errors) and $\mathcal{N}$ is large, then $P(\bp|\tbI)$ is approximately a narrow, generally anisotropic Gaussian distribution that is maximized by the true parameter values $\bp_0$:
\begin{equation}
P(\bp|\tbI) \propto \exp\!\left[-\tfrac{1}{2}(\bp-\bp_0)^{\rm T}{\bm\Sigma}^{-1}(\bp-\bp_0)\right]\!.\label{eq:MLE_P(p|I)_Gaussian}
\end{equation}
Here the covariance matrix ${\bm\Sigma}$ determines the shape and width of the distribution, and its inverse ${\bm\Sigma}^{-1}$ is the Hessian matrix of second derivatives of $\ln \nsp P(\tbI|\bp)$ evaluated at $\bp_0$. Recalling the results of the previous sections, one can see that if $P(\bp)$ is constant, then ${\bm\Sigma}^{-1}$ is equal to the observed FIM $\mathbb{J}^\text{(obs)}(\bp_0;\tbI)$, and its expected value (taken over all possible outcomes for $\tbI$) is the expected FIM $\mathbb{J}(\bp_0)$. Intuitively, a measurement with high information content, for which the FIM is large and nearly diagonal, will result in a narrow posterior distribution $P(\bp|\tbI)$, enabling a precise estimate of $\bp$. Thus, even in a Bayesian framework, the maximum likelihood estimate and the Fisher information matrix can both be shown to have clear statistical meanings.

\section{One-parameter optical MLE examples}\label{sect:MLE_examples_1param}
This section contains a series of four simple thought experiments involving one-dimensional intensity distributions $I_j(x;p_1)$ (where $j=1,2,3,4$) that depend on a single parameter $p_1$. Without loss of generality, let us assume that $p_1$ is unitless and that its range of interest is $-1\leq p_1\leq 1$. (As noted on page \pageref{footnote:MLE_parametrization}, any physical parameter can be reparametrized in this way without affecting the MLE.) The one-dimensional coordinate $x$ is also taken to be unitless. In the examples that follow, the function
\begin{equation}
\Pi(x) = 
\begin{cases}
I_0,&-1\leq x\leq 1,\\
0 & \text{otherwise},
\end{cases}
\end{equation}
where $I_0$ represents some reference intensity level, is used as a normalization factor that also serves to limit each intensity distribution to the spatial extent of the sensor (as if the beam were truncated by a hard aperture). Each intensity distribution is normalized such that it reaches a maximum value of $I_0$ over the range of interest of $p_1$. Note, however, that this does not preclude the possibility of intensities greater than $I_0$ when $|p_1|>1$.

For simplicity, suppose that the detector consists of a one-dimensional array of 9 pixels, with pixel $i$ centered at coordinate $x_i=(i-5)/4$, so that
\begin{equation}
(x_1,\ldots,x_9)=(-1,-0.75,-0.5,-0.25,0,0.25,0.5,0.75,1).\label{eq:MLE_examples_u_coords}
\end{equation}
According to Eq.~(\ref{eq:photon_pmf}), the probability of an incident photon hitting pixel $i$ is 
\begin{equation}
P_j(i|p_1) = \frac{I_j(x_i;p_1)}{\sum_i I_j(x_i;p_1)}.\label{eq:P(i|p1)}
\end{equation}
As mentioned earlier, for such a sparse array of pixels, this is a relatively poor approximation since the intensity may vary significantly over the width of each pixel. However, since the approximation is reasonable for most real applications, it is used here for instructive purposes. If desired, the exact expression for $P_j(i|p_1)$ (which is provided in footnote \ref{footnote:P(i|p)_exact} following Eq.~(\ref{eq:photon_pmf})) could be substituted into the analysis with minimal modifications required. Similarly, while the concepts of Fisher information and the Cram\'er-Rao bound are usually applied to measurements consisting of many observations (photons), the calculations below are demonstrated for measurements of just a few photons and then extended to larger sample sizes. Also note that while the following examples all involve intensity distributions over a 1D spatial coordinate, the more general 2D case can be treated in the same manner by rearranging the numerical output of the detector's 2D pixel array into a 1D array during signal processing.

The intensity distributions considered in each of the following sections are summarized in Table \ref{tbl:MLE_1param_intensity_dist}. 
\begin{table}[b]
	\renewcommand{\arraystretch}{.8}
	\begin{center} 
		\begin{tabular}{c@{\hspace{20pt}}l}
			\toprule
			Section & Intensity distribution\\
			\midrule
			\phantom{a}&\\[-8pt]
			\ref{sect:MLE_example1} & $I_1(x;p_1)=\Pi(x)(0.5+0.5\, p_1 x)$\\[8pt]
			\ref{sect:MLE_example2} & $I_2(x;p_1)=\Pi(x)(0.9+0.1\, p_1 x)$\\[8pt]
			\ref{sect:MLE_example3} & $I_3(x;p_1)=\Pi(x)\mfrac{1}{(|c|+1)^2}(p_1-cx)^2$, where $c=\text{constant}$\\[12pt]
			\ref{sect:MLE_example4} & $I_4(x;p_1)=\Pi(x)\mfrac{1}{(|d|+2)^2}(p-x-d)^2$, where $d=\text{constant}$\\
			\bottomrule
		\end{tabular}
		\caption{Intensity distributions for each example considered in Section \ref{sect:MLE_examples_1param}.}
		\label{tbl:MLE_1param_intensity_dist}
	\end{center}
\end{table}
In Section \ref{sect:MLE_example1}, an in-depth analysis is performed for a simple intensity distribution that depends linearly on $p_1$. In Section \ref{sect:MLE_example2}, the results are compared to a similar intensity distribution with a weaker linear dependence on $p_1$. Next, the commonly-used experimental configurations of null and off-null measurements are explored in Section \ref{sect:MLE_example3}. Finally, Section \ref{sect:MLE_example4} examines the case of an intensity that may be far from perfect nulling conditions, and the results are compared to the near-null case.

\subsection{Linear dependence on $p_1$}\label{sect:MLE_example1}
For the first example, consider the intensity distribution
\begin{equation}
I_1(x;p_1) = \Pi(x)\bigl(0.5+0.5\,p_1x\bigr).\label{eq:MLE_I_Ex1}
\end{equation}
The distribution is only valid when $-1\leq p_1\leq 1$ since larger parameter values would result in negative intensity values, which are not allowed. This is an extreme case of a common real-world scenario in which an approximation is made for the intensity that is only valid over some range of parameter values (for example, the quadratic approximation seen in Ref.~\cite{Vella_2018_fbs_arxiv}). In practice, for reliable parameter estimation, the range of interest of $\bp$ should be smaller than the region where the approximation is valid (within some prescribed accuracy).

\begin{figure} 
	\centering
	\includegraphics[width=\linewidth]{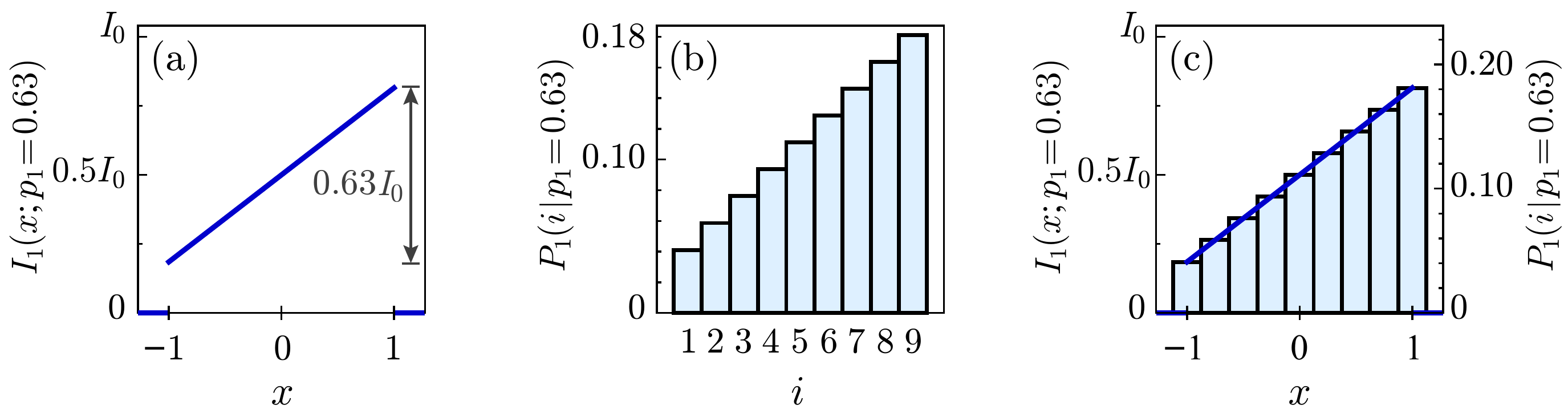}
	\caption{(a) Linear intensity distribution $I_1(x;p_1)$ and (b) the corresponding PMF for each pixel $i$, both shown for the case that $p_1=0.63$. The two plots are shown together in part (c). For practical reasons, the axis labels for $i$ are excluded from the combined plot. In all subsequent figures, the vertical axis labels are also omitted to reduce clutter.}
	\label{fig:MLE_0.63_IntPlot_Ex1}
\end{figure}

\begin{figure} 
	\centering
	\includegraphics[width=\linewidth]{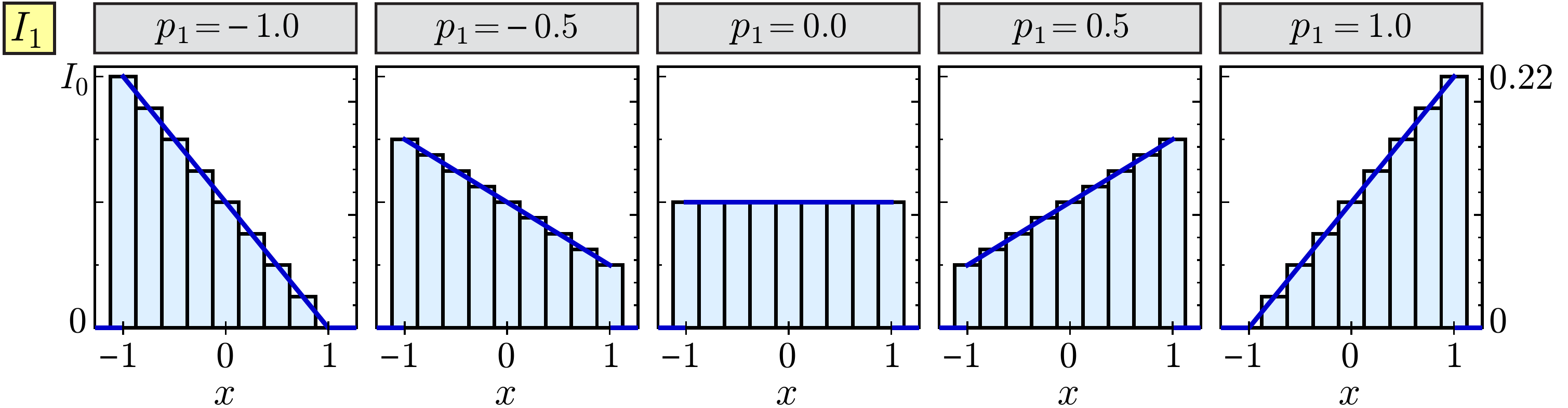}
	\caption{Plots of $I_1(x;p_1)$ (left axis) and $P_1(i|p_1)$ (right axis) for several values of $p_1$.}
	\label{fig:MLE_IntPlots_Ex1}
\end{figure}

Using Eq.~(\ref{eq:P(i|p1)}), it is straightforward to calculate the PMF for a detected photon:%
\begin{equation}
P_1(i|p_1) = \frac{1}{9}\left(1+\frac{i-5}{4}\,p_1\right)\nsp.\label{P1(i|p1)}
\end{equation}
The continuous intensity distribution $I_1(x;p_1)$ and discrete PMF $P_1(i;p_1)$ are plotted in Figs.~\ref{fig:MLE_0.63_IntPlot_Ex1}(a) and \ref{fig:MLE_0.63_IntPlot_Ex1}(b) for the case that $p_1=0.63$. To visualize the relationship between the intensity and PMF, it is useful to combine the two plots with appropriately chosen scales, as seen in Fig.~\ref{fig:MLE_0.63_IntPlot_Ex1}(c). 
The dependence of each quantity on $p_1$ is illustrated in Fig.~\ref{fig:MLE_IntPlots_Ex1}, which contains plots of $I_1(x;p_1)$ and $P_1(i|p_1)$ for five different parameter values over the range of interest.

\begin{figure} 
	\centering
	\includegraphics[width=.65\linewidth]{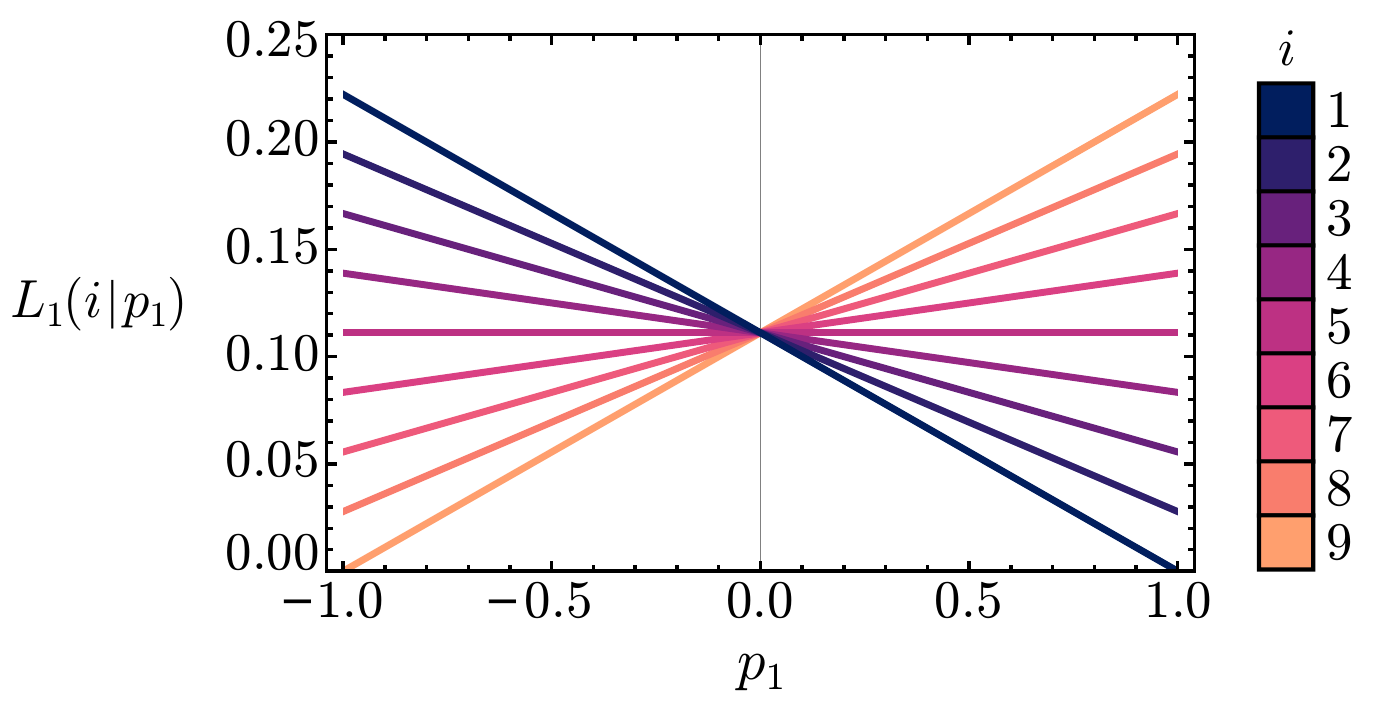}
	\caption{Likelihood functions $L_1(i|p_1)$ associated with each pixel $i$ in a measurement with theoretical intensity distribution $I_1(p_1)$.}
	\label{fig:MLE_Lplot1}
\end{figure}

As discussed previously, the likelihood function $L_1(p_1|i)$ has the same algebraic form as $P_1(i|p_1)$, but it is regarded as a continuous function of $p_1$. The likelihood functions associated with individual photons detected at each pixel $i=1,\ldots,9$ are plotted in Fig.~\ref{fig:MLE_Lplot1}.
To illustrate the procedure of calculating the MLE from the likelihood function, let us now consider a simulated measurement of the intensity for which the true parameter value is $p_1=0.63$. The simulated intensity $\tbI$ is constructed by randomly selecting individual photons according to the probability distribution $P(i|p_1\!=\!0.63)$ that was shown previously in Fig.~\ref{fig:MLE_0.63_IntPlot_Ex1}(b). For demonstrative purposes, suppose that the sensor is capable of detecting individual photons, even though this is typically not the case in real experiments where many photons accumulate within the sensor's exposure time. This will allow us to examine the influence of each photon on the likelihood and the MLE, as well as the evolution of the MLE as photons accumulate.

Suppose that the first simulated photon hits the detector at pixel 1. From Eq.~(\ref{P1(i|p1)}), the likelihood of this event is found to be $L_1(p_1|i\nsp =\nsp 1)=\frac{1}{9}(1-p_1)$. The MLE based on this single photon is obtained by maximizing the likelihood with respect to $p_1$. This example illustrates the fact that the MLE is not guaranteed to exist in general, since $L_1(p_1|i=1)$ would be unbounded if $p_1$ were allowed to take any real value. A sufficient condition for the existence of an MLE is that the parameter space is compact \cite{vanderVaart_1992,Demidenko_1999}, such as the closed interval $p_1\in[-1,1]$. Within this interval, the likelihood function is maximized by $p_1=-1$.\footnote{Note that the condition of compactness is sufficient but not necessary. In fact, in the present example, the restriction quickly becomes unnecessary as soon as multiple photons are detected at different pixels. Another example is the polarimetry application in Ref.~\cite{Ramkhalawon_2013}, in which the Stokes parameters are restricted to the interval $[-1,1]$ by definition, guaranteeing the existence of an MLE.} Notice from Fig.~\ref{fig:MLE_Lplot1} that a single photon detected at pixel 2, 3, or 4 also would have produced the same MLE, albeit with lower confidence.
 
Now suppose that a second photon is detected at pixel 7, so that the measured intensity becomes $\tbI=(1,0,0,0,0,0,1,0,0)$. The likelihood function associated with this second photon is $L_1(p_1|i=7)=\frac{1}{9}(1-\frac{1}{2}p_1)$. Using Eq.~(\ref{eq:P(tbI|p)}) (and remembering that the probability and likelihood are algebraically equivalent), the likelihood of measuring this two-photon intensity distribution is
\begin{equation}
L_1(p_1|\tbI) = \frac{2!}{1!\nsp\times\nsp 1!}L_1(p_1|i=1) L_1(p_1|i=7)=\frac{1}{81}(-p_1^2-p_1+2).
\end{equation}
It is easy to show that this function is maximized when $p_1=-0.5$, which becomes the new MLE.  Similarly, suppose that a third photon is detected, also at pixel 7, so that the measured intensity becomes $\tbI=(1,0,0,0,0,0,2,0,0)$. The likelihood of measuring this intensity distribution is
\begin{equation}
L_1(p_1|\tbI) = \frac{3!}{1!\nsp\times\nsp 2!}L_1(p_1|i=1) L_1(p_1|i=7)^2=\frac{1}{972}(-p_1^3-3p_1^2+4),
\end{equation}
which is maximized when $p_1=0$.

\begin{figure}[tb] 
	\centering
	\includegraphics[width=\linewidth]{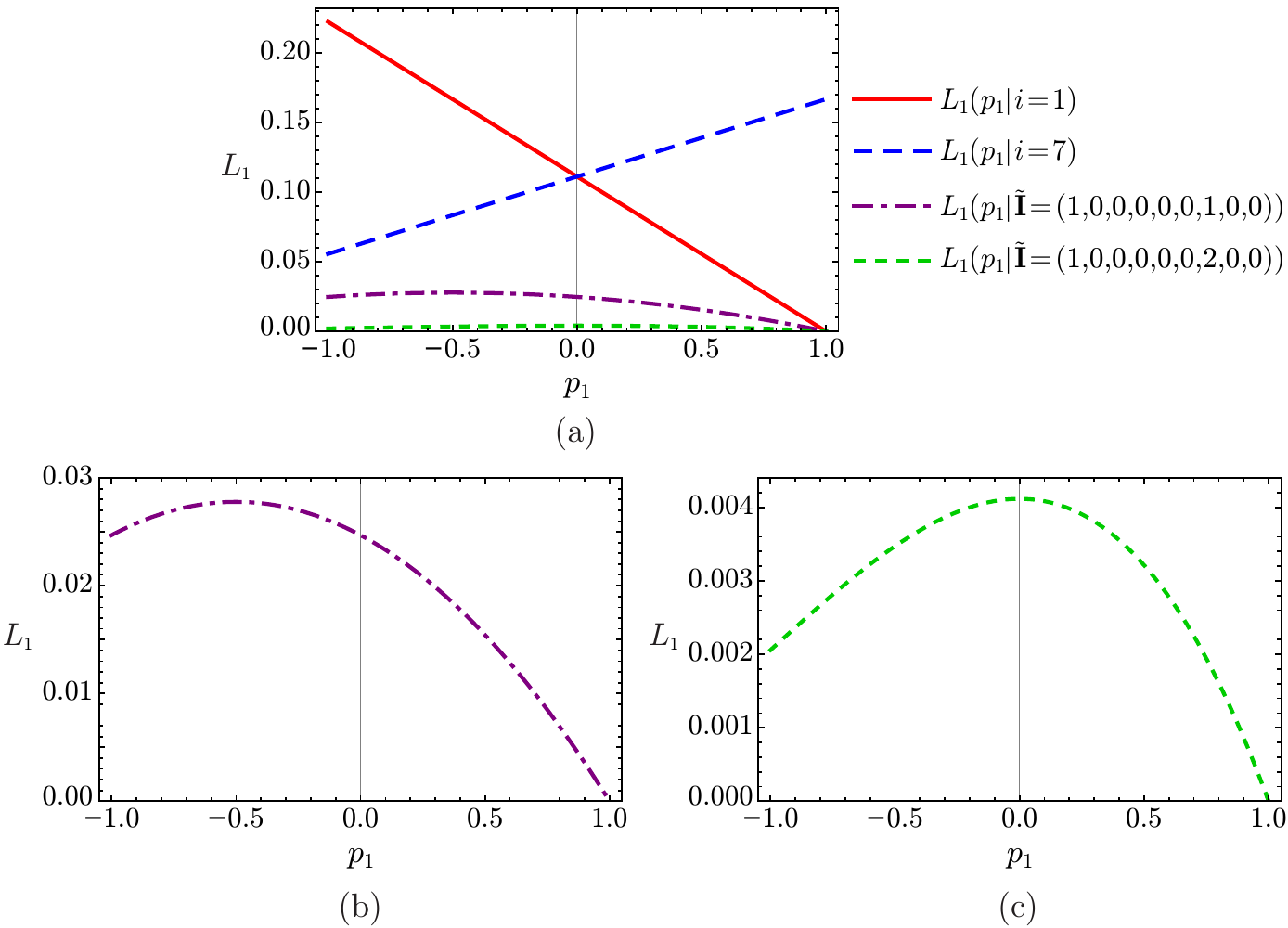}
	\caption{(a) Likelihood functions (based on intensity distribution $I_1$) for detected photons at pixels $i=1$ and $i=7$ and for intensity measurements consisting of one photon at pixel 1 and one or two photons at pixel 7. The two- and three-photon likelihoods are also shown on independent scales in plots (b) and (c).}
	\label{fig:MLE_L3ph_Ex1}
\end{figure}

The likelihood functions for individual photons at pixels 1 and 7 are plotted in Fig.~\ref{fig:MLE_L3ph_Ex1}(a), as well as the likelihoods of the two- and three-photon intensity distributions from above. The latter two functions are also plotted separately in Fig.~\ref{fig:MLE_L3ph_Ex1}(b,c). 
From these plots one can see the effect of each photon: as photons are detected at pixel 1, then pixel 7, then pixel 7 again, the peak of the likelihood function shifts from $p_1=-1$ to $p_1=-0.5$ to $p_1=0$. Additionally, the distribution becomes more sharply peaked with each accumulated photon, reducing the uncertainty in the MLE. This uncertainty can be quantified by using Eq.~({\ref{eq:MLE_Fisher_obs}) to calculate the observed Fisher information, which is a $1\times 1$ ``matrix'' (i.e., a scalar) in the one-parameter case. For example, for the three-photon measurement $\tbI=(1,0,0,0,0,0,2,0,0)$, Eq.~({\ref{eq:MLE_Fisher_obs_1der}) yields
\begin{align}
J_1^\text{(obs)}(p_1;\tbI) &= \sum_i \tI_i \left(\frac{\partial}{\partial p_1}\ln P(i|p_1)\right)^{\!2}\nonumber\\[2pt]
&= \left(\left.\frac{i-5}{4+(i-5)p_1}\right|_{i=1}\right)^{\!2} + 
 2\left(\left.\frac{i-5}{4+(i-5)p_1}\right|_{i=7}\right)^{\!2} \nonumber\\[2pt]
 &=\frac{1}{(p_1-1)^2} + \frac{2}{(p_1+2)^2},
\end{align}
which produces $J_1^\text{(obs)}=1.5$ when evaluated at the MLE $p_1=0$. In the one-parameter case, the eigenvalue of the ``matrix'' $J_1^\text{(obs)}$ is just the value of $J_1^\text{(obs)}$ itself. Therefore, the minimum expected standard deviation uncertainty of the measurement is 
$1/\sqrt{1.5}=0.816$. Considering the fact that only three photons were detected, this large uncertainty (relative to the range of interest) is not surprising.

Alternatively, using Eq.~({\ref{eq:MLE_Fisher_1der}), the minimum error for a measurement of $\mathcal{N}$ photons (independent of the specific outcome of the measurement) can be quantified by calculating the expected Fisher information
\begin{equation}
\mathcal{N}J_1(p_1)=\frac{\mathcal{N}}{36}\sum_{i=1}^{9}\frac{(i-5)^2}{4+(i-5)p_1}.\label{eq:Fisher_MLE_Ex1}
\end{equation}
For example, for a three-photon measurement with MLE $p_1=0$, the expected standard deviation error is $[3J_1(0)]^{-1/2}=0.894$. Keep in mind, however, that the expected Fisher information is not necessarily appropriate for a measurement containing very few photons. As seen in Fig.~\ref{fig:MLE_Fisher_Ex1}, $J_1(p_1)$ grows infinitely large in the limit that $|p_1|\to 1$, implying that the uncertainty approaches zero. 
\begin{figure} 
	\centering
	\includegraphics[height=2.25in]{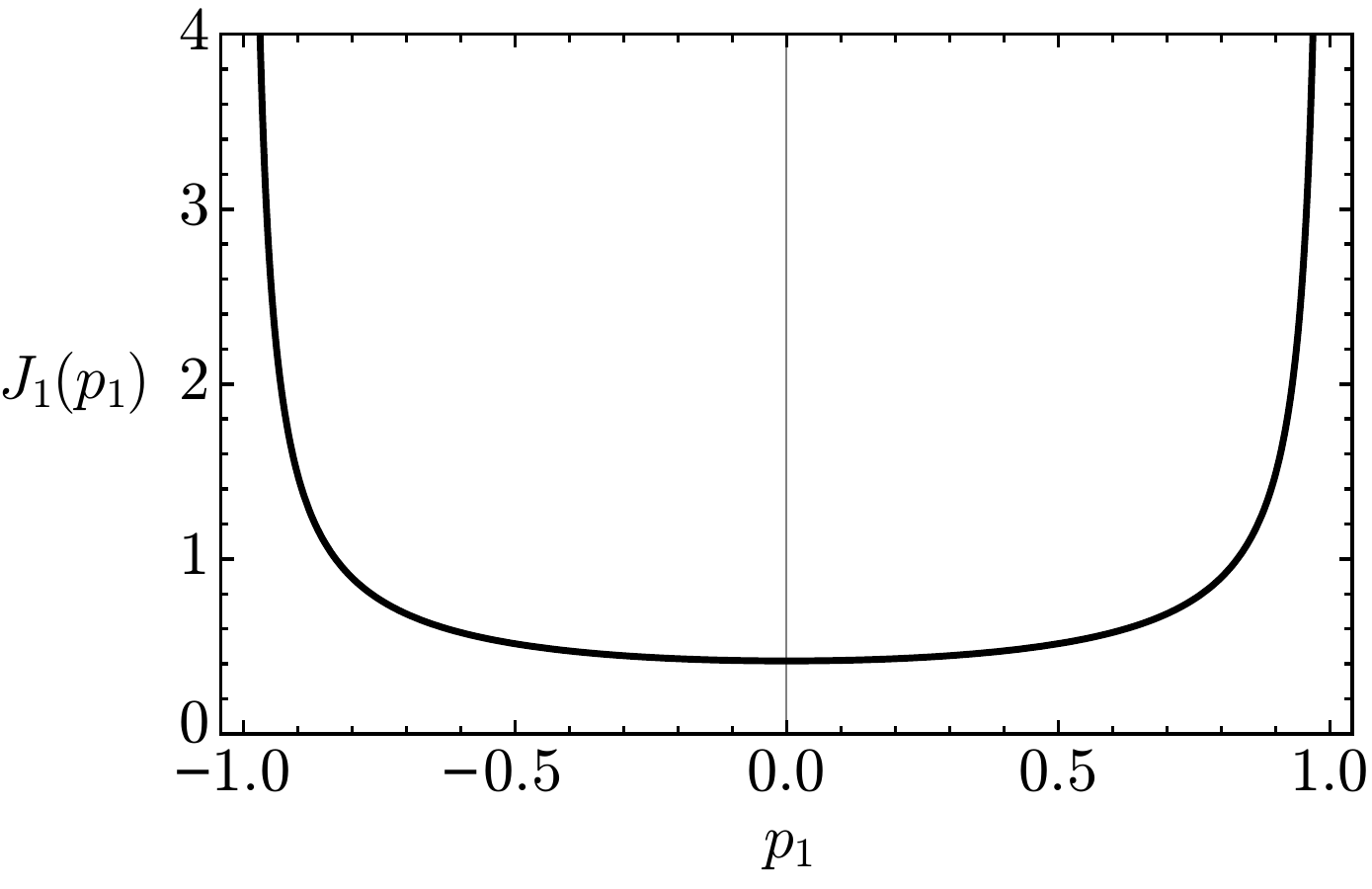}
	\caption{Expected unit Fisher information for a measurement of $I_1(x;p_1)$.}
	\label{fig:MLE_Fisher_Ex1}
\end{figure}
Although this is a meaningful limit for the case of large $\mathcal{N}$, it would clearly be nonsensical to suggest that a single photon could produce an MLE with zero uncertainty! 

To observe these concepts on a larger scale, suppose that the simulation continues until 100,000 photons have accumulated. For a single random trial of the experiment, Table \ref{tbl:MLE_ex1_photonsim} contains the measured intensities and corresponding MLEs obtained throughout the simulation for several values of $\mathcal{N}$. Notice that the MLE approaches the true parameter value ($p_1=0.63$) as $\mathcal{N}$ increases. As seen in Fig.~\ref{fig:MLE_LL_Ex1}, the log-likelihood function $\ell_1(p_1|\tbI)$ becomes increasingly narrow as photons accumulate, and its shape becomes approximately parabolic; therefore, the likelihood $L_1(p_1|\tbI)$ approaches a Gaussian distribution, i.e., an exponentiated concave-downward quadratic function. Furthermore, as observed above, the location of the peak likelihood (which by definition determines the MLE) approaches the true parameter value. The MLE is plotted against $\mathcal{N}$ in Fig.~\ref{fig:MLE_conf_Ex1}, with shaded regions representing the standard deviation confidence intervals based on the expected and observed Fisher information. Notice that as $\mathcal{N}$ increases, not only does the MLE approach the true value of $p_1$ with increasing confidence, but the expected and observed information rapidly converge. 

\begin{table}
	\begin{center} 
		\begin{tabular}{l@{\hspace{6pt}}lM{0in}l}
			\toprule
			$\mathcal{N}$ & MLE $(p_1)$ && $\tbI=(\tI_1,\ldots,\tI_9)$ \\
			\midrule
			1 & $-1.0000$ && $(1,0,0,0,0,0,0,0,0)$ \\
			2 & $-0.5000$ && $(1,0,0,0,0,0,1,0,0)$ \\
			3 & $\phantom{-}0.0000$ && $(1,0,0,0,0,0,2,0,0)$  \\
			4 & $\phantom{-}0.3187$ && $(1,0,0,0,0,0,2,1,0)$ \\
			5 & $\phantom{-}0.5024$ && $(1,0,0,0,0,0,2,1,1)$ \\
			6 & $\phantom{-}0.5429$ && $(1,0,0,0,0,1,2,1,1)$ \\
			7 & $\phantom{-}0.6187$ && $(1,0,0,0,0,1,2,2,1)$ \\
			8 & $\phantom{-}0.6727$ && $(1,0,0,0,0,1,2,3,1)$ \\
			9 & $\phantom{-}0.6916$ && $(1,0,0,0,0,2,2,3,1)$ \\
			10 & $\phantom{-}0.6646$ && $(1,0,0,1,0,2,2,3,1)$ \\
			100 & $\phantom{-}0.7114$ && $(6,1,8,9,8,9,15,19,25)$ \\
			1000 & $\phantom{-}0.6656$ && $(41,56,64,91,112,121,166,160,189)$ \\
			10000 & $\phantom{-}0.6243$ && $(413,583,784,956,1112,1262,1446,1615,1829)$ \\
			100000 & $\phantom{-}0.6329$ && $(4009,5847,7696,9460,11151,12839,14588,16160,18250)$ \\
			\bottomrule
		\end{tabular}
		\caption{Evolution of the MLE for $p_1$ and the measured intensity distribution $\tbI$ as individual photons accumulate for a simulated measurement of $I_1(x;p_1)$ with true parameter value $p_1=0.63$.}
		\label{tbl:MLE_ex1_photonsim}
	\end{center}
\end{table}

\begin{figure} 
	\centering
\includegraphics[height=2.5in]{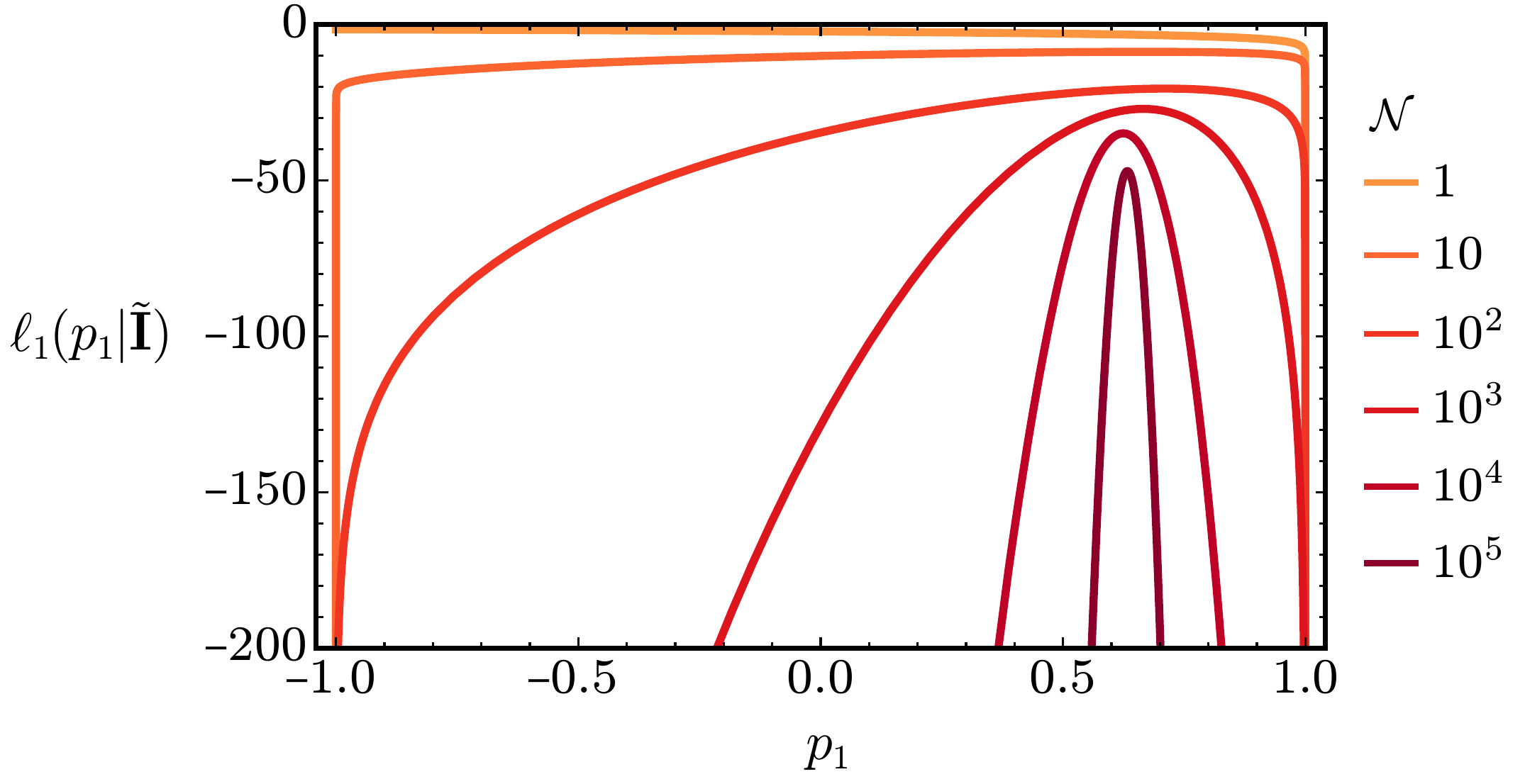}
	\caption{Log-likelihood functions associated with the simulated intensities listed in Table \ref{tbl:MLE_ex1_photonsim}.}
	\label{fig:MLE_LL_Ex1}
\end{figure}

\begin{figure} 
	\centering
	\includegraphics[width=.75\linewidth]{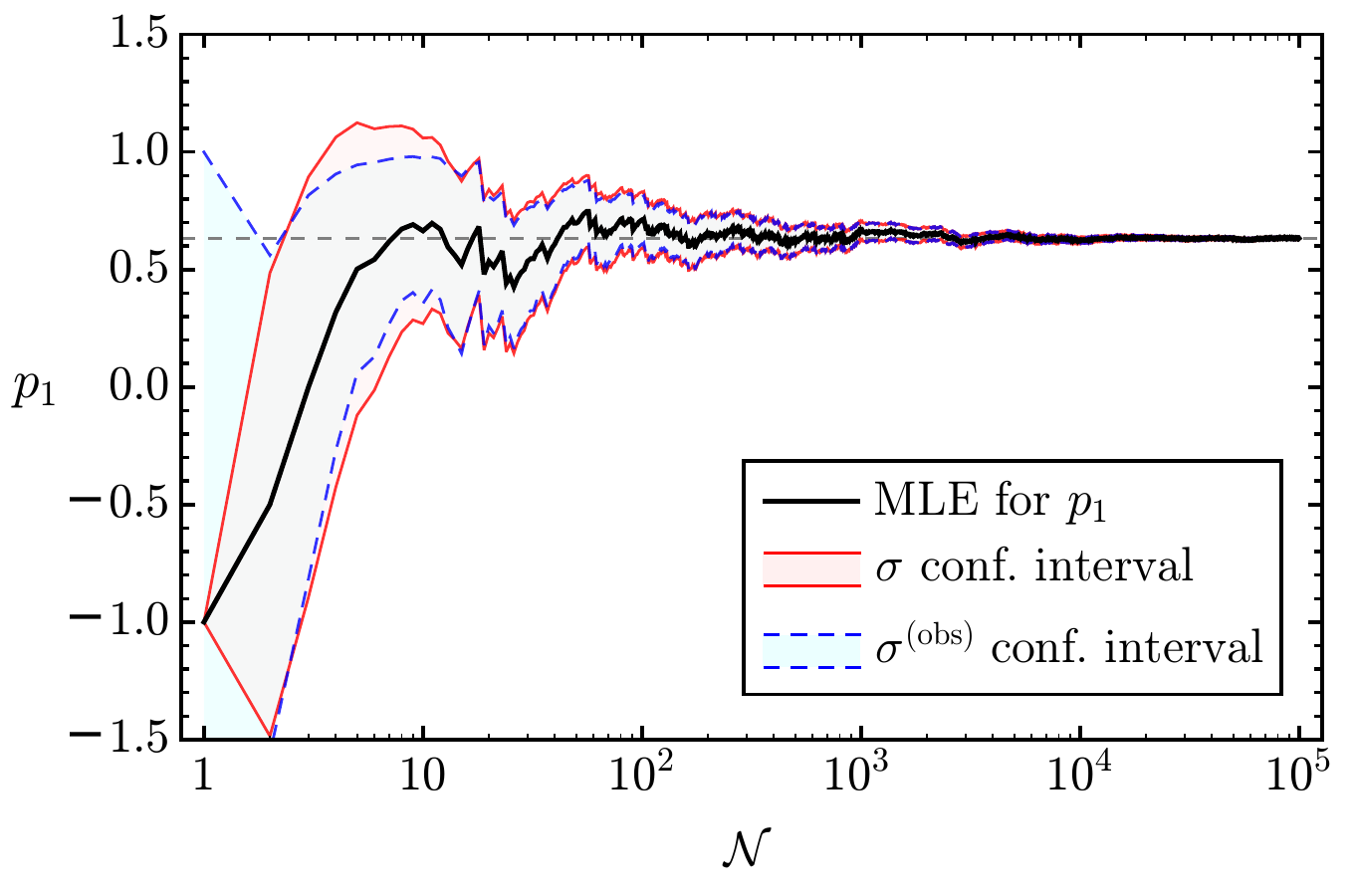}
	\caption{Evolution of the maximum likelihood estimate and standard deviation confidence interval for $p_1$ as 100,000 photons accumulate for a simulated measurement of $I_1(x;p_1)$ with true parameter value $p_1=0.63$. The solid red and dashed blue regions represent the confidence intervals based on the expected and observed Fisher information, respectively.}
	\label{fig:MLE_conf_Ex1}
\end{figure}

Although the above simulation is a representative example of the behavior of the MLE, it is merely a single observation of a random process. To gain a broader view of the statistical behavior of $I_1(x;p_1)$, a Monte Carlo simulation of 50,000 trials of a 100-photon intensity measurement was performed, first for a true parameter value of $p_1=0$ and then for $p_1=0.63$. The results of the simulations are plotted in Figs.~\ref{fig:MLE_retr_hist_Ex1}(a) and \ref{fig:MLE_retr_hist_Ex1}(b), which contain histograms showing the distribution of the MLE over all trials. 
\begin{figure} 
	\centering
	\includegraphics[width=.8\linewidth]{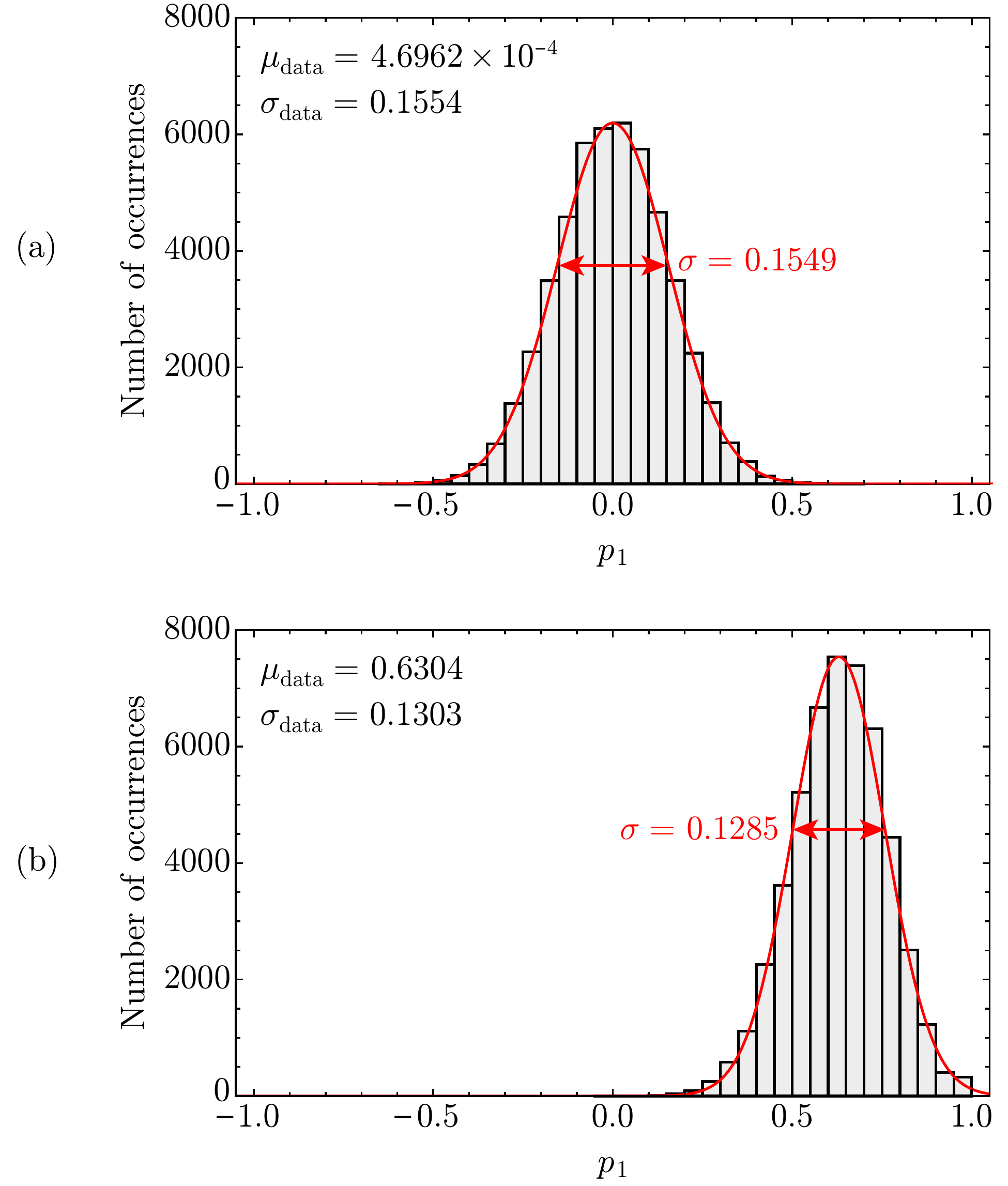}
	\caption{Histograms of the maximum likelihood estimates obtained from 50,000 trials of a 100-photon simulation of $I_1(x;p_1)$ with true parameter values (a) $p_1=0$ and (b) $p_1=0.63$. The mean ($\mu_\text{data}$) and standard deviation ($\sigma_\text{data}$) of each distribution are indicated in the upper left corner of the plot. For comparison, a normal distribution with mean $p_1$ and standard deviation $\sigma=[100J_1(p_1)]^{-1/2}$ is overlaid in red; the value of $\sigma$ is indicated alongside each curve.}
	\label{fig:MLE_retr_hist_Ex1}
\end{figure}
As seen in the upper left corner of each plot, the mean MLE over all trials differs from the true parameter value by less than 0.001. The standard deviations of the MLEs obtained for the $p_1=0$ and $p_1=0.63$ cases are 0.1554 and 0.1303, respectively. In comparison, using Eq.~(\ref{eq:Fisher_MLE_Ex1}), the expected Fisher information for the $p_1=0$ case is $100J_1(0)=41.67$, corresponding to a standard deviation error of $0.1549$. Similarly, the expected error for the $p_1=0.63$ case is found to be $0.1285$. These values closely agree with the results of the simulation. To help visualize this, a normal distribution with the expected standard deviation is overlaid in red on top of each histogram in Fig.~\ref{fig:MLE_retr_hist_Ex1}; notice that each curve almost exactly matches the distribution of MLEs over 50,000 trials.

\subsection{Weaker linear dependence on $p_1$}\label{sect:MLE_example2}
For the next example, consider the intensity distribution
\begin{equation}
I_2(x;p_1) = \Pi(x)\bigl(0.9+0.1\,p_1x\bigr),\label{eq:MLE_I_Ex2}
\end{equation}
which is valid when $-9\leq p_1\leq 9$. (However, the range of interest is still $-1\leq p_1\leq 1$.) Using Eq.~(\ref{eq:P(i|p1)}), the PMF for a single photon is
\begin{equation}
P_2(i|p_1) = \frac{1}{9}\left(1+\frac{i-5}{36}\,p_1\right)\nsp.\label{P2(i|p1)}
\end{equation}
This distribution is nearly the same as the first example except that the linear $p_1$ term is 9 times smaller. As a result, the variations in intensity, PMF, and likelihood with respect to $p_1$ have much lower contrast over the range of interest, as seen in Figs.~\ref{fig:MLE_IntPlots_Ex2} and \ref{fig:MLE_Lplot2}.
\begin{figure}[tbp] 
	\centering
	\includegraphics[width=\linewidth]{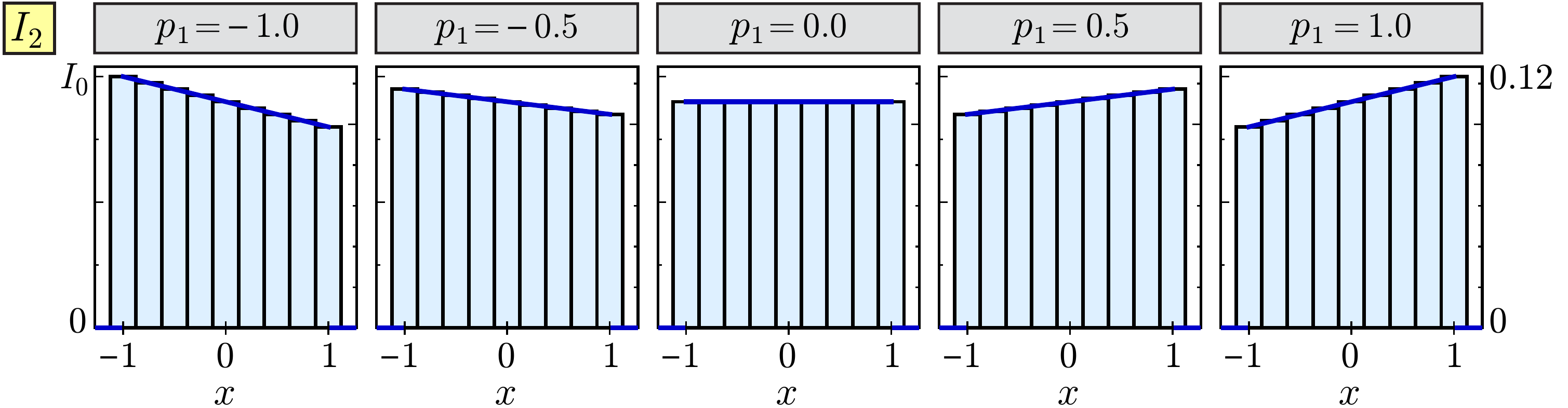}
	\caption{Plots of $I_2(x;p_1)$ (left axis) and $P_2(i|p_1)$ (right axis) for several values of $p_1$.}
	\label{fig:MLE_IntPlots_Ex2}
\end{figure}
\begin{figure}[tbp] 
	\centering
	\includegraphics[width=.65\linewidth]{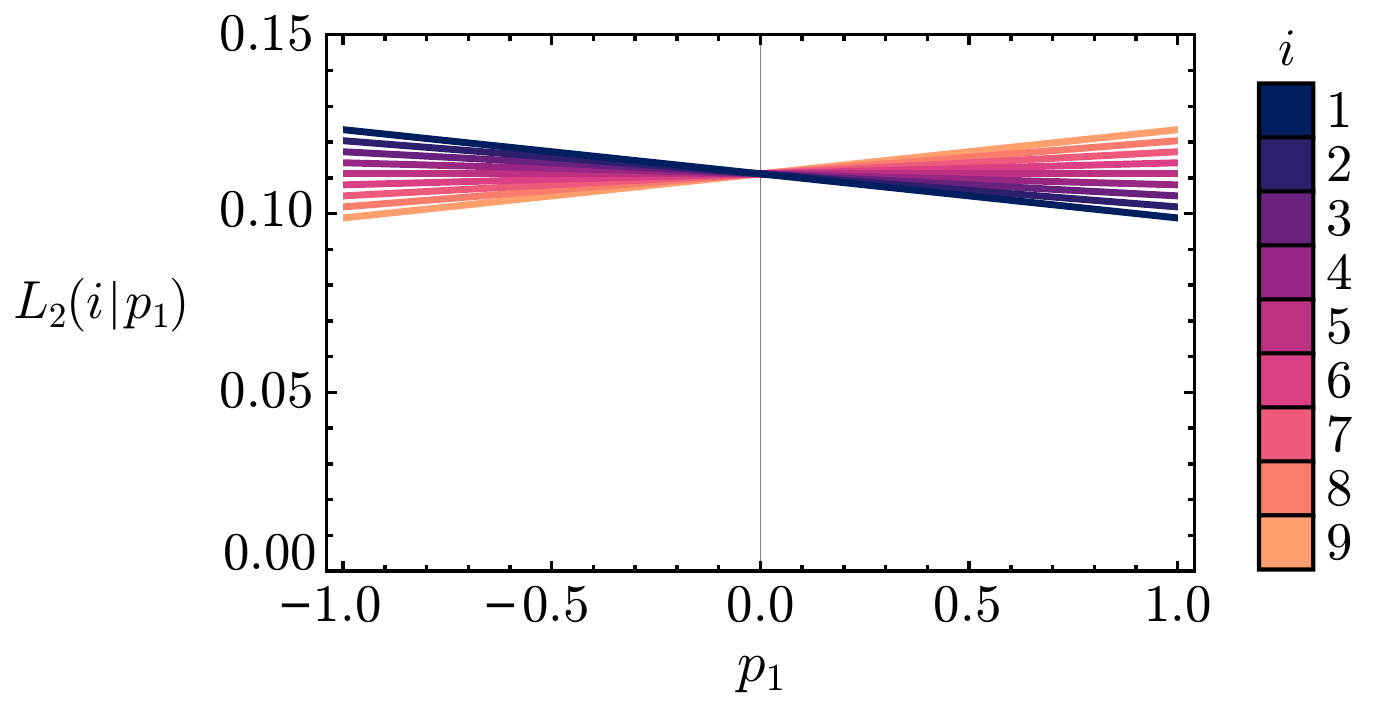}
	\caption{Likelihood functions $L_2(i|p_1)$ associated with each pixel $i$ in a measurement with theoretical intensity distribution $I_2(p_1)$.}
	\label{fig:MLE_Lplot2}
\end{figure}
Analogously to Section \ref{sect:MLE_example1}, suppose that we simulate a measurement of $I_2(x;p_1)$ and that the first three photons are again detected at pixels 1, 7, and 7. Following the same procedure as in the previous example, it can be shown that the maximum likelihood estimates after each photon detection are $p_1=-9$, $-4.5$, and $0$. The corresponding likelihood functions, shown in Fig.~\ref{fig:MLE_L3ph_Ex2}, are nearly flat, which is a sign that the MLE has a large uncertainty. 
\begin{figure} 
	\centering
	\includegraphics[width=\linewidth]{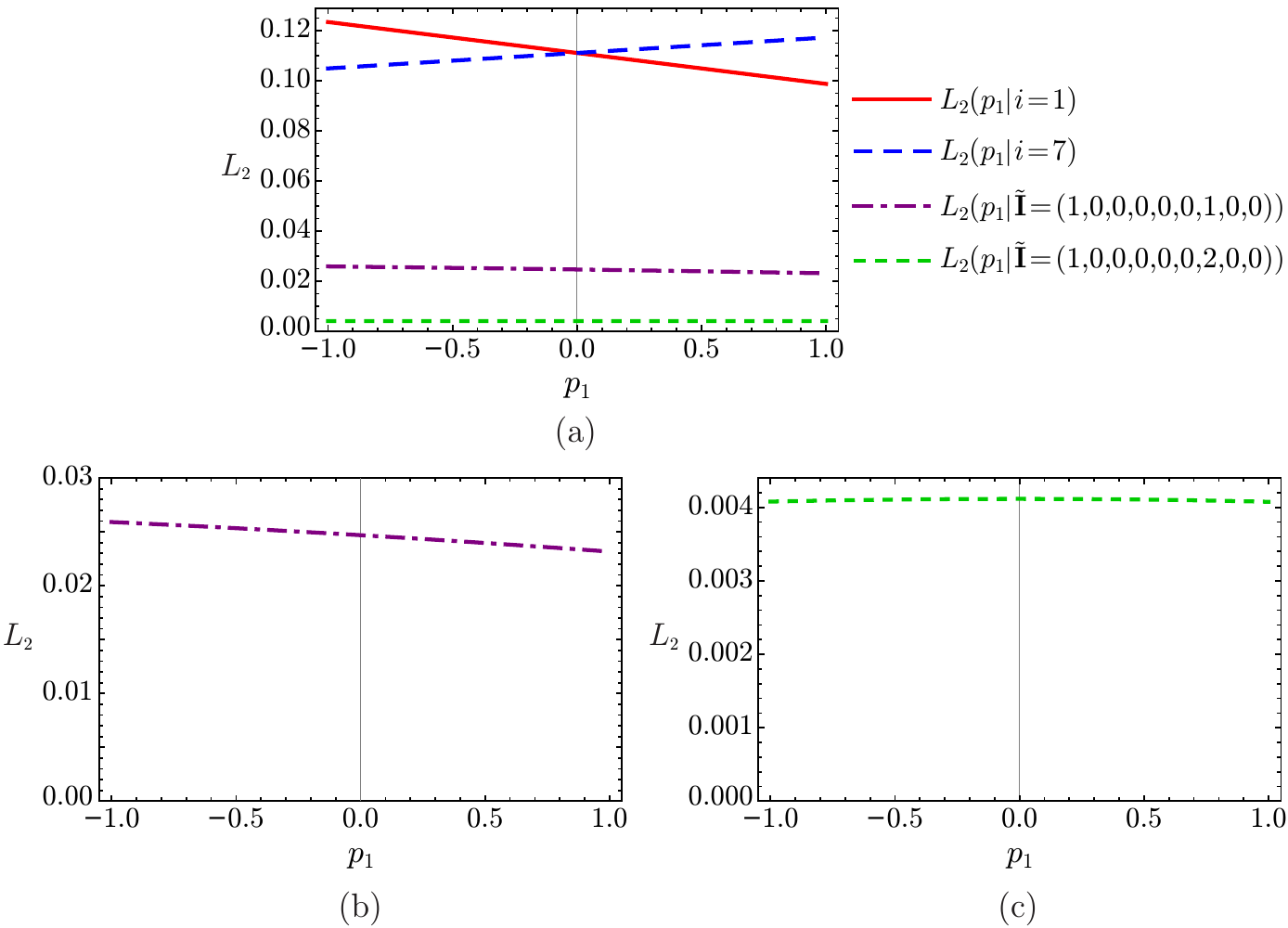}
	\caption{(a) Likelihood functions (based on intensity distribution $I_2$) for detected photons at pixels $i=1$ and $i=7$ and for intensity measurements consisting of one photon at pixel 1 and one or two photons at pixel 7. The two- and three-photon likelihoods are also plotted on independent scales in plots (b) and (c).}
	\label{fig:MLE_L3ph_Ex2}
\end{figure}
Indeed, for $\tbI=(1,0,0,0,0,0,2,0,0)$, the observed Fisher information is found to be 
\begin{equation}
J_2^\text{(obs)}(p_1;\tbI) = \frac{1}{(p_1-9)^2} + \frac{2}{(p_1+18)^2},
\end{equation}
which yields $J_1^\text{(obs)}=0.0185$ when evaluated at the MLE $p_1=0$, corresponding to a standard deviation uncertainty of $1/\sqrt{0.0185}=7.35$. Similarly, the expected Fisher information
\begin{equation}
\mathcal{N}J_2(p_1)=\frac{\mathcal{N}}{324}\sum_{i=1}^{9}\frac{(i-5)^2}{36+(i-5)p_1}\label{eq:Fisher_MLE_Ex2}
\end{equation}
for an $\mathcal{N}$-photon measurement of $I_2$ is significantly smaller than the information contained in a measurement of $I_1$, as shown in Fig.~\ref{fig:MLE_Fisher_Ex2}. For example, the expected standard deviation error for a three-photon measurement, given by 
$[3J_2(0)]^{-1/2}=8.05$, is nine times larger than it was in the previous example. The discrepancy grows even larger as $|p_1|$ increases.

\begin{figure} 
	\centering
	\includegraphics[height=2.25in]{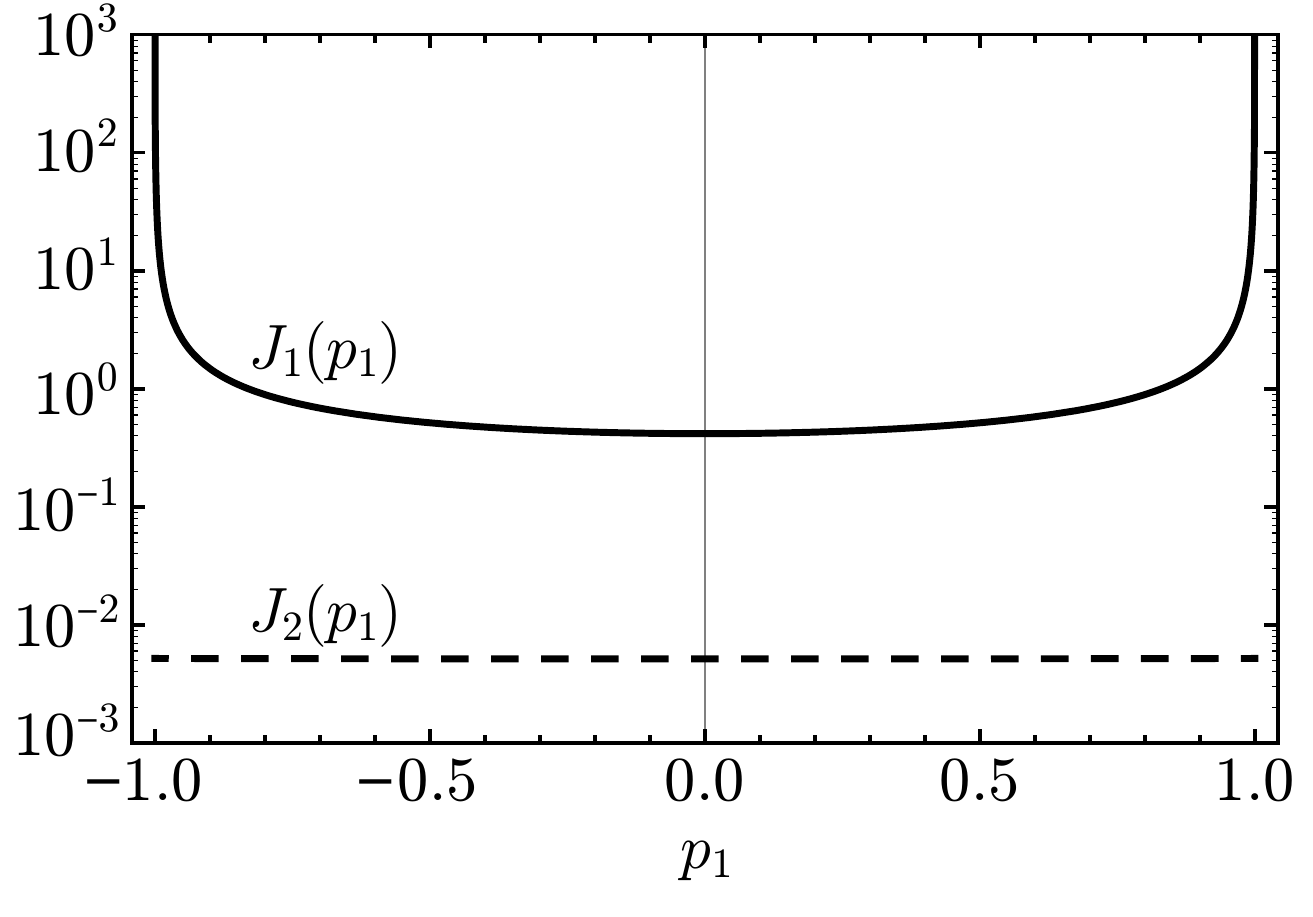}
	\caption{Expected unit Fisher information $J_1(p_1)$ and $J_2(p_1)$ for measurements of $I_1(x;p_1)$ and $I_2(x;p_1)$, respectively, plotted on a logarithmic scale.}
	\label{fig:MLE_Fisher_Ex2}
\end{figure}

Similarly to the previous section, a 100,000 photon simulation of $I_2(x;p_1)$ was performed, and the results were monitored along the way as photons accumulated. The intensities and corresponding MLEs obtained at several steps throughout the simulation are listed in Table \ref{tbl:MLE_ex2_photonsim}, and the MLE and standard deviation confidence interval are plotted as a function of $\mathcal{N}$ in Fig.~\ref{fig:MLE_conf_Ex2}. From these results, one can see that the MLE approaches the true parameter value more slowly than in the previous example, with a much larger uncertainty. (Take note of the increased scale of the plot compared to Fig.~\ref{fig:MLE_conf_Ex1}.) 

\begin{table}
	\begin{center} 
		\begin{tabular}{l@{\hspace{6pt}}lM{0in}l}
			\toprule
			$\mathcal{N}$ & MLE $(p_1)$ && $\tbI=(\tI_1,\ldots,\tI_9)$ \\
			\midrule
			1 & $-9.0000$ && $(1,0,0,0,0,0,0,0,0)$ \\
			2 & $-4.5000$ && $(1,0,0,0,0,0,1,0,0)$ \\
			3 & $\phantom{-}0.0000$ && $(1,0,0,0,0,0,2,0,0)$ \\
			4 & $-3.8285$ && $(1,1,0,0,0,0,2,0,0)$ \\
			5 & $-3.8285$ && $(1,1,0,0,1,0,2,0,0)$ \\
			6 & $-2.3629$ && $(1,1,0,0,1,1,2,0,0)$ \\
			7 & $-5.1192$ && $(1,2,0,0,1,1,2,0,0)$ \\
			8 & $-5.1192$ && $(1,2,0,0,2,1,2,0,0)$ \\
			9 & $-6.0605$ && $(1,2,0,1,2,1,2,0,0)$ \\
			10 & $-4.8152$ && $(1,2,0,1,2,2,2,0,0)$ \\
			100 & $\phantom{-}2.3159$ && $(6,6,12,17,13,11,9,13,13)$ \\
			1000 & $\phantom{-}1.8366$ && $(91,98,89,105,113,108,145,120,131)$ \\
			10000 & $\phantom{-}0.7542$ && $(1000,1044,1101,1077,1117,1088,1204,1168,1201)$ \\
			100000 & $\phantom{-}0.6331$ && $(10278,\nsp 10541,\nsp 10629,\nsp 11026,\nsp 11138,\nsp 11377,\nsp 11438,\nsp 11843,\nsp 11730)$ \\
			\bottomrule
		\end{tabular}
		\caption{Evolution of the MLE for $p_1$ and the measured intensity distribution $\tbI$ as individual photons accumulate for a simulated measurement of $I_2(x;p_1)$ with true parameter value $p=0.63$.}
		\label{tbl:MLE_ex2_photonsim}
	\end{center}
\end{table}

\begin{figure} 
	\centering
	\includegraphics[height=2.5in]{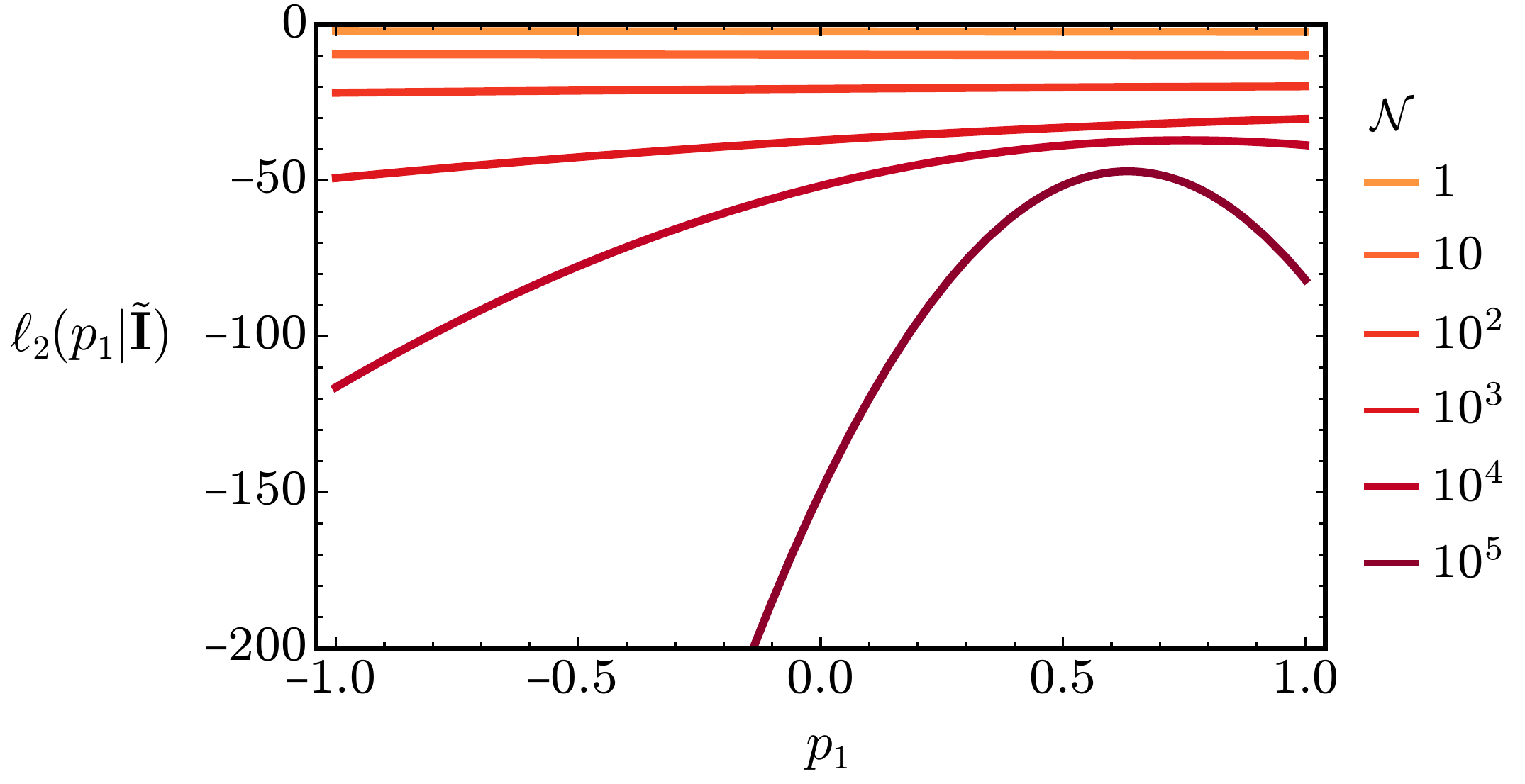}
	\caption{Log-likelihood functions associated with the simulated intensities listed in Table \ref{tbl:MLE_ex2_photonsim}.}
	\label{fig:MLE_LL_Ex2}
\end{figure}

\begin{figure} 
	\centering
	\includegraphics[width=.75\linewidth]{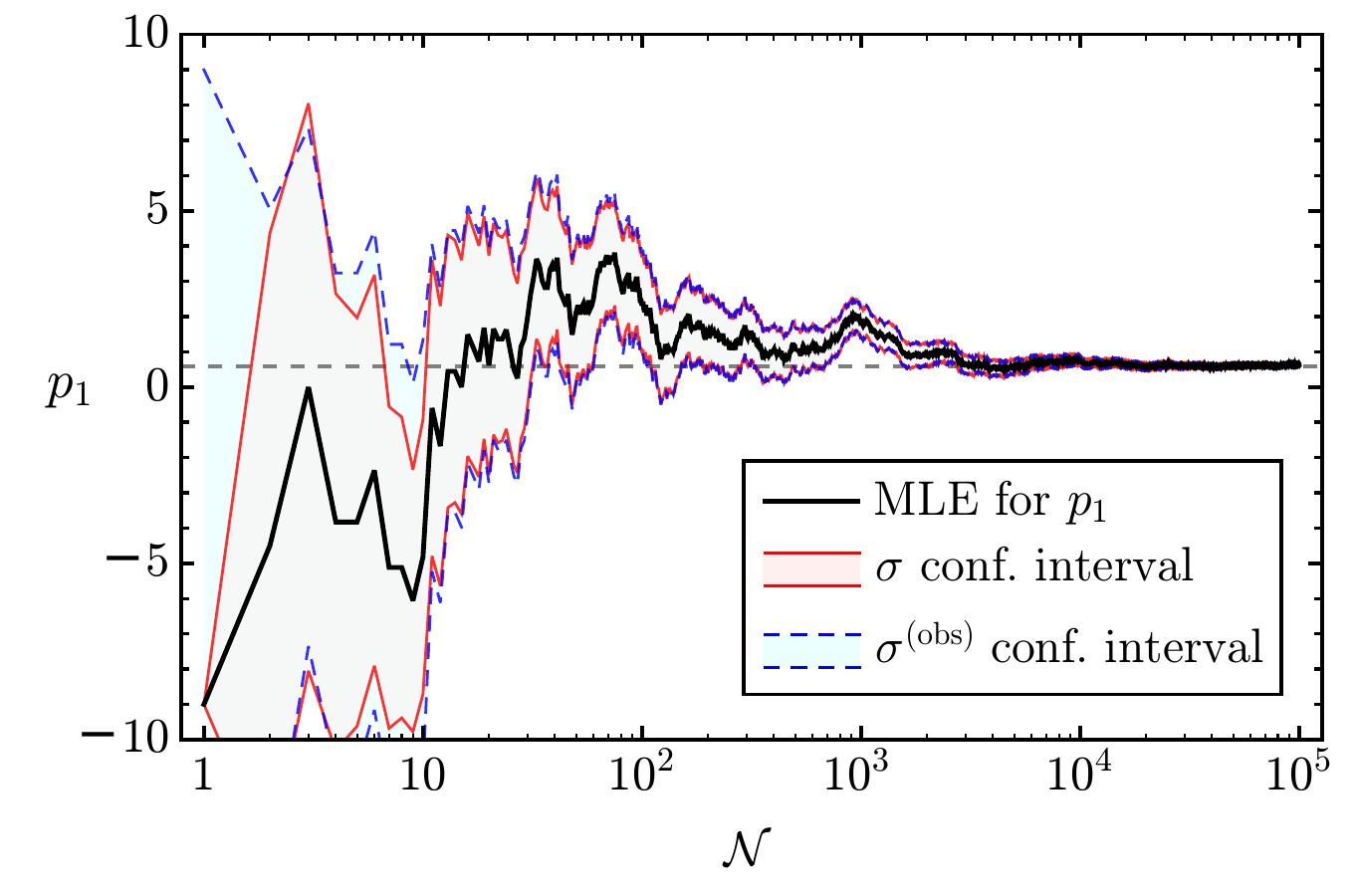}
	\caption{Evolution of the maximum likelihood estimate and standard deviation confidence interval for $p_1$ as 100,000 photons accumulate for a simulated measurement of $I_2(x;p_1)$ with true parameter value $p_1=0.63$. The solid red and dashed blue regions represent the confidence intervals based on the expected and observed Fisher information, respectively.}
	\label{fig:MLE_conf_Ex2}
\end{figure}

\widowpenalty10000

Finally, to complete the comparison to Section \ref{sect:MLE_example1}, a Monte Carlo simulation was performed for 50,000 trials of a 1000-photon measurement of $I_2(x;p_1)$. For true parameter values $p_1=0$ and $p_1=0.63$, the expected standard deviation errors are  $0.4409$ and $0.4401$, respectively. Histograms of the results of each simulation for 50,000 trials are shown in Fig.~\ref{fig:MLE_retr_hist_Ex2}; as indicated on the plots, the standard deviations of the MLEs obtained for each case are $0.4413$ and $0.4394$, closely matching expectations. 
\begin{figure} 
	\centering
	\includegraphics[width=.8\linewidth]{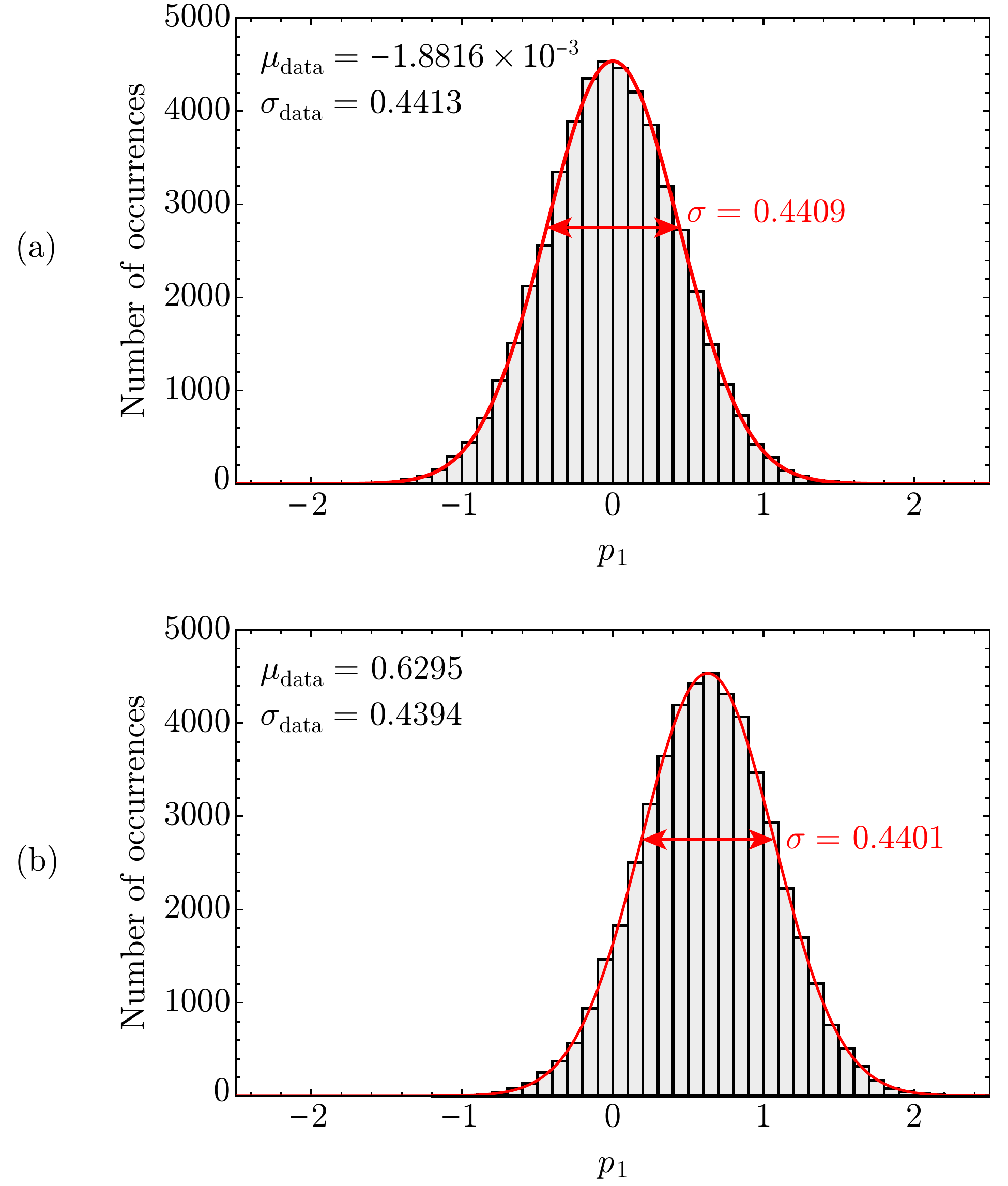}
	\caption{Histograms of the maximum likelihood estimates obtained from 50,000 trials of a 1000-photon simulation of $I_2(x;p_1)$ with true parameter values (a) $p_1=0$ and (b) $p_2=0.63$. The mean ($\mu_\text{data}$) and standard deviation ($\sigma_\text{data}$) of each distribution are indicated in the upper left corner of the plot. For comparison, a normal distribution with mean $p_1$ and standard deviation $\sigma=[1000J_1(p_1)]^{-1/2}$ is overlaid in red; the value of $\sigma$ is indicated alongside each curve.}
	\label{fig:MLE_retr_hist_Ex2}
\end{figure}
Notice that the errors are larger than they were in the previous example ($0.1554$ and $0.1303$) despite the fact that the measured intensity contains ten times as many photons. This is noteworthy because for any value of $p_1$, the total power incident on the detector (given by the sum of the intensity over all pixels) is 1.8 times larger for $I_2$ than it is for $I_1$, indicating that on average nearly twice as many photons will be measured within a given exposure time. Even so, based on the above results, we can conclude that if measurements of $I_1$ and $I_2$ were conducted with identical exposure times, then the measurement of $I_1$ (for which the output signal would contain fewer photons) would be expected to produce a more accurate parameter estimate. This is an important lesson to keep in mind when designing an experiment: the most informative measurement is not always the one with the strongest signal! On the contrary, it can be beneficial to filter out a large fraction of the light before it reaches the detector (e.g., via polarization selection) in such a way that the measured signal contains only the photons emitted from the source that provide the most information about $p_1$.\footnote{When possible, it would be preferential to encode information by rearranging the light rather than filtering it out. However, sometimes this is not possible, e.g., when measuring the coupling induced by a scattering process between a pair of specific input and output polarization states.} This idea is explored further in the next example.

\subsection{Null and off-null measurements}\label{sect:MLE_example3}
For some optical applications, it is advantageous to design the experiment so that low light levels are observed at the detector plane, resulting in increased parameter sensitivity. One notable example is off-null ellipsometry, in which polarization elements are configured to produce a high extinction ratio over the range of interest of the parameter(s) under test \cite{Arwin_1993}. The focused beam scatterometry experiment in Ref.~\cite{Vella_2018_fbs_arxiv} operates on the same principle but with a spatially-varying polarization distribution, resulting in an output intensity of the form $I\propto \left|\sum_n a_n(x)[p_n-\pbar_n(x)]\right|^2$, where the functions $a_n(x)$ characterize the sample under test and the functions $\pbar_n(x)$ (which determine the required input polarization) can be tailored to optimize the sensitivity to each parameter. As an example of this type of measurement for the one-parameter case, consider the intensity distribution
\begin{equation}
I_3(x;p_1)=\Pi(x)\frac{1}{(|c|+1)^2}(p_1-cx)^2,
\end{equation}
where $c$ is a real constant. For $c=0$, this represents a null measurement for which the (spatially uniform) intensity vanishes when $p_1=0$ and increases quadratically with $p_1$. For $c\neq 0$, the value of $p_1$ for zero intensity (i.e., the departure from perfect nulling) varies linearly with the coordinate $x$. Using Eq.~(\ref{eq:P(i|p1)}), the PMF for a detected photon is found to be
\begin{equation}
P_3(i|p_1)=\frac{(4p_1-(i-5)c)^2}{144p_1^2+60c^2}.
\end{equation}

Let us begin by examining the case of perfect nulling ($c=0$), for which the intensity $I_3(x;p_1)=\Pi(x)p_1^2$ and PMF $P_3(i|p_1)=1/9$ are plotted in Fig.~\ref{fig:MLE_IntPlots_Ex3_null}. 
\begin{figure}
	\centering
	\includegraphics[width=\linewidth]{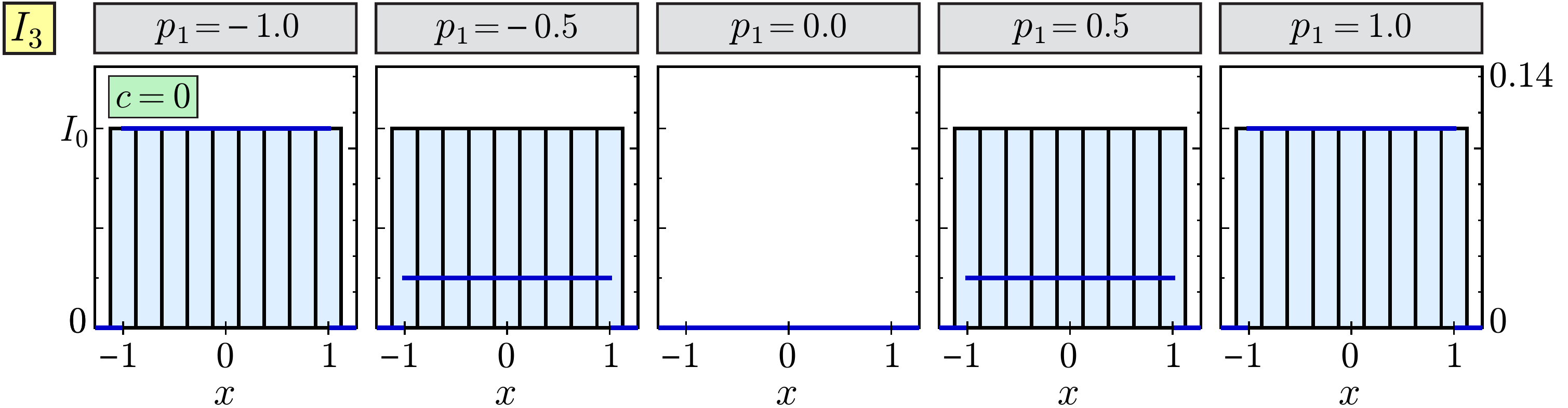}
	\caption{Plots of $I_3(x;p_1)$ (left axis) and $P_3(i|p_1)$ (right axis) for several values of $p_1$ for the case of perfect nulling.}
	\label{fig:MLE_IntPlots_Ex3_null}
\end{figure}
In contrast to the previous two examples, these plots illustrate that for a given coordinate $x_i$, the ratio between the measured intensities at two different parameter values need not be the same as the ratio between the corresponding PMF values. In fact, in this example the PMF is the same for all values of $p_1$ with the exception of $p_1=0$, for which it is undefined (due to the fact that no photons are detected). Consequently, the likelihood function is completely flat and the Fisher information is zero, implying that it is impossible to determine $p_1$ from the shape of the measured intensity distribution.\footnote{In this case, the MLE exists but it is not unique, since all values of $p_1$ within the range of interest maximize the likelihood function.} (Of course, this is also obvious from the simple fact that the PMF is independent of $p_1$.) In this situation, it would only be possible to deduce the value of $p_1$ from the total optical power incident on the detector, which is beyond the scope of the current statistical approach. Even then, it would only be possible to determine the magnitude of $p_1$ but not its sign (since $I_3$ is an even function of $p_1$), and the measurement would be susceptible to temporal fluctuation errors unless the illumination source power were very stable.

The aforementioned shortcomings of a null measurement can be avoided by designing the experiment to operate under an off-null condition, which corresponds to the choice of some constant $c\neq 0$ in the present example. The intensity and PMF are plotted in Fig.~\ref{fig:MLE_IntPlots_Ex3_offnull} for several positive values of $c\sp$; symmetric behavior is observed when $c$ is negative. Notice in each plot that the null in intensity (when one exists within the range of interest) is located at $x=p_1/c$. When $|c|=1$, the null shifts across the entire width of the sensor as $p_1$ varies from $-1$ to $1$, causing the shape of $P_3(i|p_1)$ to vary substantially over the entire parameter range. When $|c|\gg 1$, the null is confined to a narrow region near the center of the sensor, resulting in very little variation in $P_3(i|p_1)$ with respect to $p_1$. On the other hand, when $|c|\ll 1$, the null shifts away from the origin very quickly when $p_1$ is nonzero. This results in dramatic variations in $P_3(i|p_1)$ (and very low intensity levels) when $|p_1|$ is small, but much smaller changes near the edge of the parameter range.

\begin{figure}
	\centering
	\includegraphics[width=\linewidth]{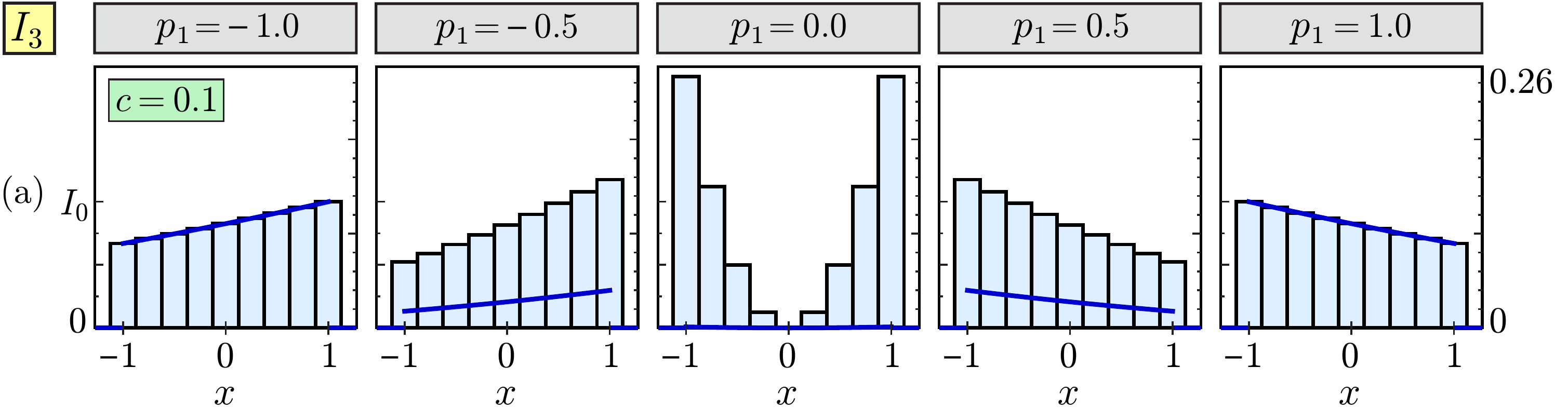}\\[4pt]
	\includegraphics[width=\linewidth]{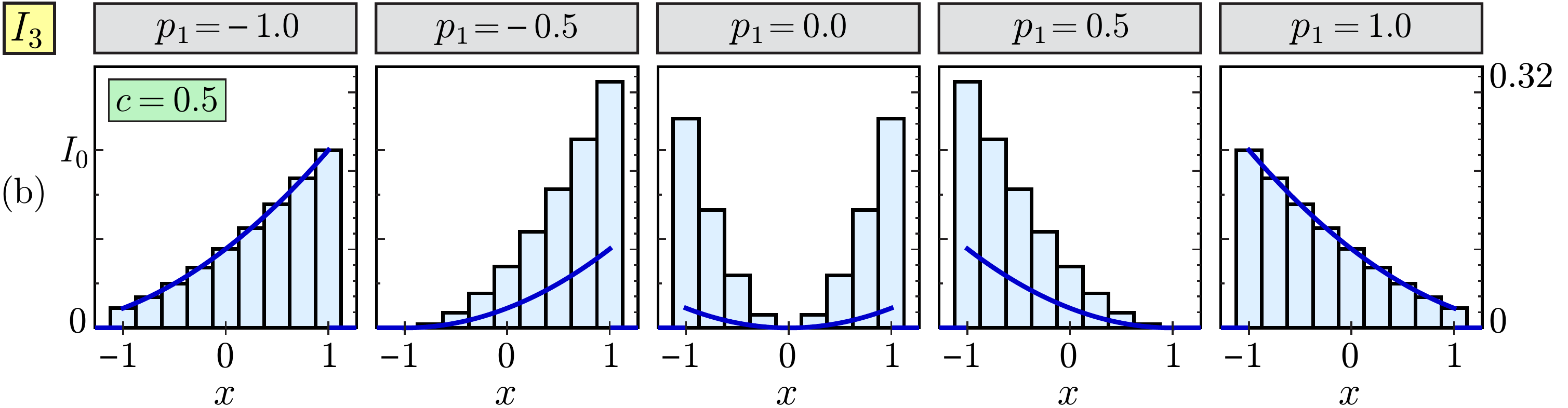}\\[4pt]
	\includegraphics[width=\linewidth]{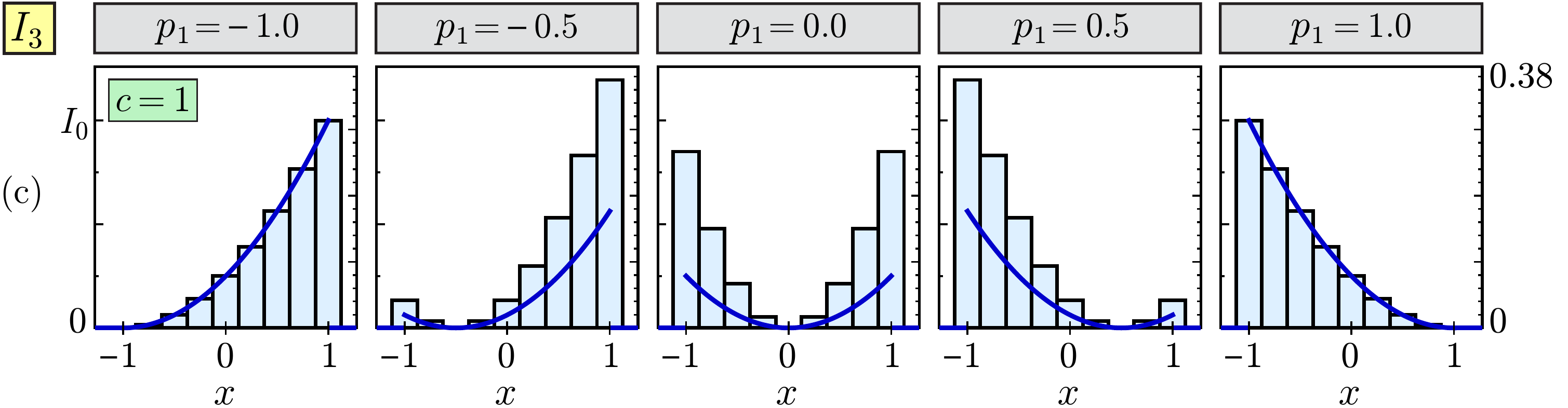}\\[4pt]
	\includegraphics[width=\linewidth]{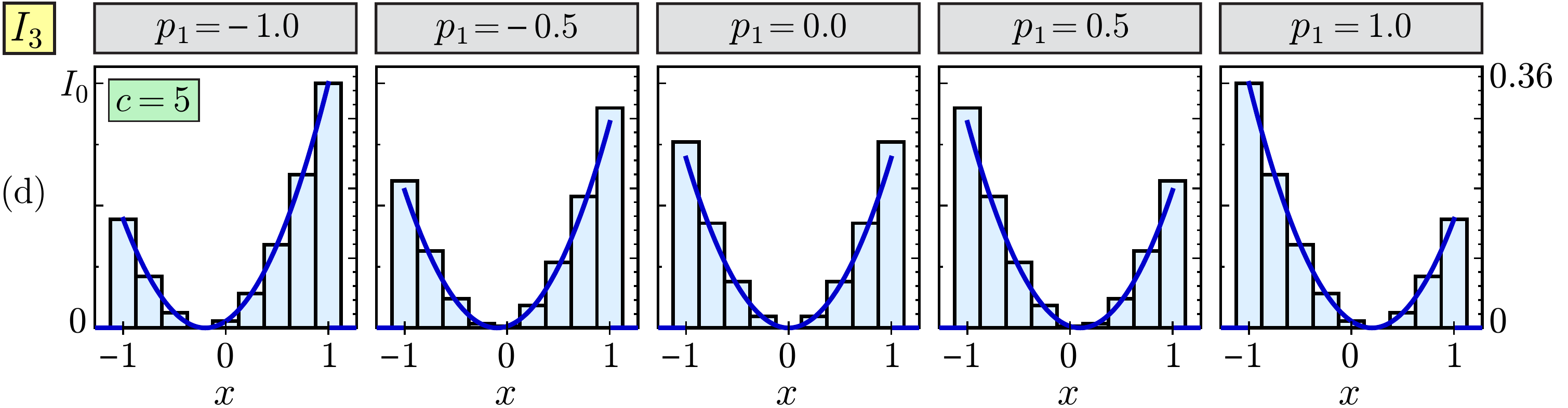}\\[4pt]
	\includegraphics[width=\linewidth]{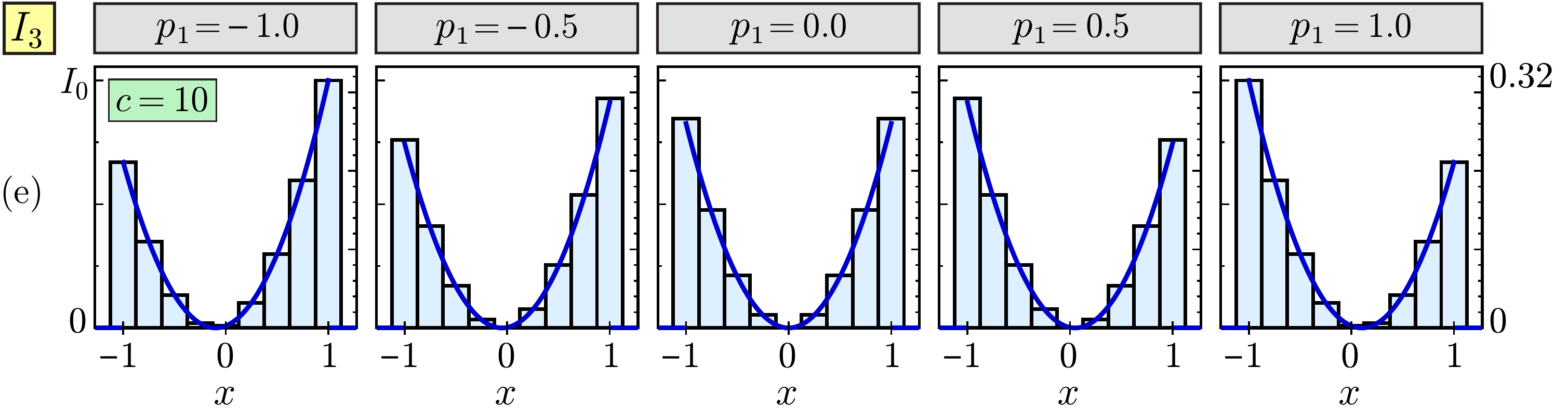}
	\caption{Plots of $I_3(x;p_1)$ (left axes) and $P_3(i|p_1)$ (right axes) for several values of $p_1$. Each row of plots corresponds to a different value of $c$, as indicated in the leftmost plot.}
	\label{fig:MLE_IntPlots_Ex3_offnull}
\end{figure}

This behavior can also be visualized by plotting the likelihood functions $L_3(i|p_1)$ for each pixel, which are shown in Fig.~\ref{fig:MLE_Lplot3}. 
\begin{figure}
	\centering
	\includegraphics[width=.95\linewidth]{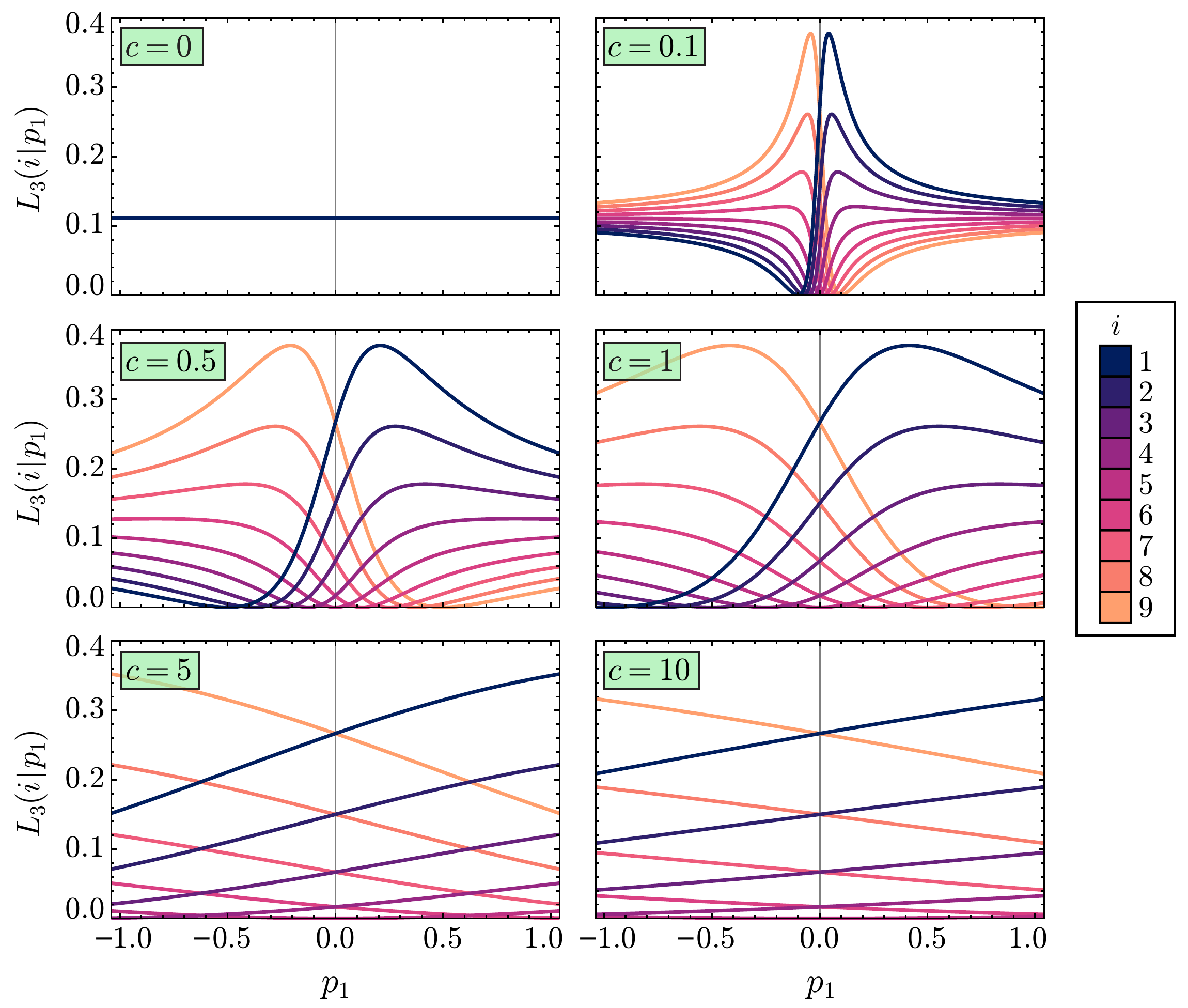}
	\vspace{-6pt}
	\caption{Likelihood functions $L_3(i|p_1)$ associated with each pixel $i$ in a measurement with theoretical intensity distribution $I_3(p_1)$, plotted for several nonnegative values of~$c$. Symmetric results are obtained for the corresponding negative values of $c$, with each plot flipped about the vertical $p_1=0$ axis.}
	\label{fig:MLE_Lplot3}
\end{figure}
From the definition of the Fisher information, recall that the magnitude of the local slope of $L_3$ is an indicator of the information content of a measurement of $p_1$. In agreement with the observations made above, for $|c|\ll 1$, the likelihood generally has a very large slope when $|p_1|$ is small (enabling a precise estimate of $p_1$), but it becomes nearly flat for larger parameter values. 
Meanwhile, for $|c|\gg 1$, the likelihood is relatively flat over the entire range of interest, making parameter estimation difficult. Qualitatively, it is evident that the best balance between these two extremes is achieved when $c$ is on the order of unity, so that $L_3(i|p_1)$ exhibits a similar amount of variation over the full range of interest \nolinebreak[4] of $p_1$.

For a measurement containing a large number of photons, the uncertainty of the MLE can be calculated from the expected unit Fisher information; a somewhat lengthy but straightforward calculation shows that
\begin{equation}
J_3(p_1)=\sum_{i=1}^{9}\frac{16c^2[5c+3(i-5)p_1^2]^2}{3(12p_1^2+5c^2)^3} = \frac{240c^2}{(12p_1^2+5c^2)^2}.
\label{eq:Fisher_MLE_Ex3}
\end{equation}
This function is plotted in Fig.~\ref{fig:MLE_Fisher_Ex3} for several values of $c$.
\begin{figure} 
	\centering
	\includegraphics[height=2.25in]{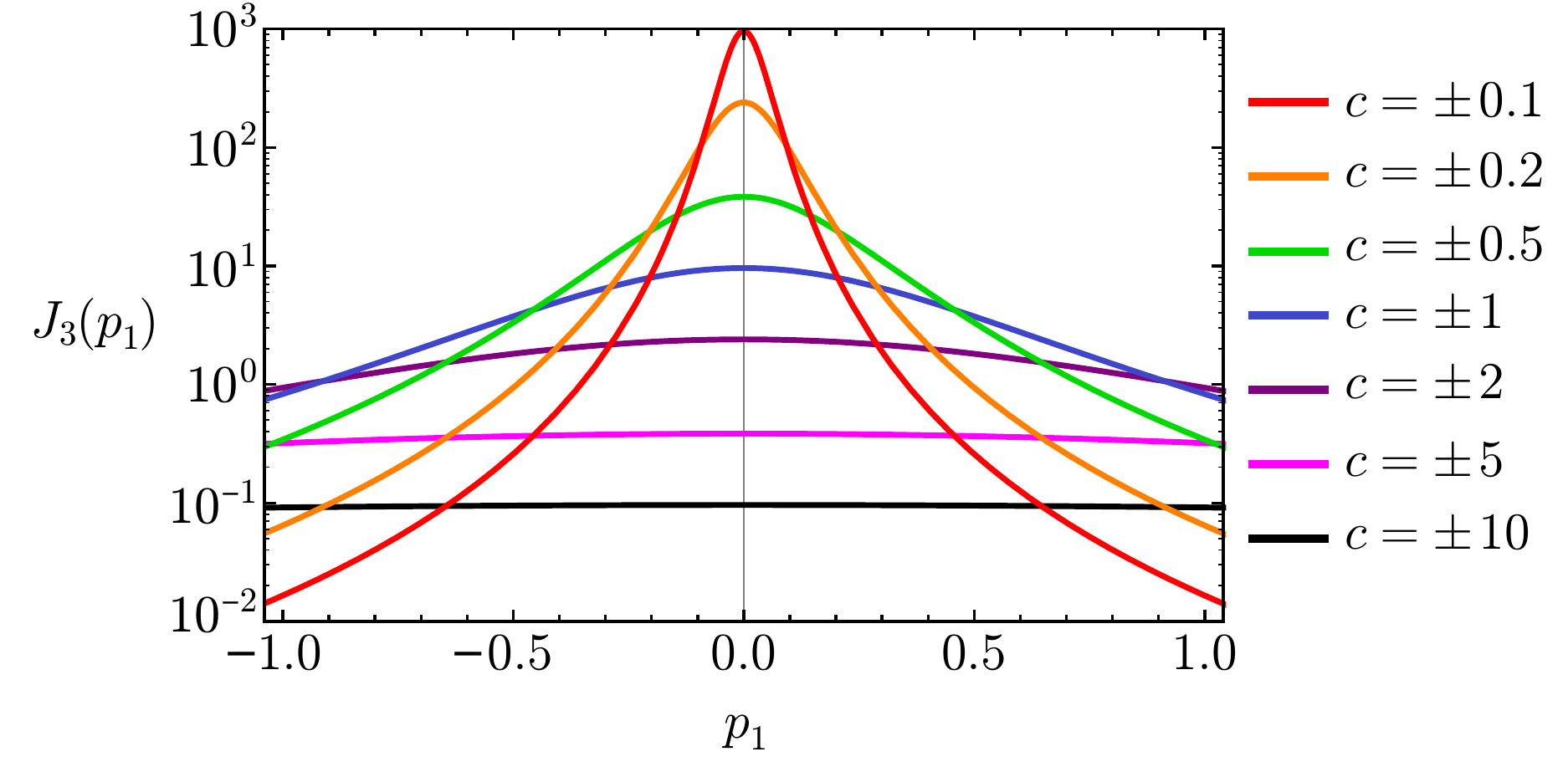}
	\vspace{-6pt}
	\caption{Expected unit Fisher information $J_3(p_1)$ for a measurement of $I_3(x;p_1)$, plotted on a logarithmic scale for several values of $c$.}
	\label{fig:MLE_Fisher_Ex3}
\end{figure}
Notice that the Fisher information is the same for positive and negative $c$; the $c=0$ case does not appear on the plot since $J_3(p_1)$ goes to zero. Suppose that we are designing an experiment where the output intensity takes the form of $I_3(x;p_1)$, and we wish to determine the optimal value of $c$ that, on average, will produce the best parameter estimate for any true value of $p_1$ within the range of interest, i.e., the smallest expected error $\sigma(p_1)=J_3(p_1)^{-1/2}$. One approach to do so is by minimizing the average value of the variance $\sigma(p_1)^2$ over the interval $p_1\in[-1,1]$, which is given by
\begin{align}
\lrangle{\sigma^2} &= \frac{1}{2}\int_{-1}^1 \sigma(p_1)^2 \ud p_1 \nonumber\\
&= \frac{1}{240c^2}\int_{-1}^1 (12p_1^2+5c^2)^2 \ud p_1 \nonumber \\[4pt]
&= \frac{5}{24}c^2 + \frac{6}{25}\frac{1}{c^2}+\frac{1}{3}.
\end{align}
This function is plotted as a solid line in Fig.~\ref{fig:MLE_sigma_vs_c_Ex3}. 
\begin{figure}
	\centering
	\includegraphics[scale=.62]{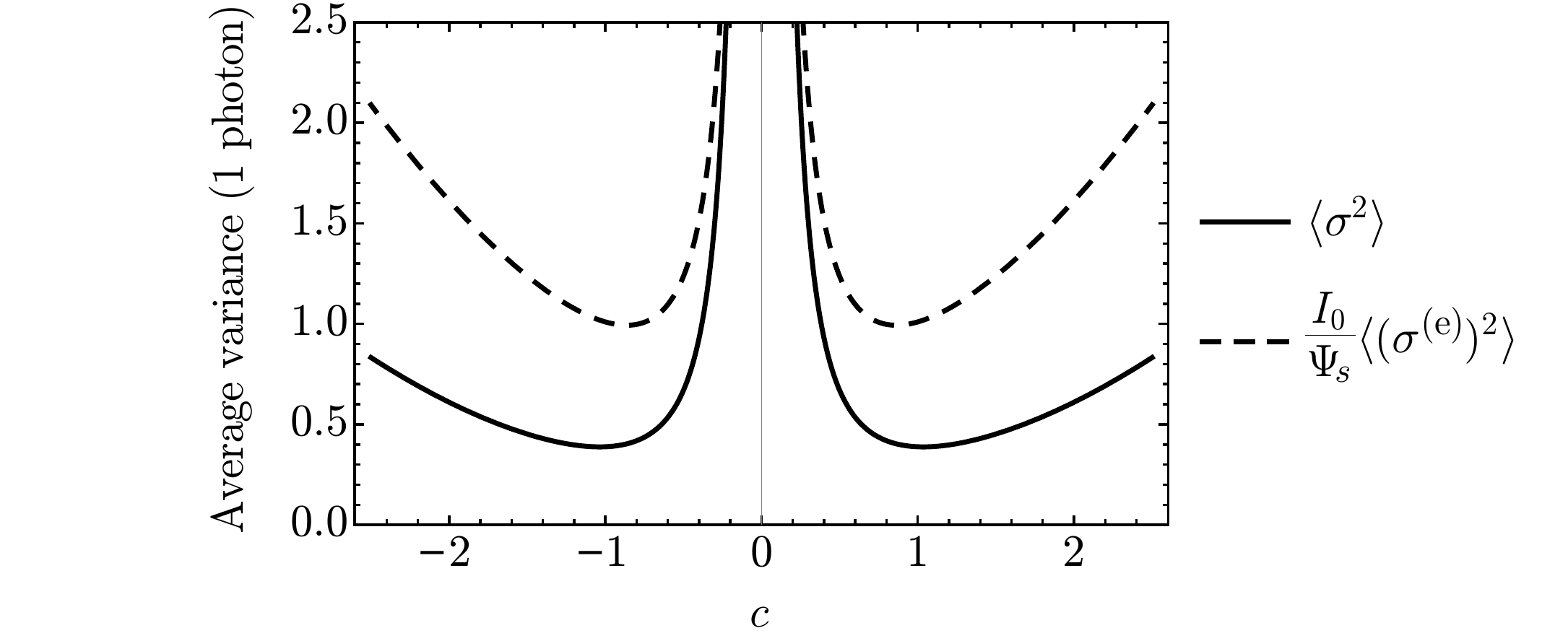}
	\caption{Expected variances (averaged over $p_1$) for parameter estimates based on measurements of $I_3(x;p_1)$ containing one detected photon (solid line) and one emitted photon (dashed line), plotted as a function of $c$. For the latter case, the error is scaled by the ratio between $I_0$ and the source power $\Psi_s$, which can be treated as a unitless quantity (see footnote \ref{footnote:I_0_power_units} on page \pageref{footnote:I_0_power_units}).}
	\label{fig:MLE_sigma_vs_c_Ex3}
\end{figure}
(The dashed line will be explained shortly). Note that for a multi-photon measurement, the variance scales as $1/\mathcal{N}$. The average error $\lrangle{\sigma^2}$ is minimized when $c=\pm(144/125)^{1/4}\approx \pm 1.036$, in close agreement with the above prediction that the optimal value of $c$ is on the order of unity.

As alluded to in the previous section, all of the statistics and performance metrics discussed thus far have pertained exclusively to photons detected by the sensor. However, the information contained in each detected photon is not the only thing to take into consideration when designing an experiment. In a typical experiment, the light source emits a constant optical power $\Psi_s$, of which some fraction reaches the detector. The power incident on the detector, which is given by
\begin{equation}
\Psi_d(p_1) = \int_{-1}^1 I_3(x;p_1)\ud x = I_0 \frac{2(3p^2+c^2)}{3(|c|+1)^2}\label{eq:MLE_Ex3_Phi}
\end{equation}
in this example\footnote{The right-hand side of Eq.~(\ref{eq:MLE_Ex3_Phi}) implicitly has units of $I_0$ times the unitless coordinate $x$ (acquired from the integration), i.e., units of power.\label{footnote:I_0_power_units}}, is usually smaller than $\Psi_s$ by some ratio that is influenced by the choice of measurement scheme (e.g., an off-null configuration). During the exposure time of the sensor, the number of detected photons is (on average) equal to $\mathcal{N}=(\Psi_d/\Psi_s)\mathcal{N}_s$, where $\mathcal{N}_s$ is the number of photons emitted by the source. If the speed of the measurement is a priority, then it is important to make efficient use of the source, i.e., to maximize the information acquired per emitted photon. To that end, let us define the expected unit Fisher information per emitted photon as
\begin{equation}
J^{({\rm e})}(p_1)=\frac{\Psi_d}{\Psi_s}J(p_1),
\end{equation}
so that the total information acquired in a given time interval is $\mathcal{N}J(p_1)=\mathcal{N}_s J^{({\rm e})}(p_1)$. (Obviously, this is not to suggest that each photon carries information about $p_1$ at the moment that it is emitted from the source; rather, $J^{({\rm e})}(p_1)$ is the average information acquired at the detector plane per photon emitted by the source.)

For the present example, using Eqs.~(\ref{eq:Fisher_MLE_Ex3}) and (\ref{eq:MLE_Ex3_Phi}), the Fisher information per emitted photon is found to be
\begin{equation}
J_3^{({\rm e})}(p_1) = \frac{I_0}{\Psi_s}\sp\frac{160c^2(3p_1^2+c^2)}{(12p_1^2+5c^2)^2(|c|+1)^2}.\label{eq:MLE_Ex3_FisherEm}
\end{equation}
This result is plotted in Fig.~\ref{fig:MLE_FisherEm_Ex3} for several values of $c$. 
\begin{figure} 
	\centering
	\includegraphics[height=2.25in]{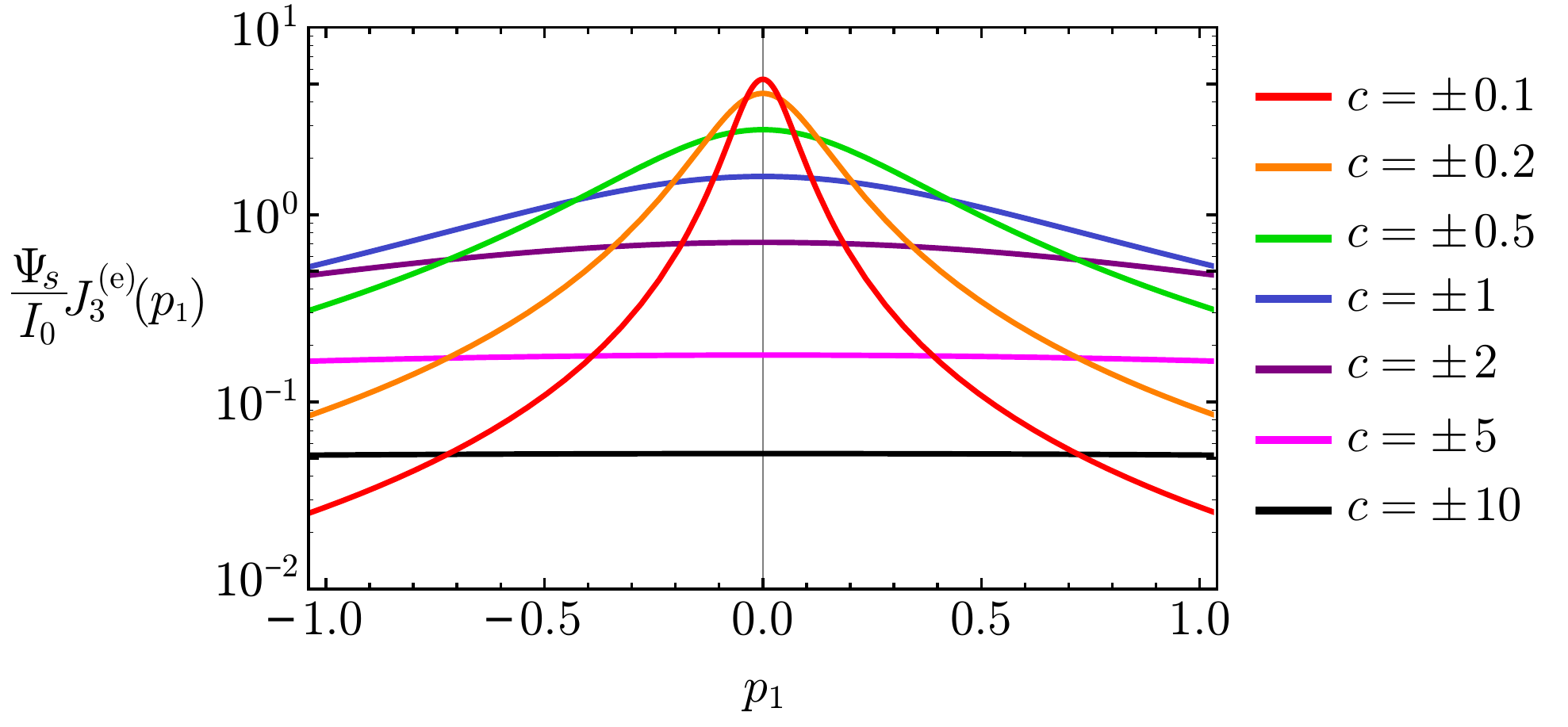}
	\caption{Expected unit Fisher information $J_3^{({\rm e})}(p_1)$ per emitted photon for a measurement of $I_3(x;p_1)$, scaled by the ratio of source power to $I_0$ and plotted on a logarithmic scale for several values of $c$.}
	\label{fig:MLE_FisherEm_Ex3}
\end{figure}
In comparison to Fig.~\ref{fig:MLE_Fisher_Ex3}, notice that the peak in $J_3^{({\rm e})}(p_1)$ when $|c|\ll 1$ is much less pronounced than that of $J_3(p_1)$. This is because as $|c|$ decreases, the amount of information per detected photon increases, but the number of detected photons decreases by nearly the same ratio. From Eq.~(\ref{eq:MLE_Ex3_FisherEm}), the minimum expected variance $\sigma^{({\rm e})}(p_1)^2=J_3^{({\rm e})}(p_1)^{-1}$ can be calculated for a measurement of one emitted photon, averaged over the range of interest of $p_1$:%
\begin{align}
\lrangle{(\sigma^{({\rm e})})^2} &= \frac{1}{2}\int_{-1}^1 \sigma^{({\rm e})}(p_1)^2 \ud p_1 \nonumber\\[4pt]
&= \frac{\Psi_s}{I_0}\frac{(|c|+1)^2}{320c^2}\int_{-1}^1 \frac{(12p_1^2+5c^2)^2}{3p_1^2+c^2} \ud p_1 \nonumber \\[4pt]
&= \frac{\Psi_s}{I_0}\frac{(|c|+1)^2}{480c^2} \left[\sqrt{3}\,c^3\arctan\Bigl(\!\mfrac{\sqrt{3}}{c}\hspace{.5pt}\Bigr) + 72 c^2 + 48 \right]\nsp.
\end{align}
This function is plotted as a dashed line in Fig.~\ref{fig:MLE_sigma_vs_c_Ex3}, shown in comparison to the average variance per detected photon derived earlier. A numerical calculation shows that the expected error per emitted photon is minimized when $c=\pm0.863$, which is slightly smaller than the optimal value $c=\pm 1.036$ for detected photons. This is due to the fact that for parameter values near $|p_1|=1$, the power on the detector is up to 10\% larger for $|c|=0.863$ than for $|c|=1.036$, compensating for the slight reduction in information per detected photon in the former case.

Recall that in this example the intensity is normalized to have a peak value of $I_0$ regardless of the value of $c$. This is not particularly realistic, since in an actual off-null measurement, a change in the (spatially varying) off-null condition is likely to be accompanied by a global scaling factor in the measured intensity. In some cases, this could result in a much more dramatic difference between the Fisher information per emitted and detected photon than in this example. On a separate note, in situations where $\sigma(p_1)^2$ and $\sigma^{({\rm e})}(p_1)^2$ cannot be calculated analytically, the integral over $p_1$ can be evaluated numerically. If the numerical integration is too computationally expensive, a simpler merit function could be constructed by summing the variance over some appropriately chosen set of parameter values.


\FloatBarrier
\subsection{Far-from-null (high intensity) measurement}\label{sect:MLE_example4}
For the final one-parameter example, consider the intensity distribution
\begin{equation}
I_4(x;p_1)=\Pi(x)\frac{1}{(|d|+2)^2}(p_1-x-d)^2,
\end{equation}
where the constant $d$ introduces a spatially uniform offset from the off-null condition considered in the previous example. When $d=0$, the intensity is identical to $I_3(x;p_1)$ with $c=1$, which was plotted previously in Fig.~\ref{fig:MLE_IntPlots_Ex3_offnull}(c). For comparison, Fig.~\ref{fig:MLE_IntPlots_Ex4} contains plots of $I_4(x;p_1)$ and the corresponding PMF for several positive values of \nolinebreak[3] $d$. (Symmetric results are obtained for negative $d$.) The likelihood functions $L_4(i|p_1)$ for each case are plotted in Fig.~\ref{fig:MLE_Lplot4}.%
\begin{figure}
	\centering
	\includegraphics[width=\linewidth]{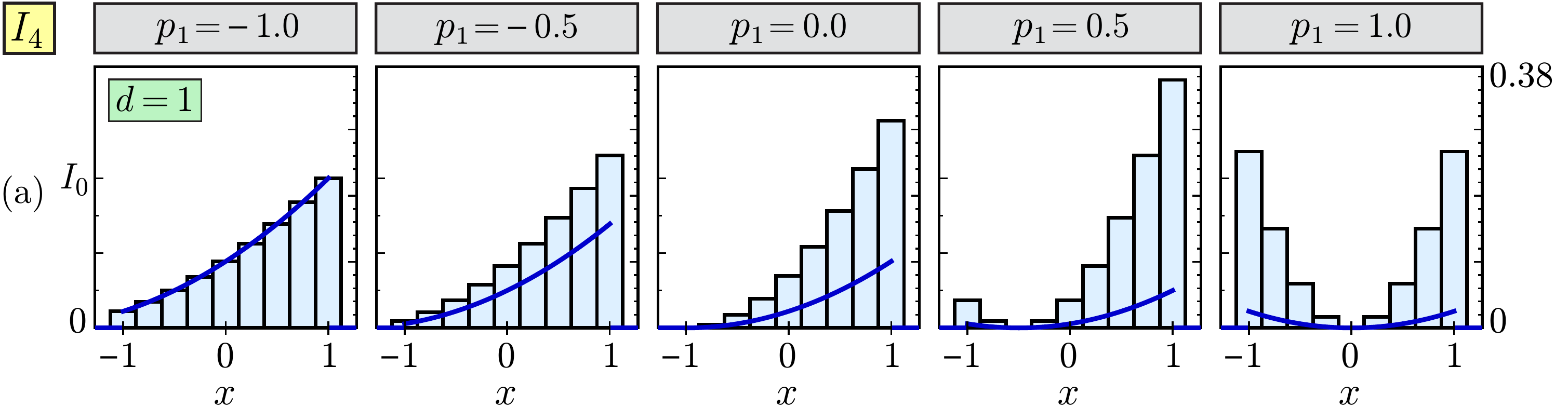}\\[2pt]
	\includegraphics[width=\linewidth]{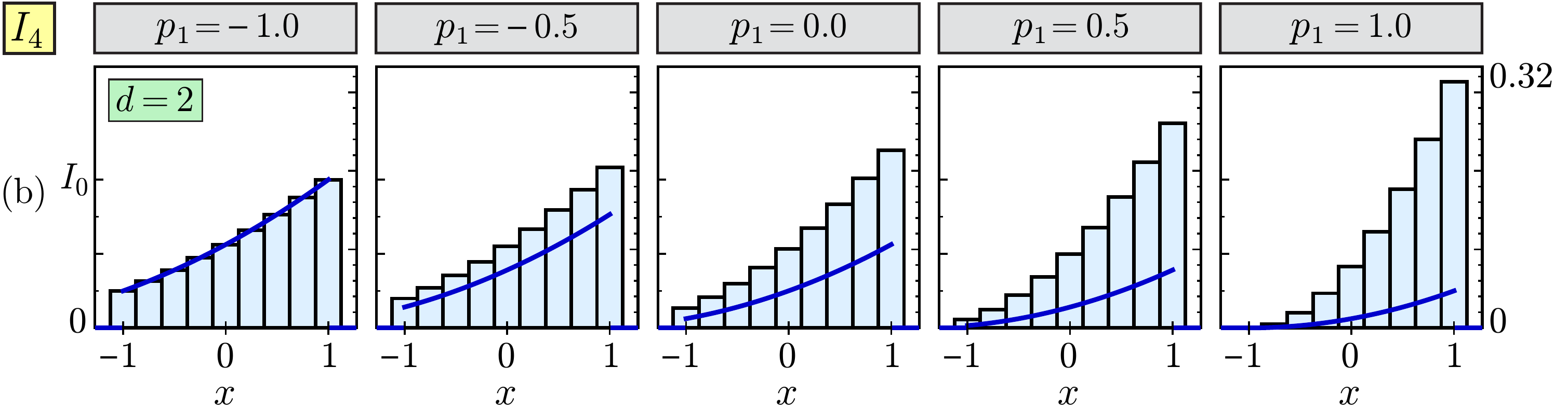}\\[2pt]
	\includegraphics[width=\linewidth]{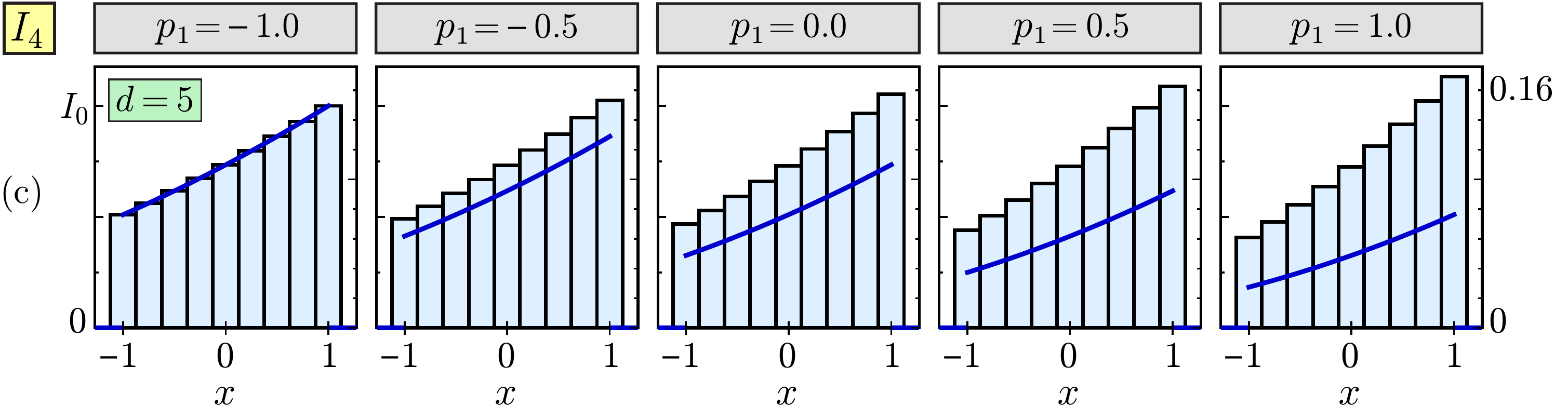}\\[2pt]
	\includegraphics[width=\linewidth]{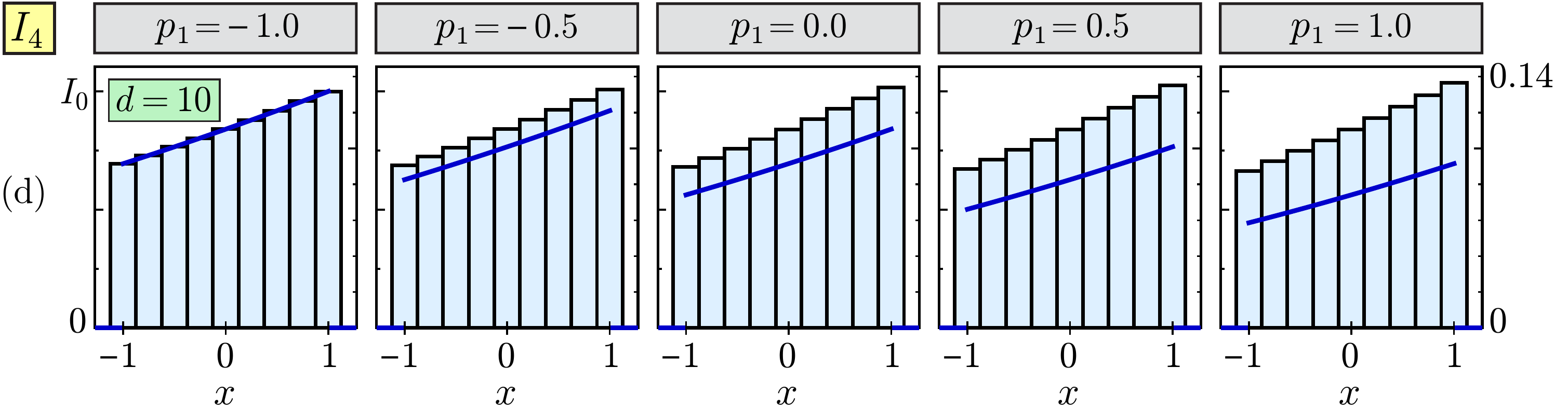}\\[2pt]
	\includegraphics[width=\linewidth]{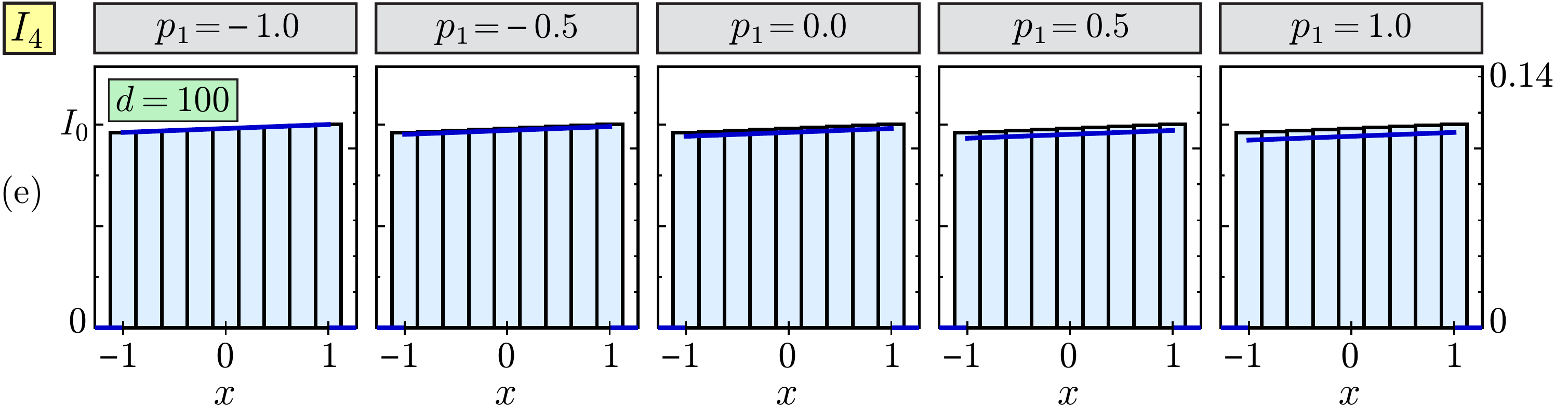}
	\caption{Plots of $I_4(x;p_1)$ (left axes) and $P_4(i|p_1)$ (right axes) for several values of $p_1$. Each row of plots corresponds to a different value of $d$, as indicated in the leftmost plot.}
	\label{fig:MLE_IntPlots_Ex4}
\end{figure}
\begin{figure}
	\centering
	\includegraphics[width=.95\linewidth]{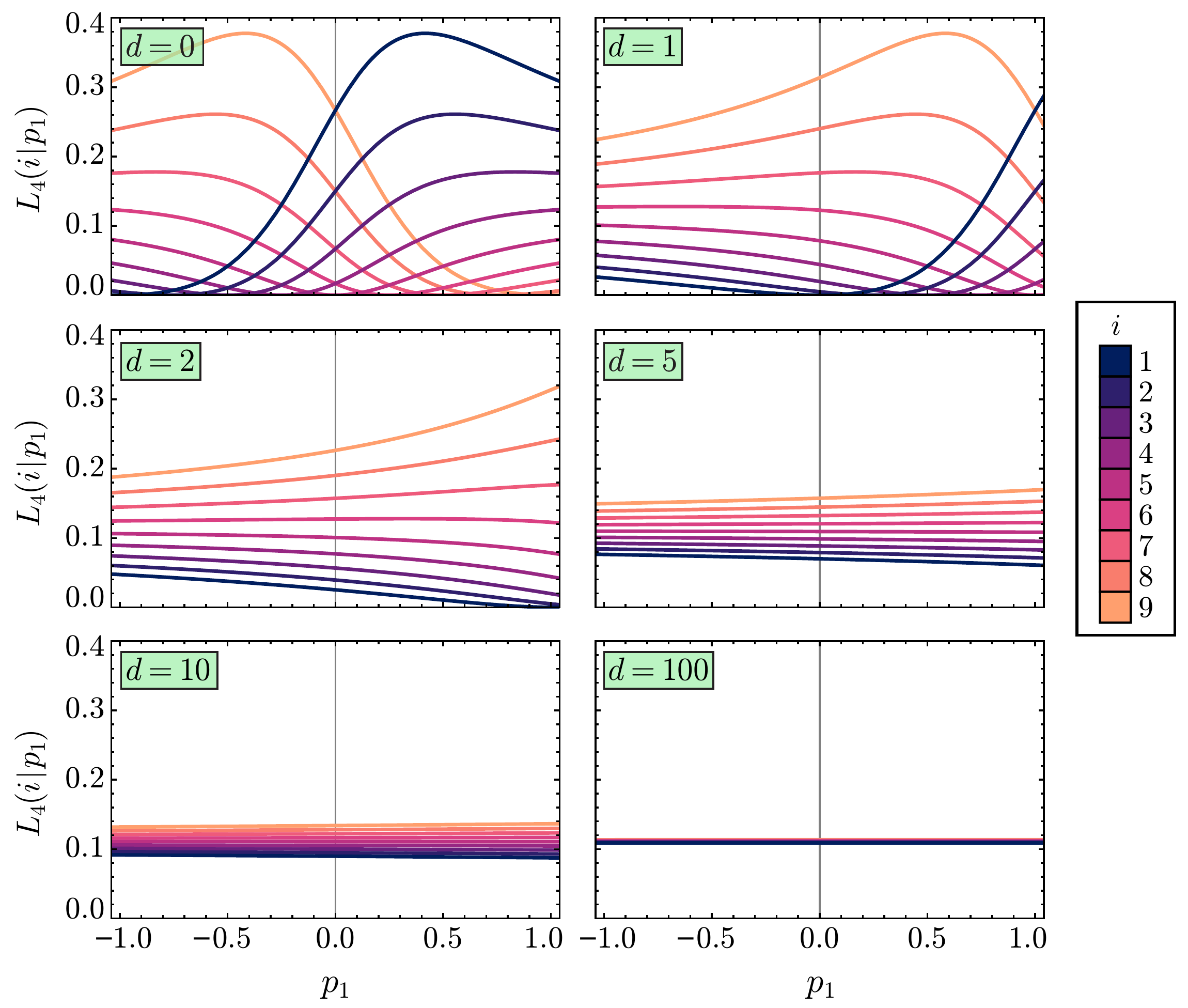}
	\caption{Likelihood functions $L_4(i|p_1)$ associated with each pixel $i$ in a measurement with theoretical intensity distribution $I_4(p_1)$, plotted for several nonnegative values of $d$. Notice that the effect of $d$ is simply a horizontal translation; when $d\gg 1$, the range of interest $p_1\in[-1,1]$ only contains a small portion of the left tail of the distribution. Symmetric results are obtained for the corresponding negative values of $d$, for which the curves are translated in the opposite direction (with respect to the $d=0$ case).}
	\label{fig:MLE_Lplot4}
\end{figure}
Observe that when $d=1$, the intensity profile and likelihood function are translated in parameter space so that they are symmetric about $p_1=1$. As $d$ increases, the distribution continues to shift farther away from the off-null condition of $I_3(x;p_1)$, so that the intensity becomes large and uniform over the range of interest of $p_1$ and the likelihood function becomes very flat. As seen in Fig.~\ref{fig:MLE_Fisher_Ex4}, the expected Fisher information per detected photon\footnote{Henceforth, all mentions of the Fisher information refer to the expected information per detected photon unless specified otherwise.} decreases rapidly as $d$ increases. 
\begin{figure} 
	\centering
	\includegraphics[height=2.25in]{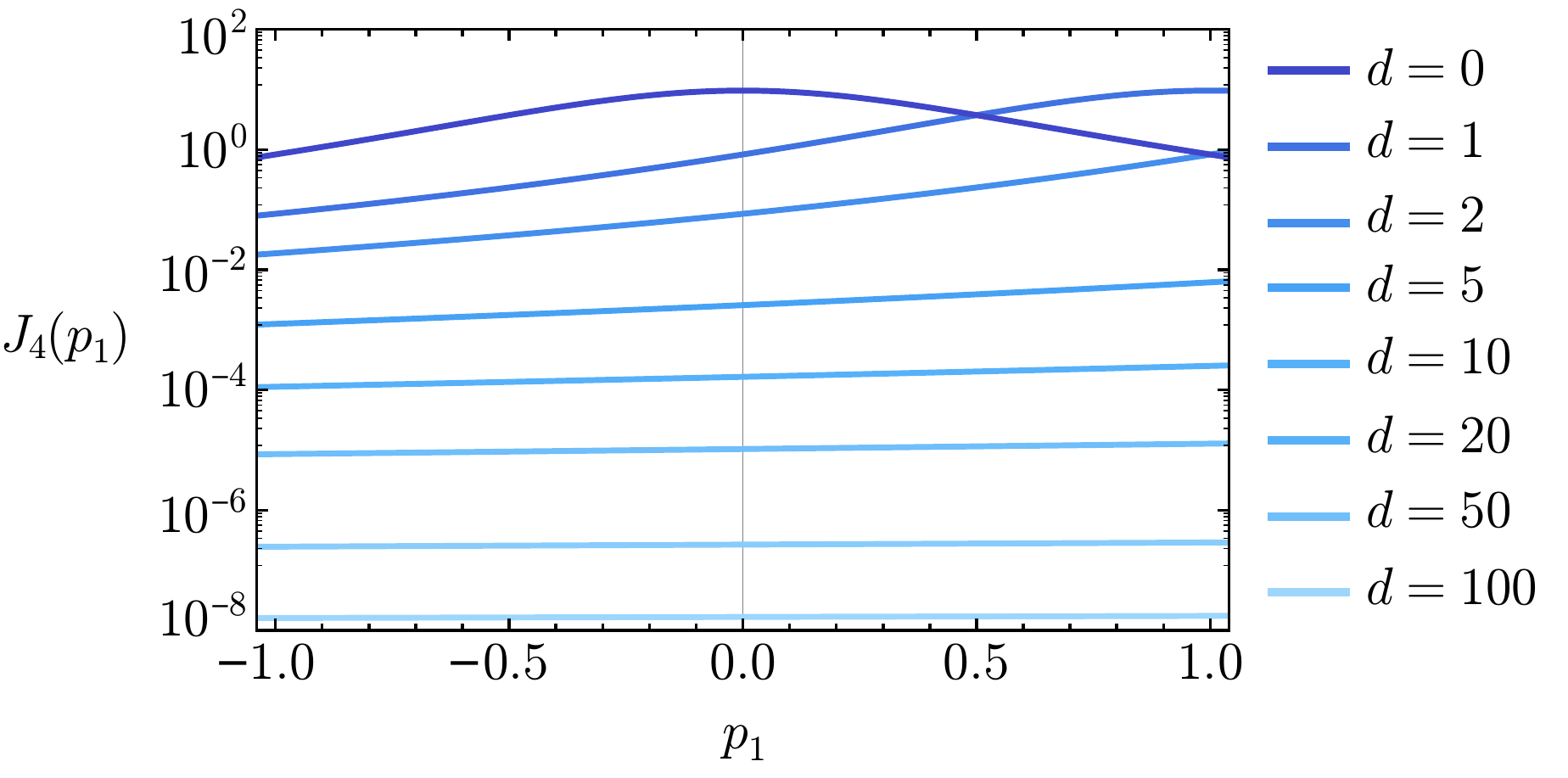}
	\caption{Expected unit Fisher information $J_4(p_1)$ for a measurement of $I_4(x;p_1)$, plotted on a logarithmic scale for several values of $d$. The $d=0$ case is identical to $J_3(p_1)$ with $c=1$ (see Fig.~\ref{fig:MLE_Fisher_Ex3}). For negative values of $d$, each curve is flipped about the vertical $p_1=0$ axis.}
	\label{fig:MLE_Fisher_Ex4}
\end{figure}
Following the same procedure as in the previous example, it can be shown that the average estimation error over the parameter range is minimized when $d=0$. (This holds true when optimizing for detected or emitted photons, though as noted before, the latter result is in part due to the choice of normalization of the intensity.)

The takeaway from this example is that it illustrates the statistical advantage of off-null measurements over a ``far-from-null'' experimental configuration in which the parameter of interest causes a small fractional change in the output intensity. Although the parameter estimation technique outlined in Section \ref{sect:MLE_optics} is only useful for imaging experiments where the off-null condition (and thus the output intensity) varies with position, by looking at Fig.~\ref{fig:MLE_IntPlots_Ex4} one can also appreciate the principle of traditional off-null ellipsometry, in which only the total power is measured. In that case, the off-null configuration greatly increases the contrast of the variation in power with respect to $p_1$, enabling a more accurate measurement while placing less stringent requirements on the fidelity of the sensor. More generally, a similar argument can be made for a broader class of optical experiments that are applications of the weak measurement formalism in quantum mechanics \cite{Aharonov_1988,Tamir_2013,Svensson_2013}, wherein preselected and postselected states are chosen to enhance the sensitivity to small variations of an unknown parameter. Examples of such applications include the measurement of small optical beam shifts \cite{Hosten_2008,Dennis_2012} and the focused beam scatterometry experiment discussed in Ref.~\cite{Vella_2018_fbs_arxiv}.

\section{Two-parameter optical MLE examples}\label{sect:MLE_examples_2param}
To illustrate the use of MLE in the multiple-parameter case, this section contains several intensity distributions that depend on two parameters $\bp=(p_1,p_2)$. The procedures for calculating the PMF, FIM, and expected error are fundamentally the same as in the one-parameter case, although the algebra is more complicated. Rather than dwelling on the mathematical details, numerical results are presented in the following discussion. This is representative of most real-world applications, in which MLE techniques are typically implemented numerically.

The intensity distributions discussed in Sections \ref{sect:MLE_example5} through \ref{sect:MLE_example10} are summarized in Table \ref{tbl:MLE_2param_intensity_dist}. 
\begin{table}
	\renewcommand{\arraystretch}{.8}
	\begin{center} 
		\begin{tabular}{c@{\hspace{20pt}}l}
			\toprule
			Section & Intensity distribution\\
			\midrule
			\phantom{a}&\\[-8pt]
			\ref{sect:MLE_example5} & $I_5(x;\bp) = 0.563\sp\Pi(x)[2 + p_1 x + p_2 \sin(\pi x)]$\\[10pt]
			\ref{sect:MLE_example6} & $I_6(x;\bp) = 0.250\sp\Pi(x)[2 + p_1 x + p_2 \cos(\pi x)]$\\[8pt]
			\ref{sect:MLE_example7} & $I_7(x;\bp) = 
			\begin{cases}
			0.5\Pi(x)(1 + p_1 x),&x<0\\
			0.5\Pi(x)(1 + p_2 x),&x\geq 0
			\end{cases}$\\[20pt]
			\ref{sect:MLE_example8} & $I_8(x;\bp) = 
			\begin{cases}
			0.5\Pi(x)\left[1 + 2p_1 (x+0.625)\right],&x<-0.125\\
			0.5\Pi(x),&-0.125\leq x< 0.125\\
			0.5\Pi(x)\left[1 + 2p_2 (x-0.625)\right],&x\geq 0.125
			\end{cases}$\\[26pt]
			\ref{sect:MLE_example9} & $I_9(x;\bp) = 0.125\Pi(x)\bigl[(p_1-x)^2 + (p_2-\cos(\pi x))^2\sp\bigr]$\\[10pt]
			\ref{sect:MLE_example10} & $I_{10}(x;\bp) = 0.320\Pi(x)\!\left[(p_1-0.25x)^2 + (p_2-0.25\cos(\pi x))^2\right]$\\
			\bottomrule
		\end{tabular}
		\caption{Intensity distributions for each example considered in Section \ref{sect:MLE_examples_2param}.}
		\label{tbl:MLE_2param_intensity_dist}
	\end{center}
\end{table}
Similarly to the one-parameter examples, each intensity distribution is normalized so that it attains a maximum value of $I_0$ over the region of interest $-1\leq p_1,p_2\leq 1$. The distributions considered in Sections \ref{sect:MLE_example5} and \ref{sect:MLE_example6} each have a $p_1$ term with linear spatial variation and a $p_2$ term with sinusoidal spatial variation, serving as simple examples for the two-parameter case. 
Sections \ref{sect:MLE_example7} and \ref{sect:MLE_example8} contain two thought-provoking (albeit unrealistic) examples that illustrate the mathematical mechanisms that can lead to statistical correlations between the parameter estimates for $p_1$ and $p_2$. Finally, a pair of two-parameter off-null measurements are discussed in Sections \ref{sect:MLE_example9} and \ref{sect:MLE_example10}.

\FloatBarrier
\subsection{Linear and sinusoidal variations (case 1)}\label{sect:MLE_example5}
For the first two-parameter example, consider the intensity distribution
\begin{equation}
I_5(x;\bp) = 0.563\sp\Pi(x)[2 + p_1 x + p_2 \sin(\pi x)]\sp,\label{eq:MLE_I_Ex5}
\end{equation}
which is valid over the region of interest $-1\leq p_1,p_2\leq 1$. Similarly to the first example in Section \ref{sect:MLE_examples_1param}, $I_5(x;\bp)$ depends linearly on the product of $p_1$ and $x$. The dependence on $p_2$ is also linear, but this additional term varies sinusoidally across the sensor. Therefore, variations in $p_1$ and $p_2$ result in distinct changes in the shape of the intensity $I_5(x;\bp)$ and the PMF $P_5(i|\bp)$, as shown in Fig.~\ref{fig:MLE_IntPlots_Ex5}. 
\begin{figure}
	\centering
	\includegraphics[width=\linewidth]{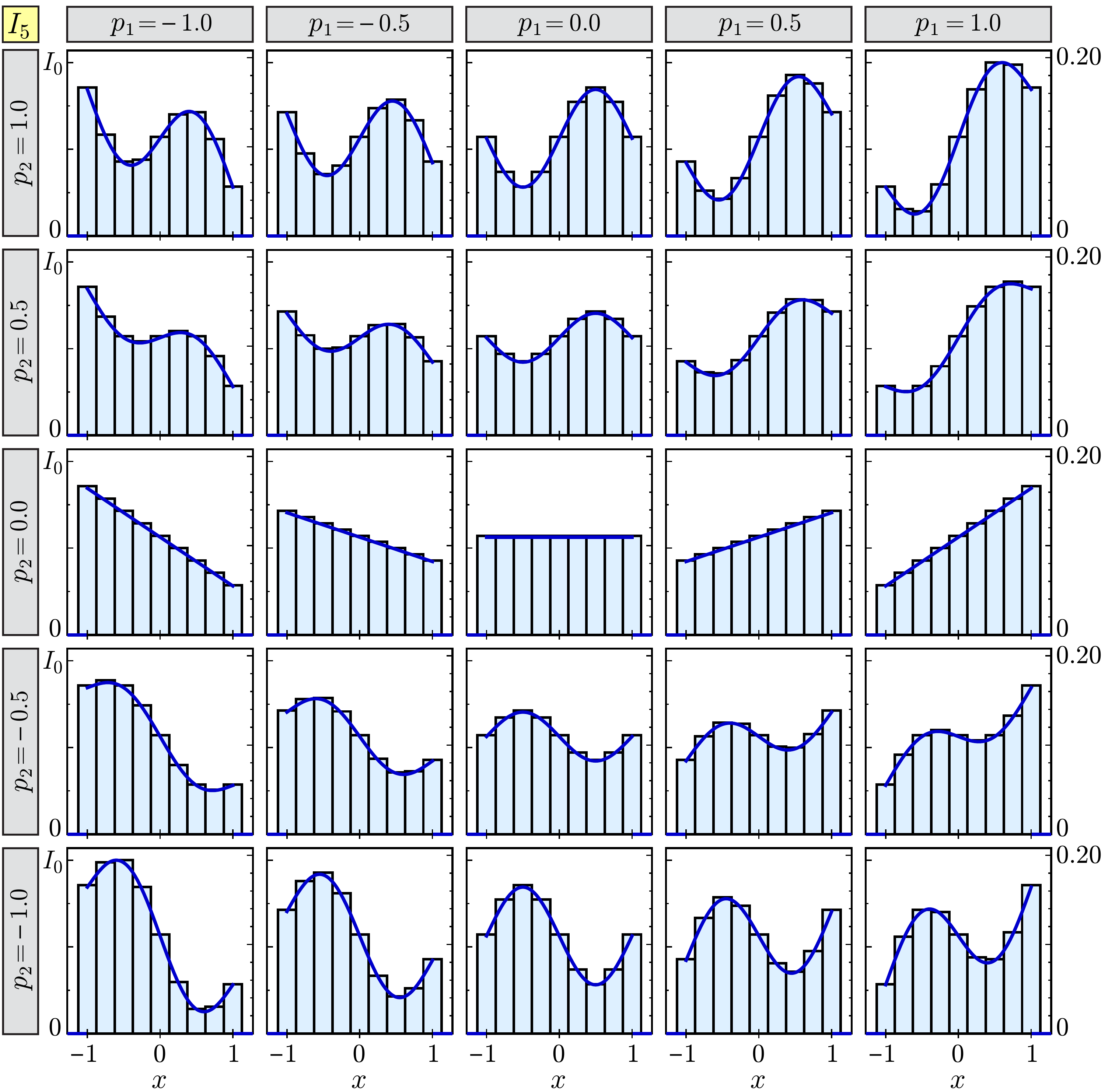}
	\caption{Plots of $I_5(x;\bp)$ (left axes) and $P_5(i|\bp)$ (right axes) for several values of $p_1$ and $p_2$.}
	\label{fig:MLE_IntPlots_Ex5}
\end{figure}
For instance, when $p_2=0$ (the third row of plots), the intensity is strictly a linear function of $x$ with slope $p_1$. When $p_1=0$ (the third column of plots), it is a sine function with a DC offset. For all other cases, the intensity is a linear combination of the two.

For the two-parameter case, the likelihood $L_5(\bp|i)=P_5(i|\bp)$ can be plotted in two dimensions as a function of $p_1$ and $p_2$. The likelihood functions associated with each pixel are shown in Fig.~\ref{fig:MLE_L2_Ex5}, with contour lines drawn as a visual aid to identify paths of constant likelihood.
\begin{figure}
	\centering
	\includegraphics[width=\linewidth]{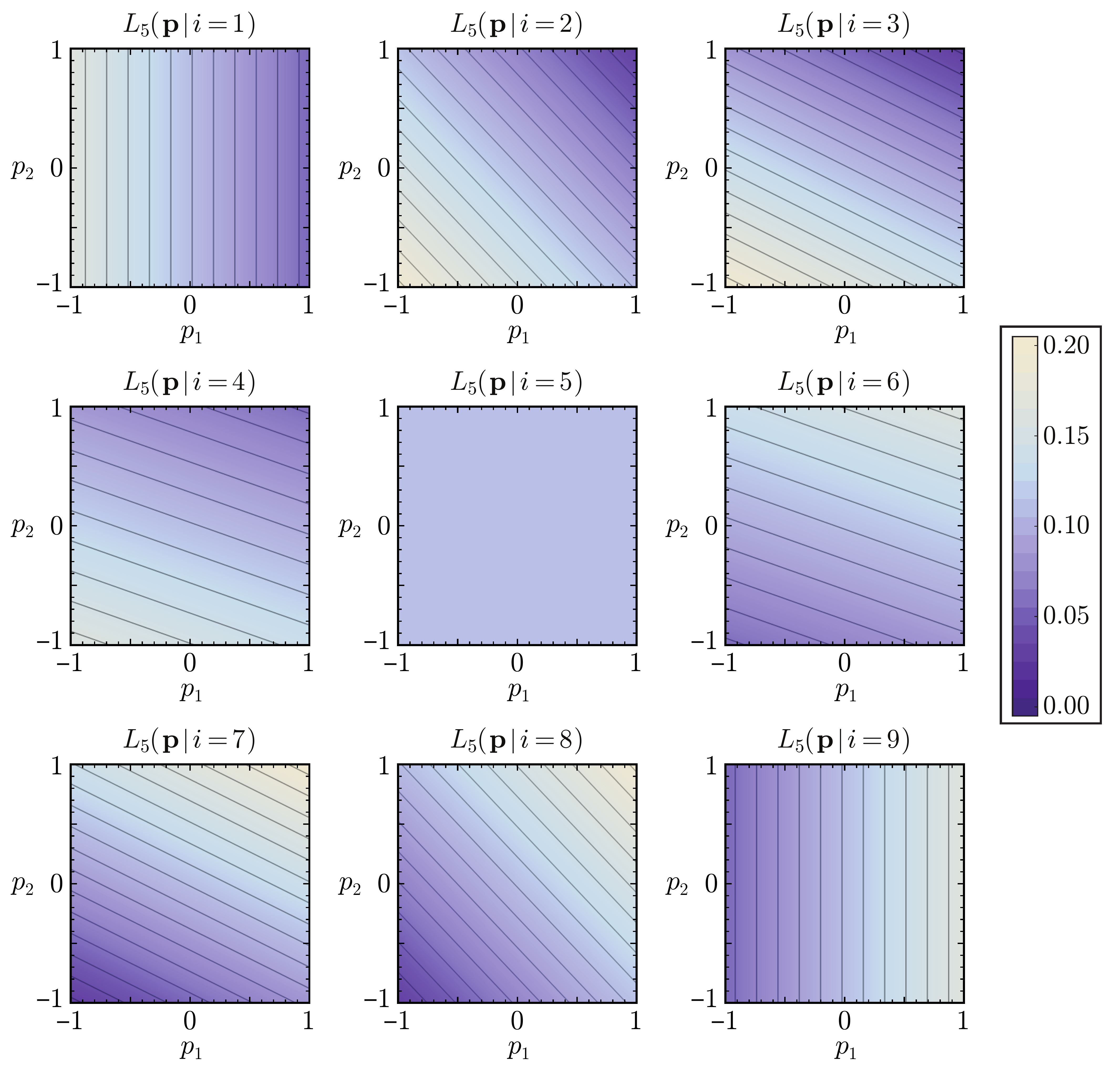}
	\caption{Likelihood functions $L_5(\bp|i)$ associated with each pixel $i$ for a measurement of $I_5(x;\bp)$. Contour lines are shown in increments of $0.01$.}
	\label{fig:MLE_L2_Ex5}
\end{figure}
These plots have several interesting features. First, notice that $L_5(\bp|i=5)$ is constant, meaning that pixel 5 provides no useful information about $p_1$ and $p_2$. (Incidentally, this was also the case for the one-parameter intensity distributions $I_1$ and $I_2$. Since the signal from pixel 5 has no effect on the MLE, it can be ignored.) Secondly, the likelihood functions for pixels 1 and 9 are independent of $p_2$ (as evident from the vertical contour lines) since $\sin(\pi x)=0$ for $x=\pm 1$. In contrast, the likelihood functions associated with pixels 4 and 6 depend more strongly on $p_2$ than $p_1$ as a consequence of the fact that $\sin(\pi x)$ has a larger slope near the center of the sensor than the linear term $x$. Lastly, note that the paths of constant likelihood generally have negative (or vertical) slopes in parameter space. Roughly speaking, this means that if $p_1$ increases and $p_2$ decreases by a similar amount (or if $p_2$ increases and $p_1$ decreases), the likelihood function will only change slightly, making it difficult to distinguish linear combinations of parameters along this direction. On the other hand, a simultaneous increase (or simultaneous decrease) in $p_1$ and $p_2$ will tend to cause a more significant change in the likelihood function, making it easier to distinguish this type of variation in $\bp$.

The patterns described above can be quantified by calculating the estimation error based on the $2\times 2$ expected Fisher information matrix, whose elements may be computed using either form of Eq.~(\ref{eq:MLE_Fisher}). For a measurement of $\mathcal{N}=1000$ photons with true parameter values $\bp=(0,0)$, the FIM and its inverse are found to be
\begin{equation}
\renewcommand{\arraystretch}{.8}
\mathcal{N}\mathbb{J}_5 = \left[\!
\begin{array}{rr}
104.2 & 67.1 \\ 67.1 & 111.1
\end{array}
\!\right]\nsp,
\qquad\quad
(\mathcal{N}\mathbb{J}_5)^{-1} = \left[\!
\begin{array}{rr}
0.0157 & -0.0095 \\ -0.0095 & 0.0147
\end{array}
\!\right]\nsp.\label{eq:MLE_Ex5_Fisher}
\end{equation}
As discussed in Section \ref{sect:MLE_overview}, $(\mathcal{N}\mathbb{J}_5)^{-1}$ places a lower limit on the covariance matrix for a 1000-photon measurement of $p_1$ and $p_2$. Since its off-diagonal elements are fairly large in relation to its diagonal elements, a strong coupling between parameters (i.e., large covariance) is expected. Indeed, the principal axes of the error ellipse are given by the eigenvectors $[0.69;0.72]$ and $[0.72;-0.69]$, and the axis lengths (the square roots of the corresponding eigenvalues) are $0.076$ and $0.157$, respectively. Thus, the major axis of the ellipse is oriented at approximately $-45^\circ$ in parameter space, and the standard deviation error is about twice as large along the $-45^\circ$ direction as the $+45^\circ$ direction.\footnote{It is only meaningful to refer to angles in parameter space when $p_1$ and $p_2$ have the same units and are normalized to their respective ranges of interest, as they are in this discussion.} In this example, it turns out that similar results are obtained for all values of $\bp$ within the region of interest. The error ellipses for a selection of true parameter values are plotted in Fig.~\ref{fig:MLE_ellipses_Ex5}.

\begin{figure}
	\centering
	\includegraphics[width=.553\linewidth]{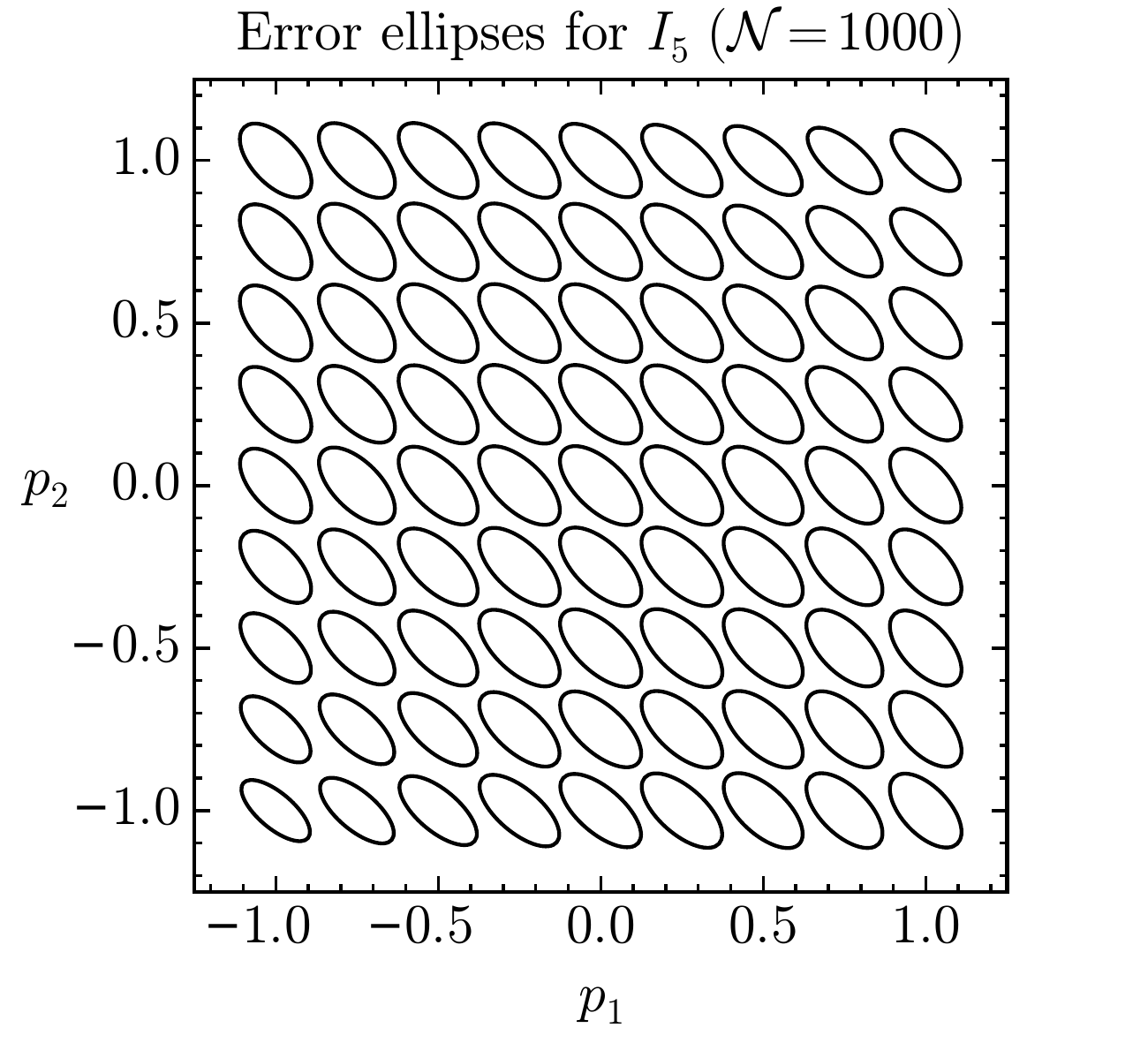}
	\caption{Ellipses representing the expected standard deviation error of a \mbox{1000-photon} measurement of $I_5(x;\bp)$ with true parameter values $p_1$ and $p_2$, sampled over a $9\times 9$ grid in parameter space.}
	\label{fig:MLE_ellipses_Ex5}
\end{figure}

Given a measured intensity $\tbI$, the magnitude and orientation of the uncertainty of the MLE are also manifested in the shape of the likelihood function $L_5(\bp|\tbI)$ and its logarithm $\ell_5(\bp|\tbI)$. Fig.~\ref{fig:MLE_LL2_1000ph_Ex5} contains two examples of the log-likelihood functions obtained for simulated 1000-photon measurements with true parameter values $\bp=(0,0)$ and $\bp=(0.63,-0.25)$. 
\begin{figure}
	\centering
	\includegraphics[width=\linewidth]{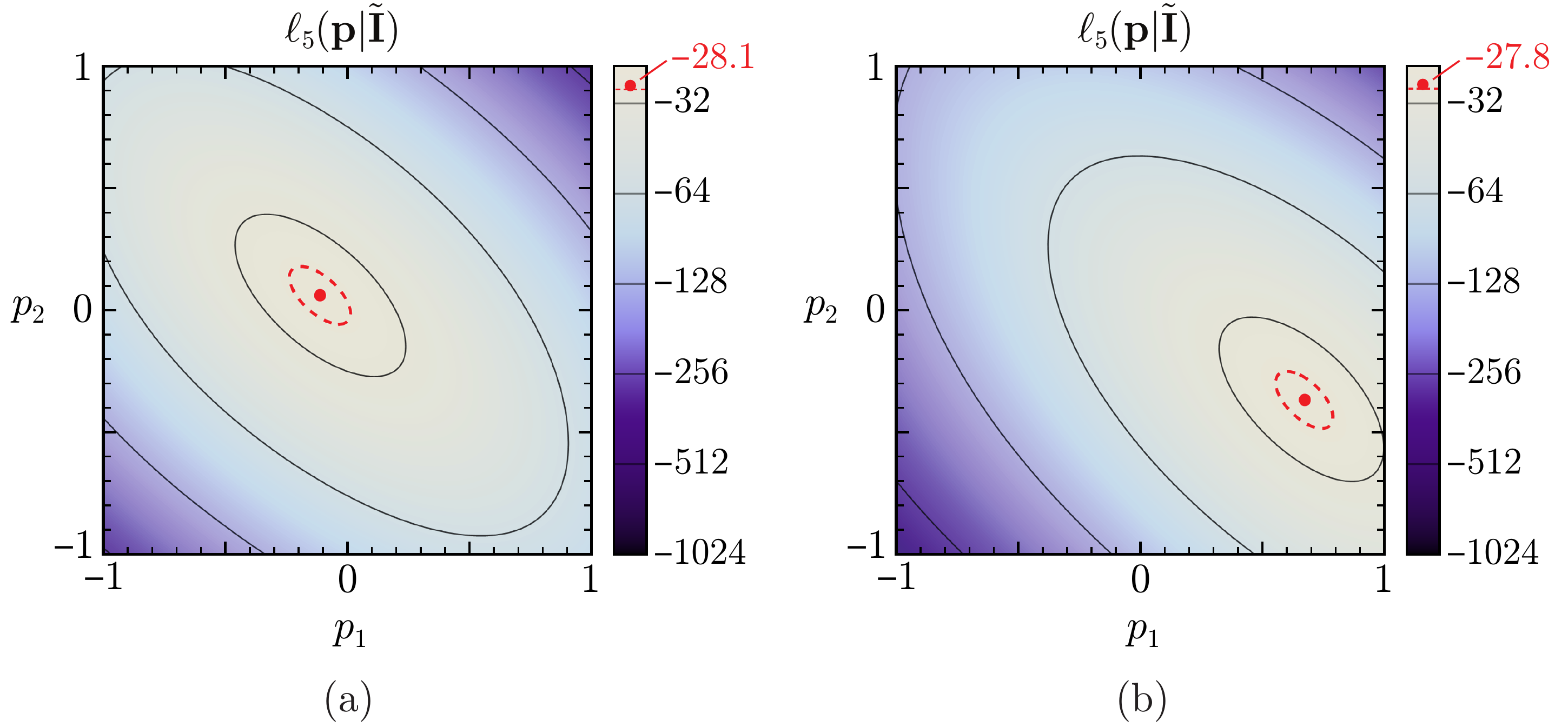}
	\caption{Log-likelihood functions $\ell_5(\bp|\tbI)$ for simulated 1000-photon measurements of $I_5(x;\bp)$ with true parameter values (a) $\bp\nsp=\nsp(0,0)$ and (b) $\bp\nsp=\nsp(0.63,-0.25)$. The plots are shaded on a logarithmic scale with solid contour lines drawn at powers of 2, as indicated in the legend. The peak of each distribution is marked with a red dot. The locations of these maxima (i.e., the MLEs for each measurement) are $\bp=(-0.115,0.064)$ and $\bp=(0.673,-0.366)$, respectively. The dashed contour line indicates where the likelihood $L_5(\bp|\tbI)$ drops to $1/\sqrt{e}$ times its peak value, representing the standard deviation confidence interval for the MLE.}
	\label{fig:MLE_LL2_1000ph_Ex5}
\end{figure}
Again, these plots contain several interesting features. First, notice that the contours of equal likelihood are approximately elliptical. This behavior is characteristic of a bivariate Gaussian distribution $f(\bp)=f_0\exp(-\frac{1}{2}\bp^{\rm T}\bm{\Sigma}^{-1}\sp\bp)$ with covariance matrix $\bm{\Sigma}$, for which the locus of points satisfying $\bp^{\rm T}\bm{\Sigma}^{-1}\sp\bp=\kappa^2$ (for some constant $\kappa$) traces out an ellipse \cite{Friendly_2013}. Thus, the shape of $\ell_5(\bp|\tbI)$ supports the claim made earlier (see Eq.~(\ref{eq:MLE_P(p|I)_Gaussian})) that the posterior probability distribution $P(\bp|\tbI)$, which is a scaled version of the likelihood if no prior distribution is assumed, closely approximates a Gaussian distribution when a large number of photons are measured. Comparing Figs.~\ref{fig:MLE_ellipses_Ex5} and \ref{fig:MLE_LL2_1000ph_Ex5}, one can also see that the likelihood function is elongated along the direction with the largest expected estimation error. In Section \ref{sect:MLE_examples_1param} it was noted that the estimation error is largest when the likelihood function is nearly flat; for the multiple-parameter case, it can be further specified that the error is largest along the \emph{direction} where the likelihood function is flattest, i.e., the direction perpendicular to the local gradient of $\ell$ with respect to $\bp$.

Each plot in Fig.~\ref{fig:MLE_LL2_1000ph_Ex5} contains a red dot representing the MLE for the measurement, i.e., the location of the peak of $\ell_5(\bp|\tbI)$. The estimated parameter values (which are listed in the figure caption) differ considerably from the true values, with errors as large as $\sim\! 0.11$ for each parameter. The standard deviation confidence interval for the MLE, which is outlined by a red dashed line, consists of the region where the likelihood function $L_5(\bp|\tbI)$ is greater than or equal to $1/\sqrt{e}$ times its peak value.\footnote{For the Gaussian distribution $f(\bp)$ mentioned above, the $\kappa=1$ ellipse encloses one standard deviation. Along this contour, the function value drops to $f_0\exp(-\frac{1}{2})=f_0/\sqrt{e}$.} This is equivalent to an additive decrease in the log-likelihood by $\ln(e^{-1/2})=-0.5$. Notice that this region is elliptical, and its size and shape are virtually identical to the nearest ellipse in Fig.~\ref{fig:MLE_ellipses_Ex5}. In fact, by evaluating the expected FIM at the MLE with $\mathcal{N}=1000$, an extremely close agreement is found between the predicted covariance matrix $(\mathcal{N}\mathbb{J}_5)^{-1}$ and the standard deviation confidence interval of $\ell_5(\bp|\tbI)$. (When plotted together, the ellipses are virtually indistinguishable even when zoomed in.) In general, the correlation between the two grows stronger as the number of photons increases. In this example, 1000 photons are sufficient to obtain a very close agreement; in an experiment with smaller expected error, fewer photons would be required.

To conclude this example, similarly to Sections \ref{sect:MLE_example1} and \ref{sect:MLE_example2}, a Monte Carlo simulation was performed for 50,000 trials of a 1000-photon simulated measurement of $I_5(x;\bp)$ for which the true parameter values are given by $\bp=(0,0)$. A histogram of the maximum likelihood estimates obtained in all trials is shown in Fig.~\ref{fig:MLE_hist2_Ex5}(a); an overhead view of the distribution is also shown in Fig.~\ref{fig:MLE_hist2_Ex5}(b). 
\begin{figure}
	\centering
	\includegraphics[width=\linewidth]{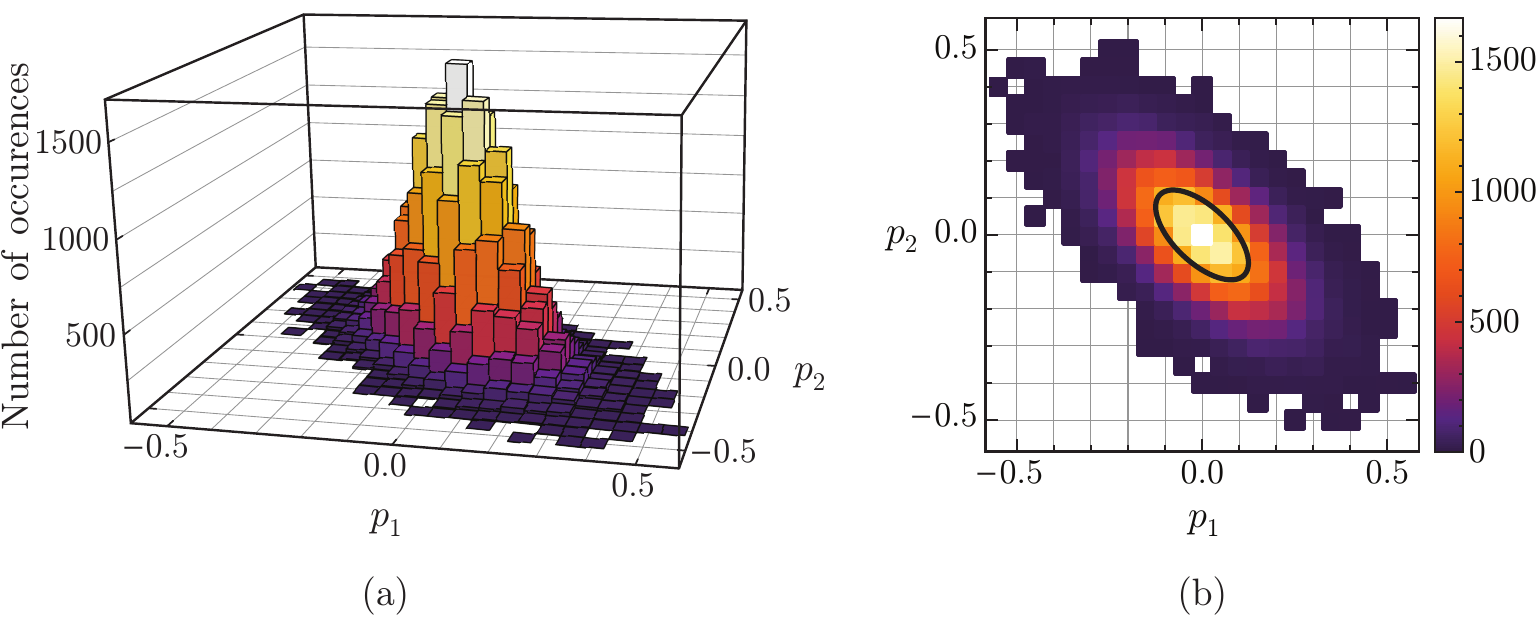}
	\caption{(a) Histogram of the maximum likelihood estimates obtained from 50,000 trials of a simulated 1000-photon measurement of $I_5(x;\bp)$ with true parameter value $\bp=(0,0)$. (b) Overhead view of the distribution shown in plot (a), with the color of each pixel indicating the number of trials for which the MLE was within a given interval. The black ellipse at the center of the plot represents the expected standard deviation error based on the Fisher information matrix.}
	\label{fig:MLE_hist2_Ex5}
\end{figure}
The data closely resembles a Gaussian distribution with the same orientation as the expected error ellipse, which is shown in black in the overhead view. The statistical covariance matrix of the data matches the matrix $(\mathcal{N}\mathbb{J}_5)^{-1}$ given in Eq.~(\ref{eq:MLE_Ex5_Fisher}) to within three significant digits.

\subsection{Linear and sinusoidal variations (case 2)}\label{sect:MLE_example6}
For the second two-parameter example, consider the intensity distribution
\begin{equation}
I_6(x;\bp) = 0.250\sp\Pi(x)[2 + p_1 x + p_2 \cos(\pi x)]\sp,
\end{equation}
which is similar to $I_5(x;\bp)$, but with the sine term replaced by a cosine. The intensity and PMF are plotted for several parameter values in Fig.~\ref{fig:MLE_IntPlots_Ex6}, and the likelihood functions for each pixel are shown in Fig.~\ref{fig:MLE_L2_Ex6}.
\begin{figure}
	\centering
	\includegraphics[width=\linewidth]{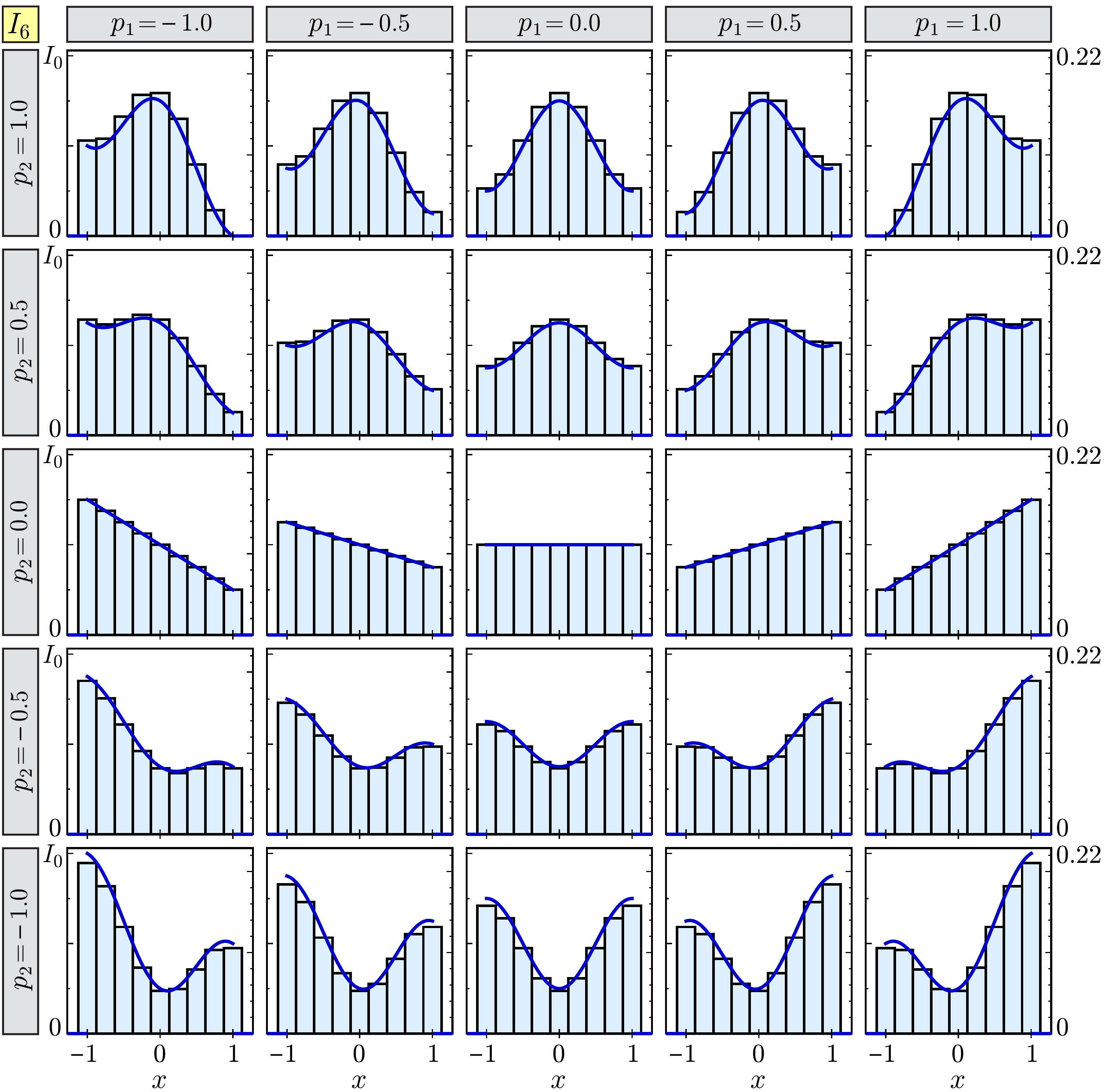}
	\caption{Plots of $I_6(x;\bp)$ (left axes) and $P_6(i|\bp)$ (right axes) for several values of $p_1$ and $p_2$.}
	\label{fig:MLE_IntPlots_Ex6}
\end{figure}
\begin{figure}
	\centering
	\includegraphics[width=\linewidth]{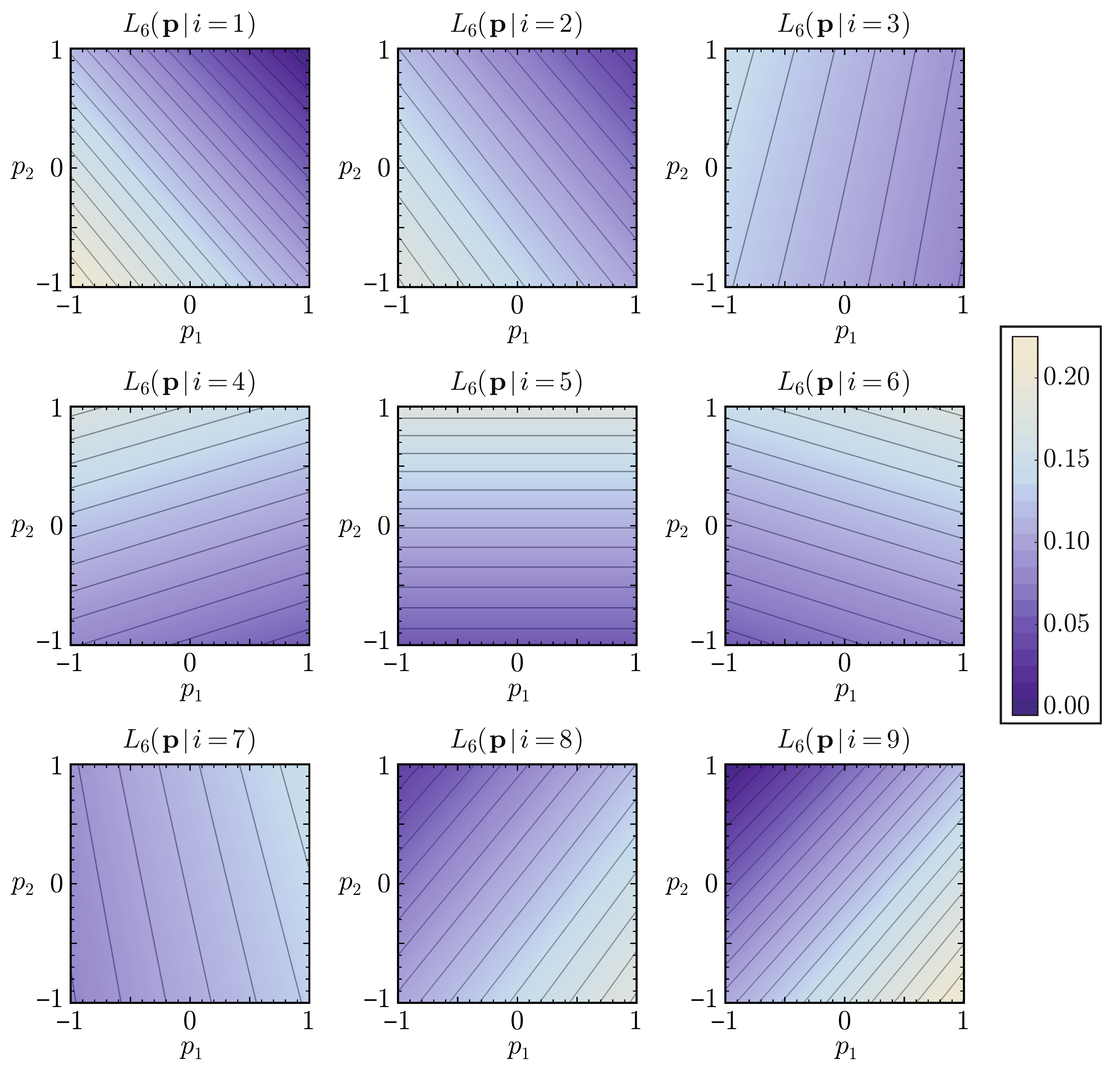}
	\caption{Likelihood functions $L_6(\bp|i)$ associated with each pixel $i$ for a measurement of $I_6(x;\bp)$. Contour lines are shown in increments of $0.01$.}
	\label{fig:MLE_L2_Ex6}
\end{figure}
In this example, it can be seen\nolinebreak[3] that the paths of constant likelihood have different orientations for each pixel. This implies, for instance, that a simultaneous increase in $p_1$ and $p_2$ will cause a significant change in $L_6(\bp|i\!=\!1)$, but very little change in $L_6(\bp|i\!=\!9)$; meanwhile, a simultaneous increase in $p_1$ and decrease in $p_2$ will do just the opposite. The reason for this can be understood by examining the plots of $x$, $\sin(\pi x)$, and $\cos(\pi x)$ shown in Fig.~\ref{fig:MLE_sincos}. 
\begin{figure}
	\centering
	\includegraphics[height=2.05in]{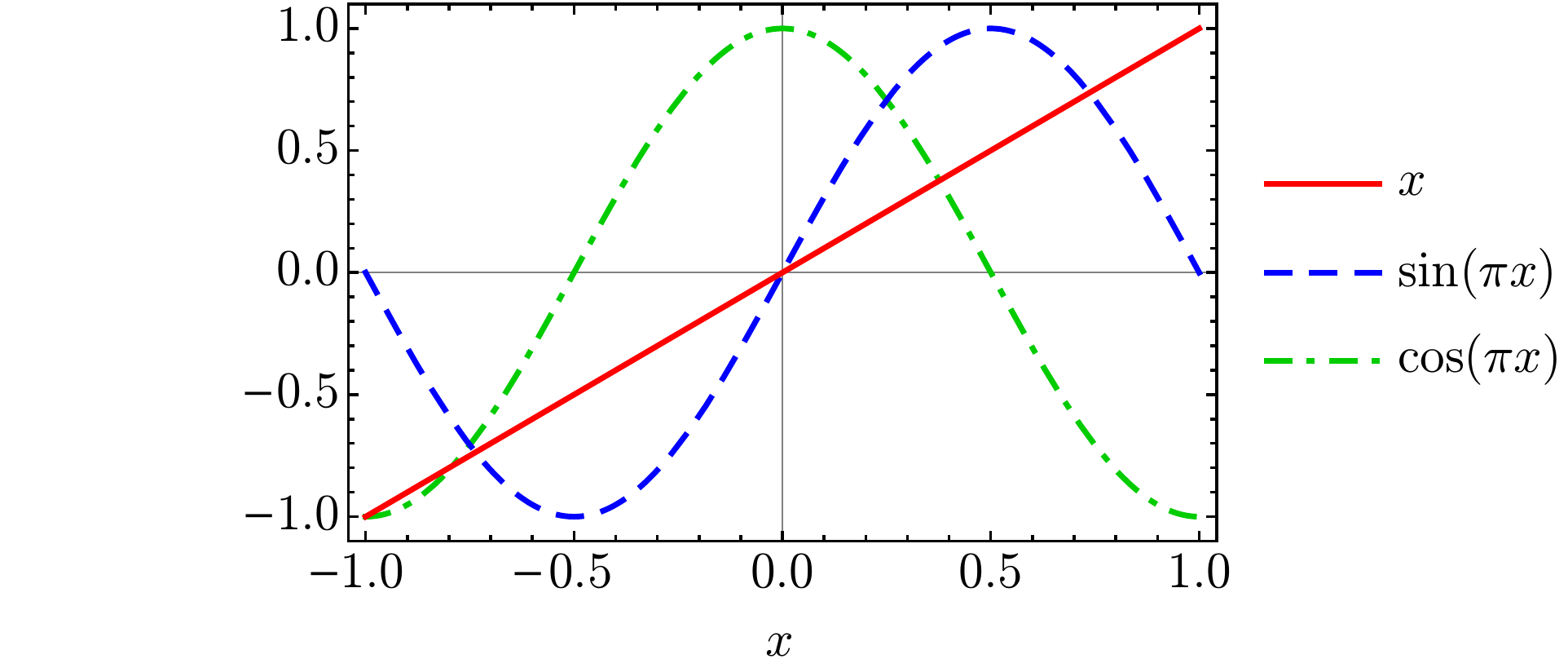}
	\caption{Spatial variations of each term appearing in intensity distributions $I_5(x;\bp)$ and $I_6(x;\bp)$.}
	\label{fig:MLE_sincos}
\end{figure}
Whereas $x$ and $\sin(\pi x)$ always have the same sign, this is not the case for $x$ and $\cos(\pi x)$. Therefore, for the intensity distribution $I_5(x;\bp)$, an increase in $p_1$ can be compensated (to a certain extent) by a decrease in $p_2$. The distribution $I_6(x;\bp)$ is less prone to this situation since any linear combination of $p_1$ and $p_2$ produces distinct fluctuations at different pixels. However, correlations can still arise in cases where very few photons are incident on one or more pixels (for example, when $p_1=p_2=1$), since the contributions of each pixel to the log-likelihood function $\ell_6(\bp|\tbI)$ associated with a measured intensity $\tbI$ may be imbalanced.

Based on the above observations, one can reasonably expect there to be a smaller correlation between the estimated parameters from a measurement of $I_6(x;\bp)$ than in the previous example. As a matter of fact, for $\bp=(0,0)$, the FIM and its inverse are diagonal, indicating that there is zero covariance:
\begin{equation}
\renewcommand{\arraystretch}{.8}
\mathcal{N}\mathbb{J}_6 = \left[\!
\begin{array}{cc}
104.2 & 0 \\ 0 & 135.8
\end{array}
\!\right]\nsp,
\qquad\quad
(\mathcal{N}\mathbb{J}_6)^{-1} = \left[\!
\begin{array}{cc}
0.0096 & 0 \\ 0 & 0.0074
\end{array}
\!\right]\nsp,\label{eq:MLE_Ex6_Fisher}
\end{equation}
where $\mathcal{N}=1000$. The eigenvectors of $(\mathcal{N}\mathbb{J}_6)^{-1}$ are $[1;0]$ and $[0;1]$, and the square roots of the corresponding eigenvalues are 0.098 and 0.086, respectively. Thus, the error ellipse is nearly circular, with its principal axes oriented along the $p_1$ and $p_2$ axes. The error ellipses for a selection of parameter values are shown in Fig.~\ref{fig:MLE_ellipses_Ex6}. 
\begin{figure}
	\centering
	\includegraphics[width=.553\linewidth]{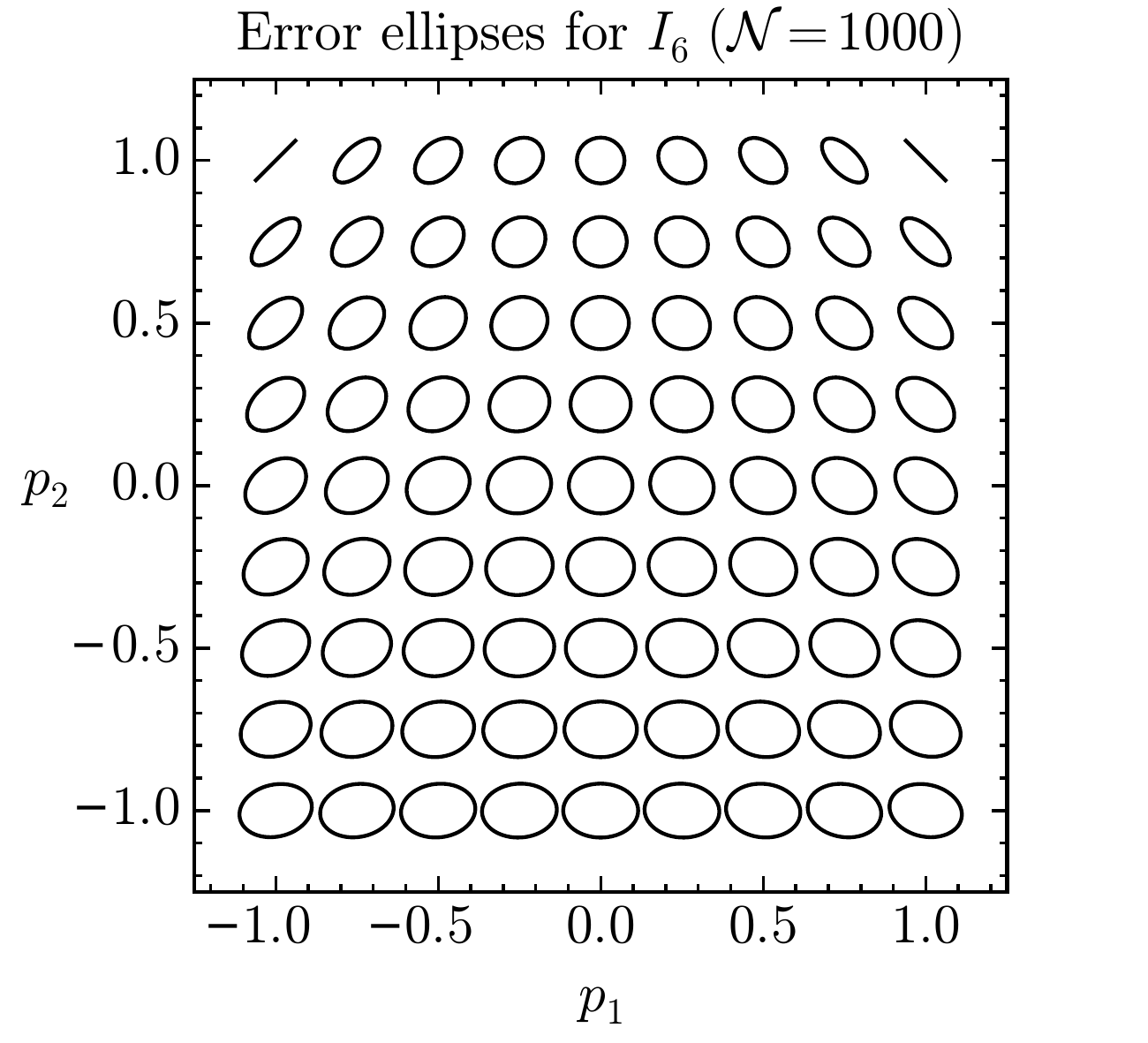}
	\caption{Ellipses representing the expected standard deviation error of a \mbox{1000-photon} measurement of $I_6(x;\bp)$ with true parameter values $p_1$ and $p_2$, sampled over a $9\times 9$ grid in parameter space.}
	\label{fig:MLE_ellipses_Ex6}
\end{figure}
As seen in the plot, the expected error is relatively uniform over the entire parameter range, with the smallest error occurring when $p_2$ is close to 1. The covariance between $p_1$ and $p_2$ is also generally small, with one notable exception: as $|p_1|\to 1$ and $p_2\to 1$, the two parameters become highly correlated. At the far upper corners of the region of interest, the error ellipse resembles a straight line, indicating complete correlation between $p_1$ and $p_2$. (Even so, the magnitude of the uncertainty of each parameter is still smaller than the expected errors for other parameter values.) From the uppermost plots in Fig.~\ref{fig:MLE_IntPlots_Ex6}, it can be seen that this correlation arises when the intensity drops to zero at either edge of the sensor (near pixel 1 or pixel 9). This happens because the intensity distribution and the likelihood functions $L(\bp|i)$ are distributed such that the remaining pixels cannot easily distinguish between all possible combinations of $p_1$ and $p_2$, as alluded to in the previous paragraph.\footnote{The astute reader might wonder why the expected error is asymmetric with respect to $p_2$ despite the fact that the last term of $I_6(x;\bp)$ exhibits symmetry with respect to both $p_2$ and $x$. The answer is that the asymmetry is a sampling artifact of the 9-pixel array, since pixels 1 and 9 sample the periodic function $\cos(\pi x)$ at points that are offset by $2\pi$ radians. This causes the total measured intensity to vary with $p_2$ despite the fact that $\int_{-1}^1\cos(\pi x)\ud x=0$. As is often the case, the error is smallest in this example when the total intensity is minimized, which occurs when $p_2=1$.}

The log-likelihood functions $\ell_6(\bp|\tbI)$ for simulated 1000-photon measurements of $I_6(x;\bp)$ with true parameter values $\bp=(0,0)$ and $\bp=(0.63,-0.25)$ are shown in Fig.~\ref{fig:MLE_LL2_1000ph_Ex6}. 
\begin{figure}[t]
	\centering
	\includegraphics[width=\linewidth]{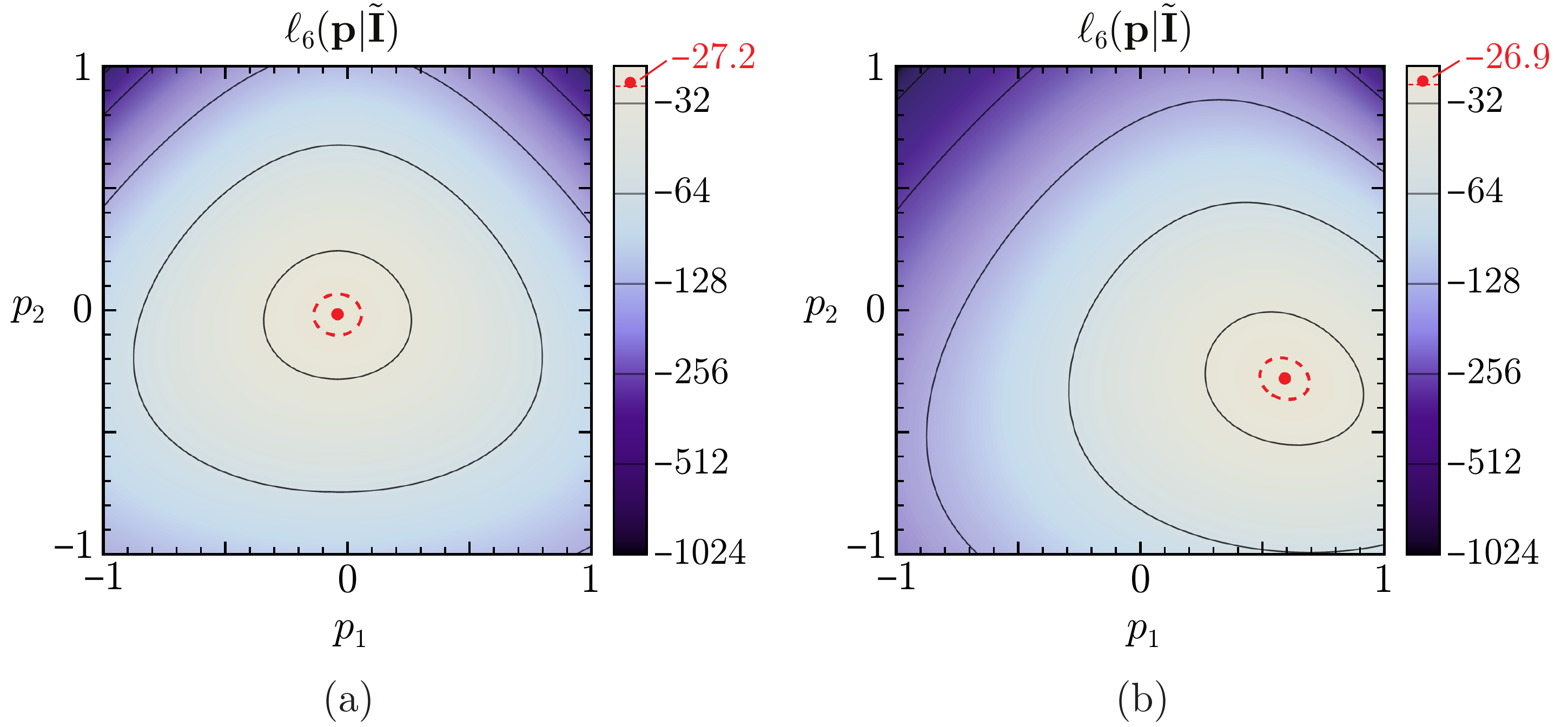}
	\caption{Log-likelihood functions $\ell_6(\bp|\tbI)$ for simulated 1000-photon measurements of $I_6(x;\bp)$ with true parameter values (a) $\bp\nsp=\nsp(0,0)$ and (b) $\bp\nsp=\nsp(0.63,-0.25)$. The plots are shaded on a logarithmic scale with solid contour lines drawn at powers of 2, as indicated in the legend. The peak of each distribution is marked with a red dot. The locations of these maxima (i.e., the MLEs for each measurement) are $\bp=(-0.043,-0.014)$ and $\bp=(0.591,-0.278)$, respectively. The dashed contour line indicates where the likelihood $L_6(\bp|\tbI)$ drops to $1/\sqrt{e}$ times its peak value, representing the standard deviation confidence interval for the MLE.}
	\label{fig:MLE_LL2_1000ph_Ex6}
\end{figure}
As in the previous example, the contours of equal likelihood are highly elliptical near the peak, indicating that the likelihood is approximately a Gaussian distribution. The Gaussian approximation weakens away from the peak, with the contours of $\ell_6(\bp|\tbI)$ becoming slightly distorted. Compared to $\ell_5(\bp|\tbI)$, the distribution is much more symmetric due to the small covariance between $p_1$ and $p_2$ (for these particular true parameter values). The standard deviation confidence interval, indicated by the dashed red line, is also highly symmetric and slightly narrower than it was in the previous example, matching the expected error based on the FIM. The uncertainty is also reflected in the distribution of the MLEs obtained from 50,000 trials of a 1000-photon measurement of $I_6(\bp|\tbI)$, as shown in Fig.~\ref{fig:MLE_hist2_Ex6}}. The diagonal elements of the covariance matrix of the simulated data agree with the matrix $(\mathcal{N}\mathbb{J}_6)^{-1}$ given in Eq.~(\ref{eq:MLE_Ex6_Fisher}) to within two significant digits; the off-diagonal elements of the matrix are very close to zero (approximately 500 times smaller than the diagonal elements). 

\begin{figure}
	\centering
	\includegraphics[width=\linewidth]{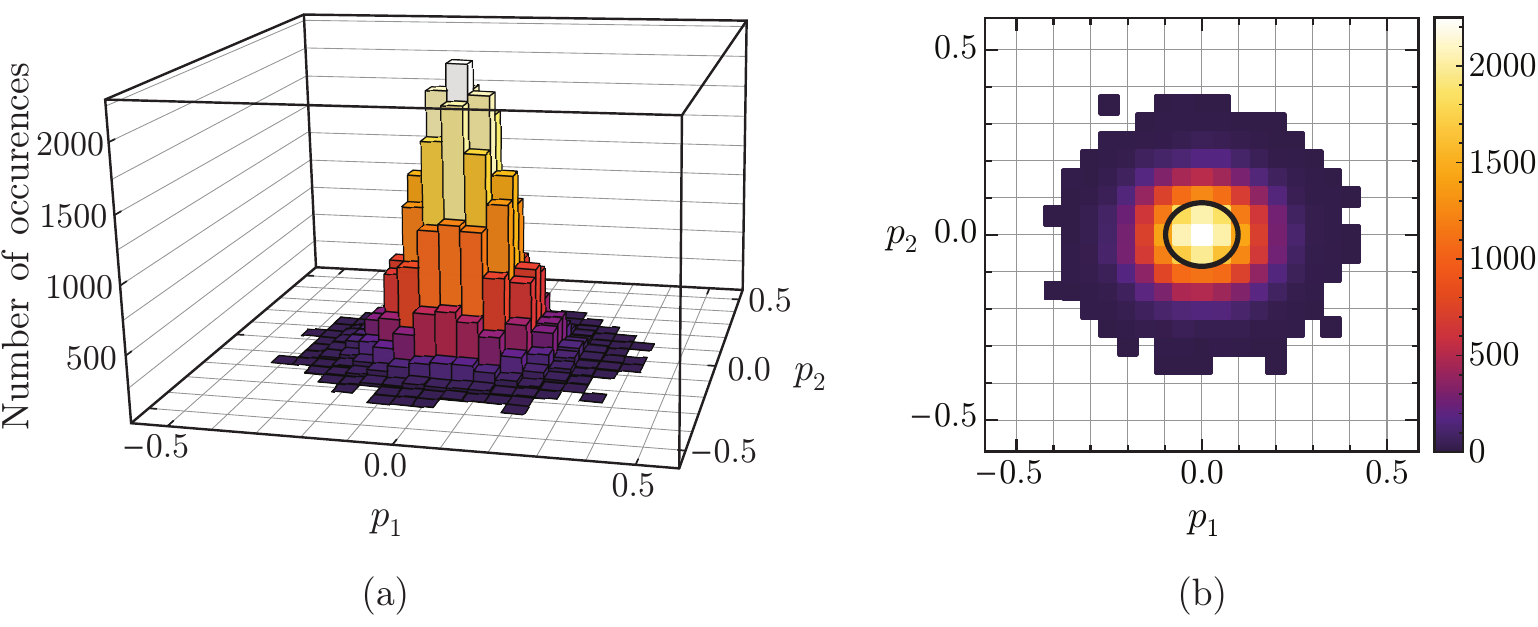}
	\caption{(a) Histogram of the maximum likelihood estimates obtained from 50,000 trials of a simulated 1000-photon measurement of $I_6(x;\bp)$ with true parameter value $\bp=(0,0)$. (b) Overhead view of the distribution shown in plot (a), with the color of each pixel indicating the number of trials for which the MLE was within a given interval. The black ellipse at the center of the plot represents the expected standard deviation error based on the Fisher information matrix.}
	\label{fig:MLE_hist2_Ex6}
\end{figure}

\subsection{Piecewise linear dependence (nonzero covariance)}\label{sect:MLE_example7}
The next two examples involve intensity distributions for which fluctuations due to $p_1$ and $p_2$ occur in completely separate portions of the sensor. Although this is not a particularly common real-world scenario, some interesting insight can be gained from the analysis. First, consider the piecewise intensity distribution
\begin{equation}
I_7(x;\bp) = 
\begin{cases}
0.5\Pi(x)(1 + p_1 x),&x<0,\\
0.5\Pi(x)(1 + p_2 x),&x\geq 0,
\end{cases}
\end{equation}
which is plotted in Fig.~\ref{fig:MLE_IntPlots_Ex7}. 
\begin{figure}
	\centering
	\includegraphics[width=\linewidth]{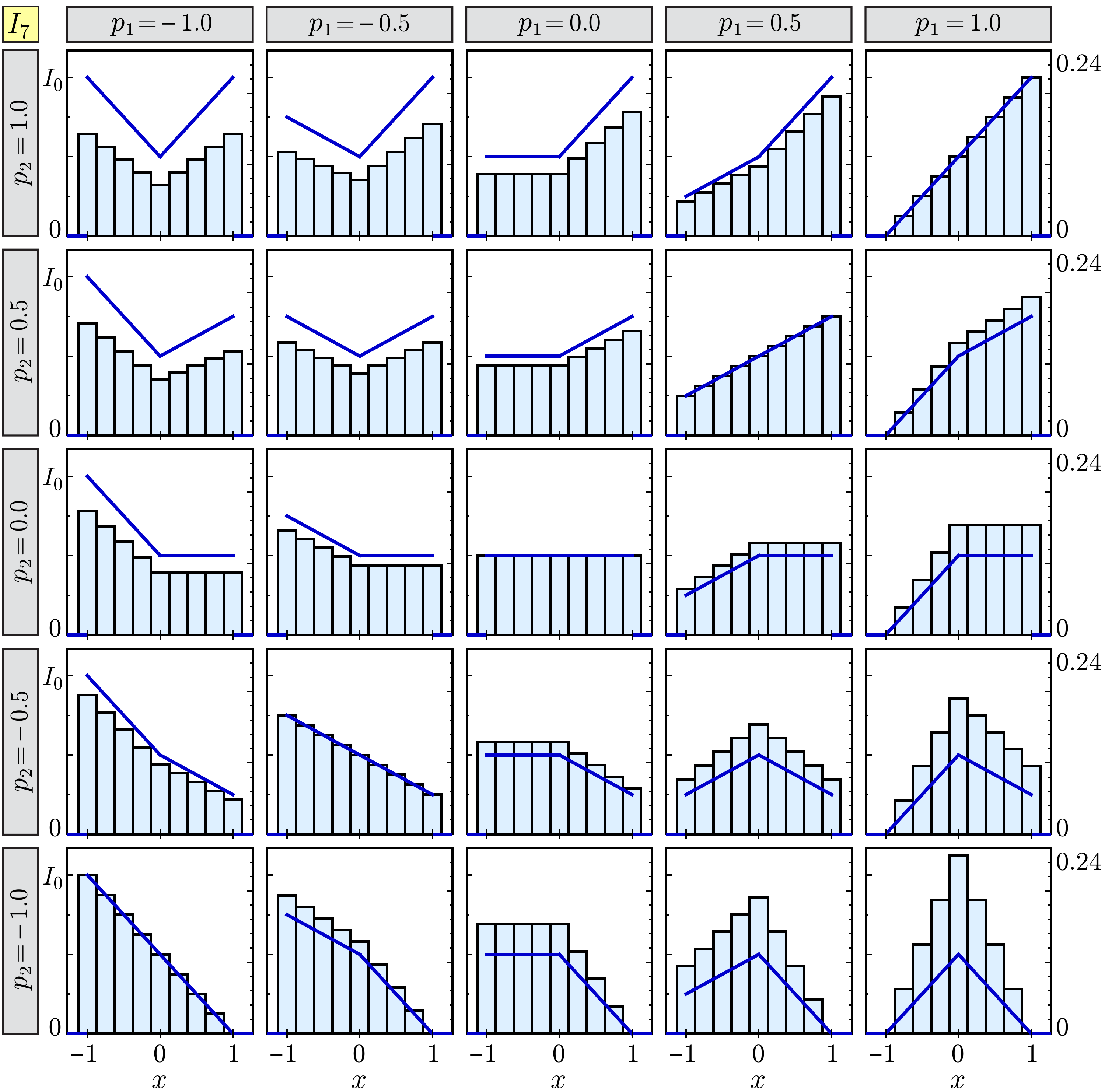}
	\caption{Plots of $I_7(x;\bp)$ (left axes) and $P_7(i|\bp)$ (right axes) for several values of $p_1$ and $p_2$.}
	\label{fig:MLE_IntPlots_Ex7}
\end{figure}
This distribution is similar to the one-parameter linear intensity profile $I_1(x;p_1)$, except that the slopes on the left and right halves of the sensor are proportional to $p_1$ and $p_2$, respectively. Since the intensities on each half of the sensor only depend on a single parameter, one would expect the parameters to be completely uncoupled, enabling an estimate with zero covariance. However, this turns out not to be the case when applying the MLE approach outlined in Section \ref{sect:MLE_optics}. (Note: the MLE formalism only requires the PMF to be twice differentiable with respect to $\bp$, so the discontinuity in the derivative of $I_7(\bx;\bp)$ with respect to $\bx$ is not problematic.) As established previously, this treatment relies on the information contained in the shape of the intensity distribution, that is, the relative intensity or the PMF. Clearly, the value of $p_1$ impacts the probability $P_7(i|\bp)$ of detecting a photon at each pixel on the left half of the sensor ($i=1,\ldots 5$); what is perhaps less obvious, however, is that it also affects the probabilities for pixels 6 through 9. Indeed, within any given row of Fig.~\ref{fig:MLE_IntPlots_Ex7} (for which $p_2$ has a fixed value), the intensity on the right half of the sensor is always the same, yet the PMF changes depending on the value of $p_1$. This is possible because the total intensity $\sum_i I_7(x_i|\bp)$, which appears in the denominator of $P_7(i|\bp)$, varies with $p_1$ and $p_2$ so that each parameter affects the relative number of photons incident on every pixel $i$. Therefore, the estimates for $p_1$ and $p_2$ based on the PMF will generally be correlated to some degree. (In this particular example, the best workaround is to treat the signals from each half of the detector as completely separate measurements --- more on this later.)

As usual, these effects can also be visualized by plotting the likelihood functions $L_7(i|\bp)$ for each pixel, which are shown in Fig.~\ref{fig:MLE_L2_Ex7}. 
\begin{figure}
	\centering
	\includegraphics[width=\linewidth]{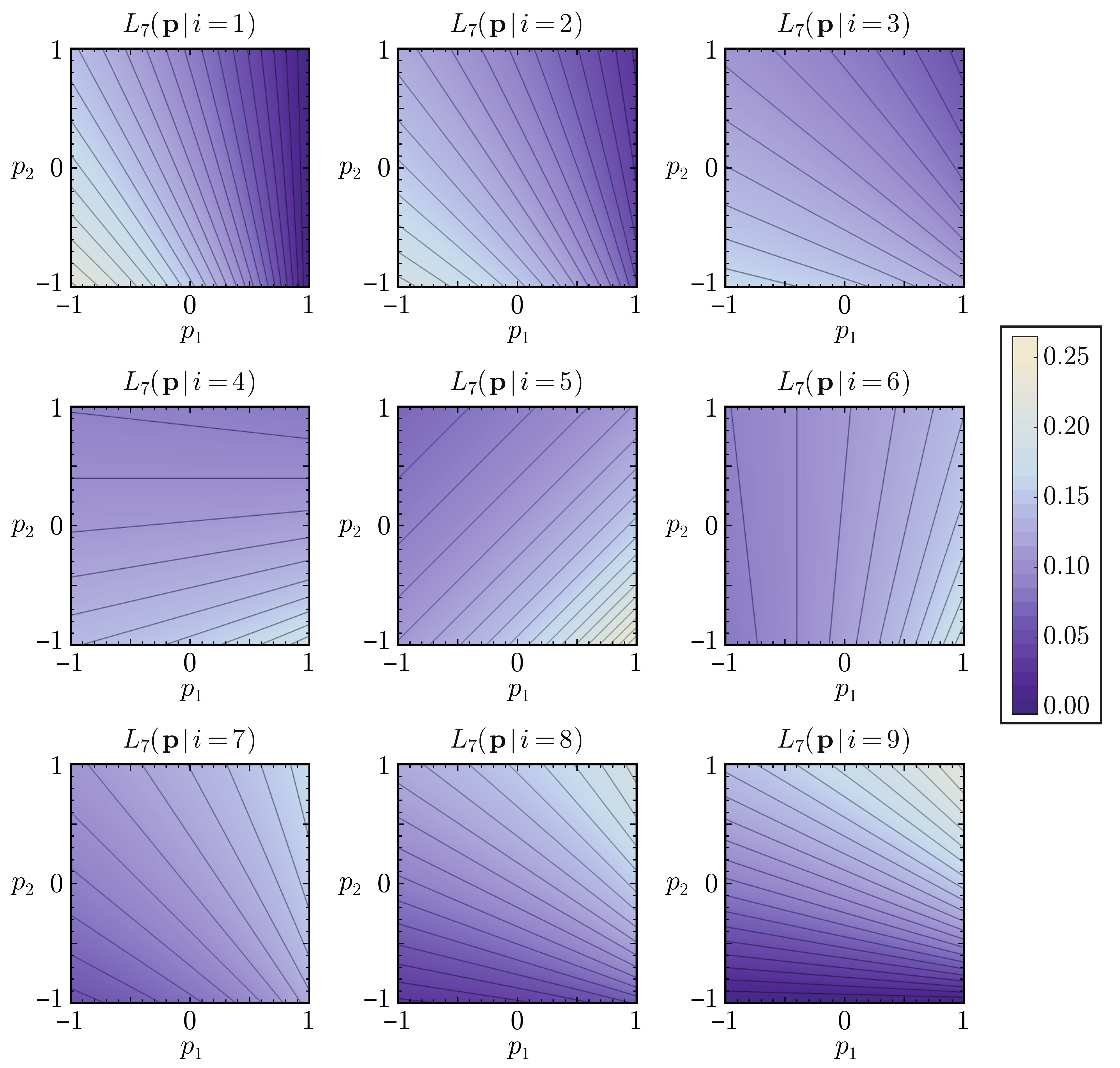}
	\caption{Likelihood functions $L_7(\bp|i)$ associated with each pixel $i$ for a measurement of $I_7(x;\bp)$. Contour lines are shown in increments of $0.01$.}
	\label{fig:MLE_L2_Ex7}
\end{figure}
Notice that the likelihood function for pixel 1 is most heavily influenced by $p_1$, while that of pixel 9 is mostly influenced by $p_2$. Nevertheless, every pixel contains information about both $p_1$ and $p_2$, since the partial derivatives of $\ell_7(i|\bp)$ with respect to each parameter are nonzero. Interestingly, this even implies that photons measured at pixel 5 (the center of the sensor, where $I(x_5|\bp)=0.5$ for any $\bp$) provide information about $p_1$ and $p_2$ when considered in relation to the number of photons measured at the other eight pixels.

The error ellipses for several values of $p_1$ and $p_2$ are shown in Fig.~\ref{fig:MLE_ellipses_Ex7}. 
\begin{figure}[t]
	\centering
	\includegraphics[width=.553\linewidth]{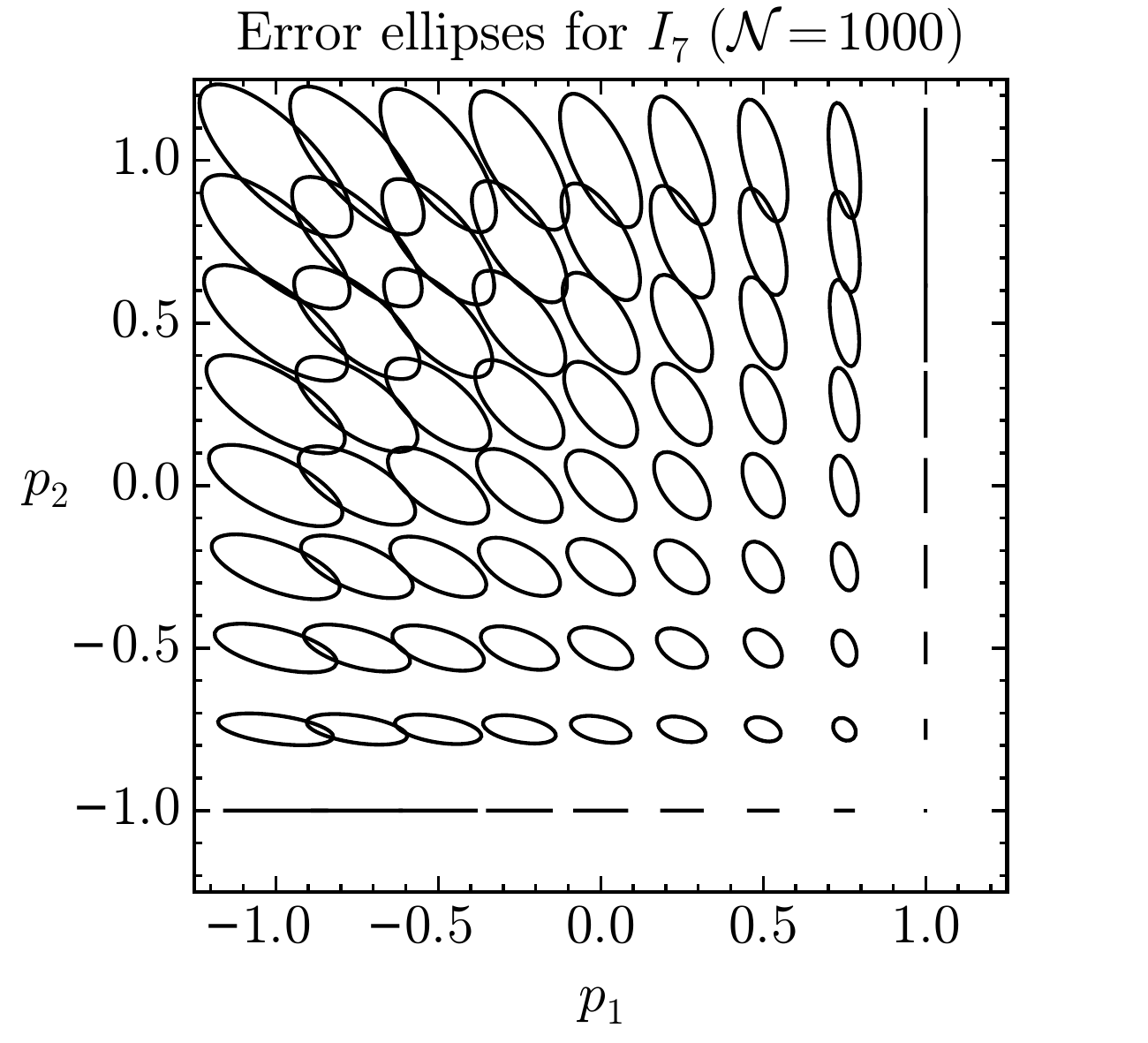}
	\caption{Ellipses representing the expected standard deviation error of a \mbox{1000-photon} measurement of $I_7(x;\bp)$ with true parameter values $p_1$ and $p_2$, sampled over a $9\times 9$ grid in parameter space.}
	\label{fig:MLE_ellipses_Ex7}
\end{figure}
Unlike the prior two examples, the expected estimation error for a measurement of $I_7(x;\bp)$ is strongly dependent on $\bp$, with the largest error (and substantial covariance between $p_1$ and $p_2$) occurring in the upper left quadrant where $p_1<0$ and $p_2>0$. The distributions of the log-likelihood functions obtained for two 1000-photon measurements with different true parameter values, shown in Fig.~\ref{fig:MLE_LL2_1000ph_Ex7}, are consistent with this trend. 
\begin{figure}[h]
	\centering
	\includegraphics[width=\linewidth]{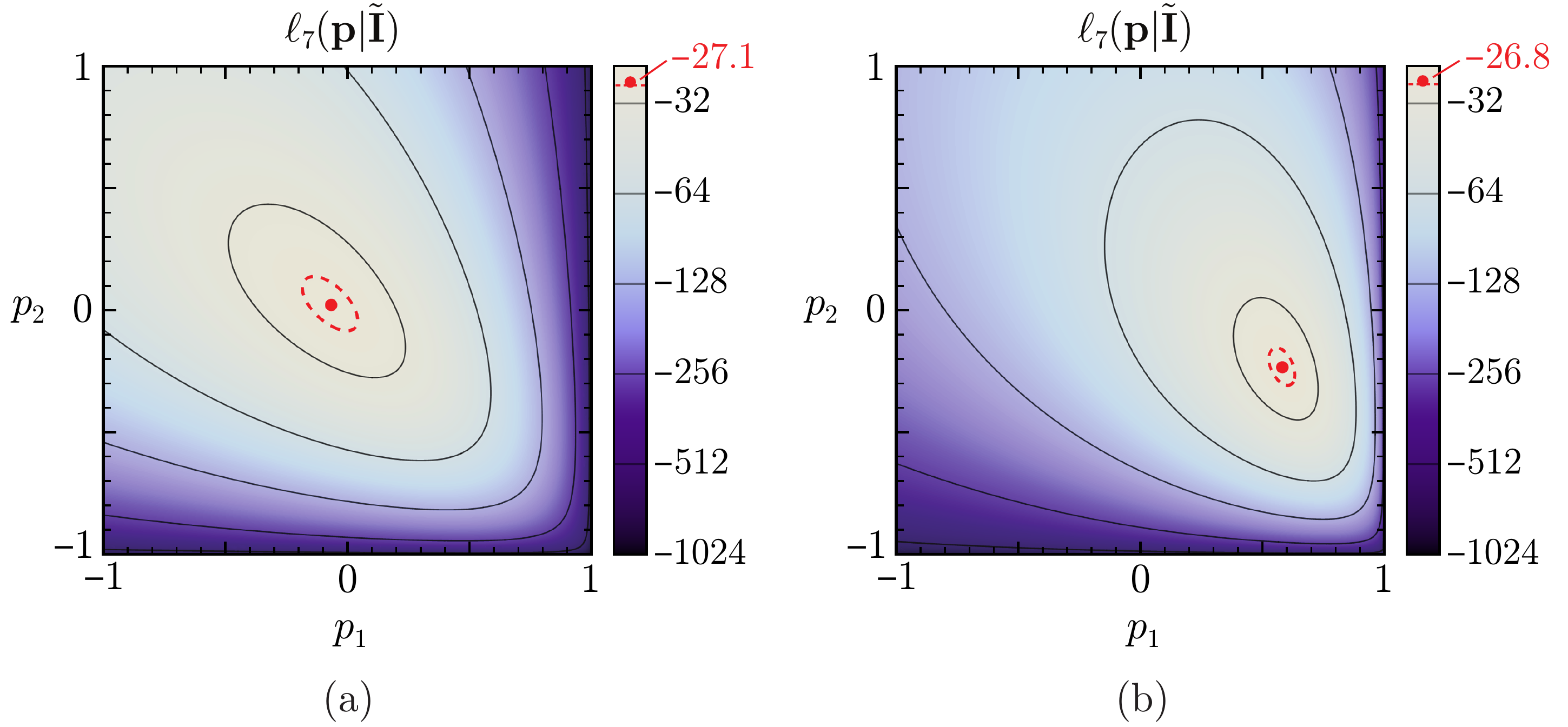}
	\caption{Log-likelihood functions $\ell_7(\bp|\tbI)$ for simulated 1000-photon measurements of $I_7(x;\bp)$ with true parameter values (a) $\bp\nsp=\nsp(0,0)$ and (b) $\bp\nsp=\nsp(0.63,-0.25)$. The plots are shaded on a logarithmic scale with solid contour lines drawn at powers of 2, as indicated in the legend. The peak of each distribution is marked with a red dot. The locations of these maxima (i.e., the MLEs for each measurement) are $\bp=(-0.067,0.024)$ and $\bp=(0.582,-0.232)$, respectively. The dashed contour line indicates where the likelihood $L_7(\bp|\tbI)$ drops to $1/\sqrt{e}$ times its peak value, representing the standard deviation confidence interval for the MLE.}
	\label{fig:MLE_LL2_1000ph_Ex7}
\end{figure}
The magnitude of the expected error is inversely proportional to the total intensity $\sum_i I_7(x_i|\bp)$, which is minimized when $p_1=1$ and $p_2=-1$. Not coincidentally, the errors in $p_1$ and $p_2$ approach zero as $p_1\to 1$ and $p_2\to -1$, respectively. (As in Section \ref{sect:MLE_example1}, this expectation of zero error is only meaningful in the limit of large $\mathcal{N}$.)
The dramatic variations in error with respect to $\bp$ can also be understood by revisiting Fig.~\ref{fig:MLE_L2_Ex7}, in which the contours of equal likelihood for each pixel tend to be most closely spaced in the lower right quadrant (where $p_1>0$ and $p_2<0$), indicating high information content. Pixel 5 in particular provides extremely useful information in this quadrant, not only due to the large slope of $L_7(\bp|i\!=\!5)$, but also because the direction of maximum variation (i.e., the gradient with respect to $\bp$) opposes that of pixels 1 and 9. In contrast, pixel 5 is nearly useless in the upper left quadrant of the parameter space since the likelihood changes very slowly with respect to $\bp$.

As mentioned before, in practice, the best way to deal with an intensity distribution such as $I_7(x;\bp)$ would be to treat it as two separate measurements: one involving pixels 1 through 5 (for which the intensity only depends on $p_1$), and another involving pixels 5 through 9 (for which the intensity only depends on $p_2$). The MLE approach could then be applied separately to each set of data, producing independent estimates for each parameter. In general, whenever it is possible to set up an experiment such that independent measurements can be made in this manner, it is probably best to do so, at least from a statistical standpoint. However, in cases where one does not have this luxury, the above example illustrates how subtle interactions between parameters (of either a physical or mathematical nature) can affect the accuracy of the measurement. Therefore, extra care should be taken to design the experiment such that the error obtained using the chosen statistical method is minimized.

\FloatBarrier
\subsection{Piecewise linear dependence (zero covariance)}\label{sect:MLE_example8}
Next, in comparison to the previous example, consider the intensity distribution
\begin{equation}
I_8(x;\bp) = 
\begin{cases}
0.5\Pi(x)\left[1 + 2p_1 (x+0.625)\right],&x<-0.125,\\
0.5\Pi(x),&-0.125\leq x< 0.125,\\
0.5\Pi(x)\left[1 + 2p_2 (x-0.625)\right],&x\geq 0.125,
\end{cases}
\end{equation}
which is plotted in Fig.~\ref{fig:MLE_IntPlots_Ex8}. 
\begin{figure}
	\centering
	\includegraphics[width=\linewidth]{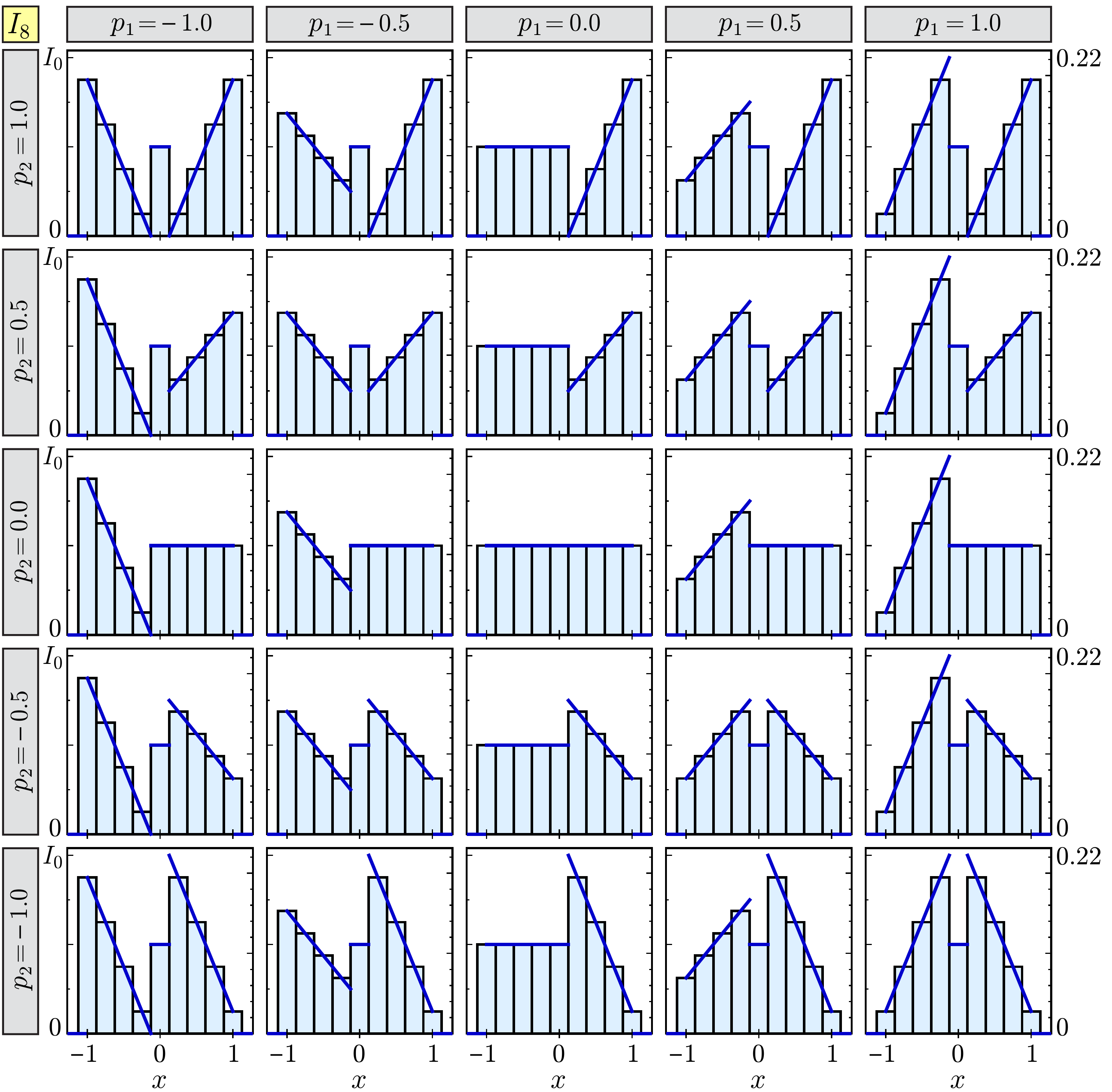}
	\caption{Plots of $I_8(x;\bp)$ (left axes) and $P_8(i|\bp)$ (right axes) for several values of $p_1$ and $p_2$.}
	\label{fig:MLE_IntPlots_Ex8}
\end{figure}
As with $I_7(x;\bp)$, this intensity varies linearly with $p_1$ or $p_2$ in either half of the sensor. The key difference in this example is that $I_8(x;\bp)$ is contrived in such a way that the total intensity $\sum_i I_8(x_i|\bp)$ is independent of $\bp$. As a result, the PMF (relative intensity) $P_8(i|\bp)$ only depends on $p_1$ on the left half of the sensor and $p_2$ on the right half of the sensor. Naturally, the same is true of the likelihood function $L_8(\bp|i)$, as seen in Fig.~\ref{fig:MLE_L2_Ex8}.
\begin{figure}
	\centering
	\includegraphics[width=\linewidth]{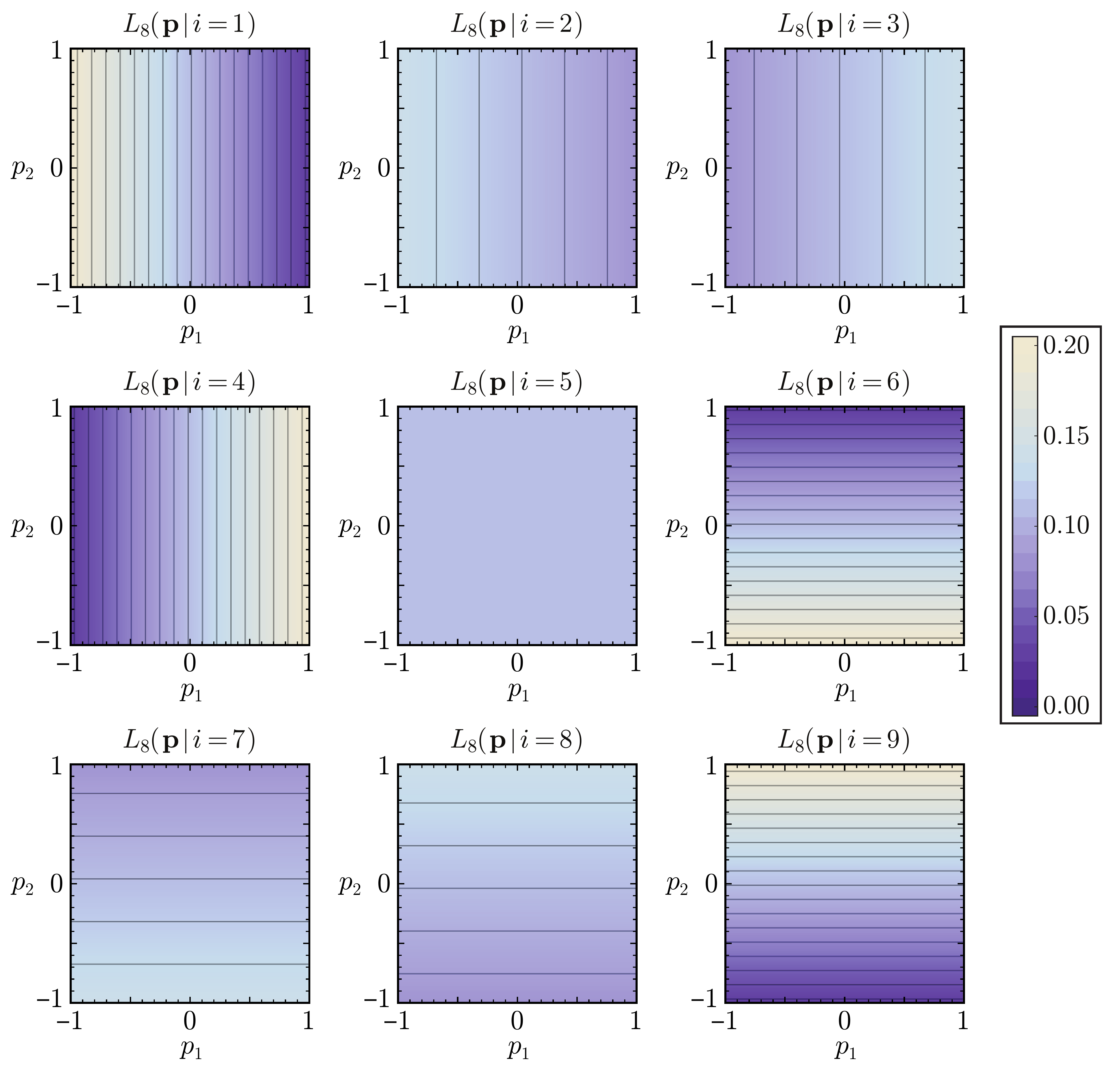}
	\caption{Likelihood functions $L_8(\bp|i)$ associated with each pixel $i$ for a measurement of $I_8(x;\bp)$. Contour lines are shown in increments of $0.01$.}
	\label{fig:MLE_L2_Ex8}
\end{figure}
Since the gradient of $L_8(\bp|i)$ always points along $p_1$ or $p_2$ (when it is nonzero), the FIM and its inverse are always diagonal, indicating that there is zero covariance between the parameters. For any value of $\bp$, the principal axes of the error ellipse are oriented along the $p_1$ and $p_2$ axes, as seen in Fig.~\ref{fig:MLE_ellipses_Ex8}. 
\begin{figure}
	\centering
	\includegraphics[width=.553\linewidth]{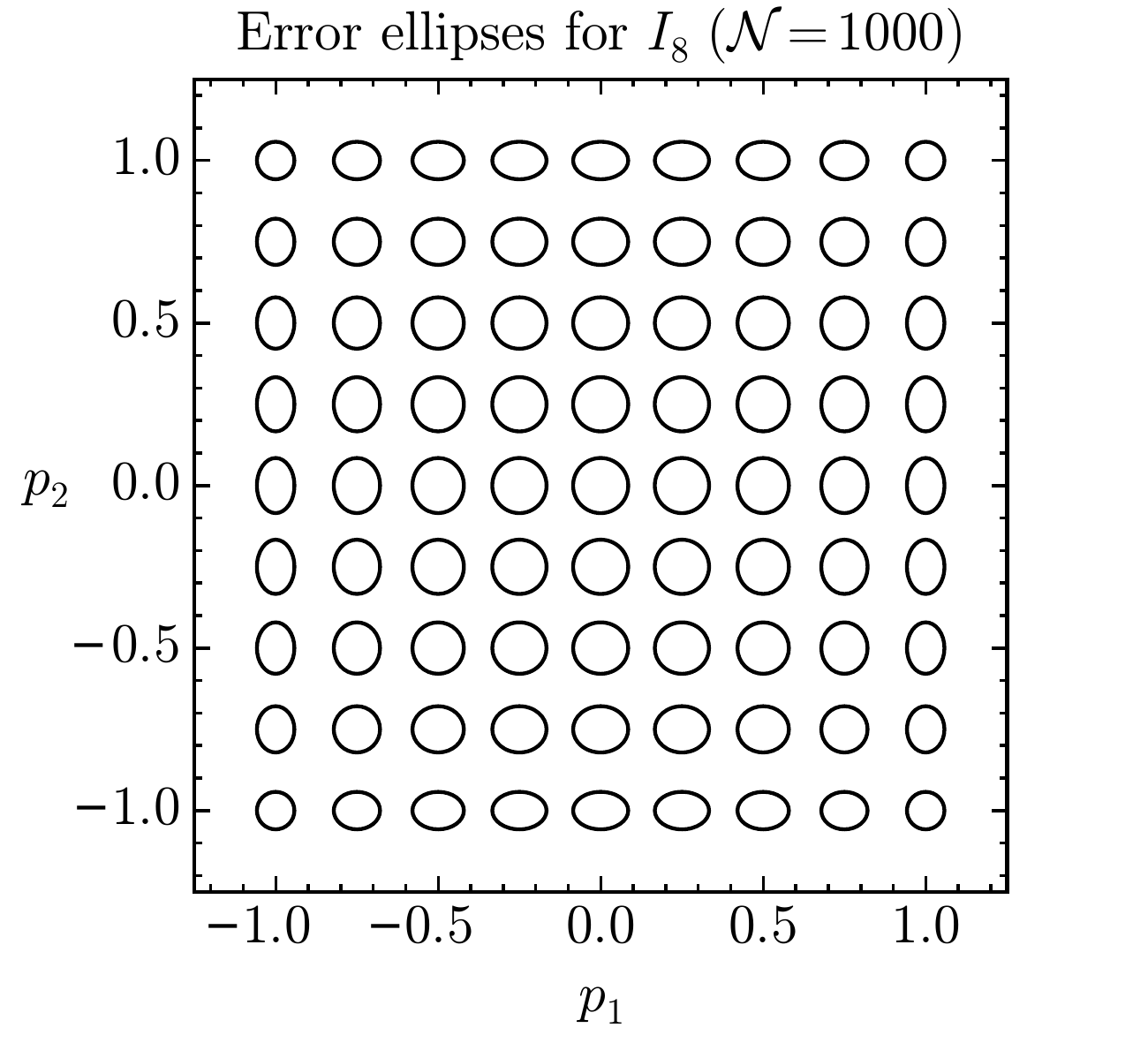}
	\caption{Ellipses representing the expected standard deviation error of a \mbox{1000-photon} measurement of $I_8(x;\bp)$ with true parameter values $p_1$ and $p_2$, sampled over a $9\times 9$ grid in parameter space.}
	\label{fig:MLE_ellipses_Ex8}
\end{figure}
When $\bp=(0,0)$, the error ellipse is circular, meaning that the expected error is identical for each parameter. For other values of $\bp$, the relative errors of the two parameters vary in a symmetric fashion over the region of interest. Fig.~\ref{fig:MLE_LL2_1000ph_Ex8} contains plots of the log-likelihood functions $\ell_8(\bp|\tbI)$ for simulated 1000-photon measurements of $I_8(x;\bp)$ with true parameter values $\bp=(0,0)$ and $\bp=(0.63,-0.25)$. In light of the above observations, it should come as no surprise that the distribution is highly symmetric about the MLE in each case.

\begin{figure}[tb]
	\centering
	\includegraphics[width=\linewidth]{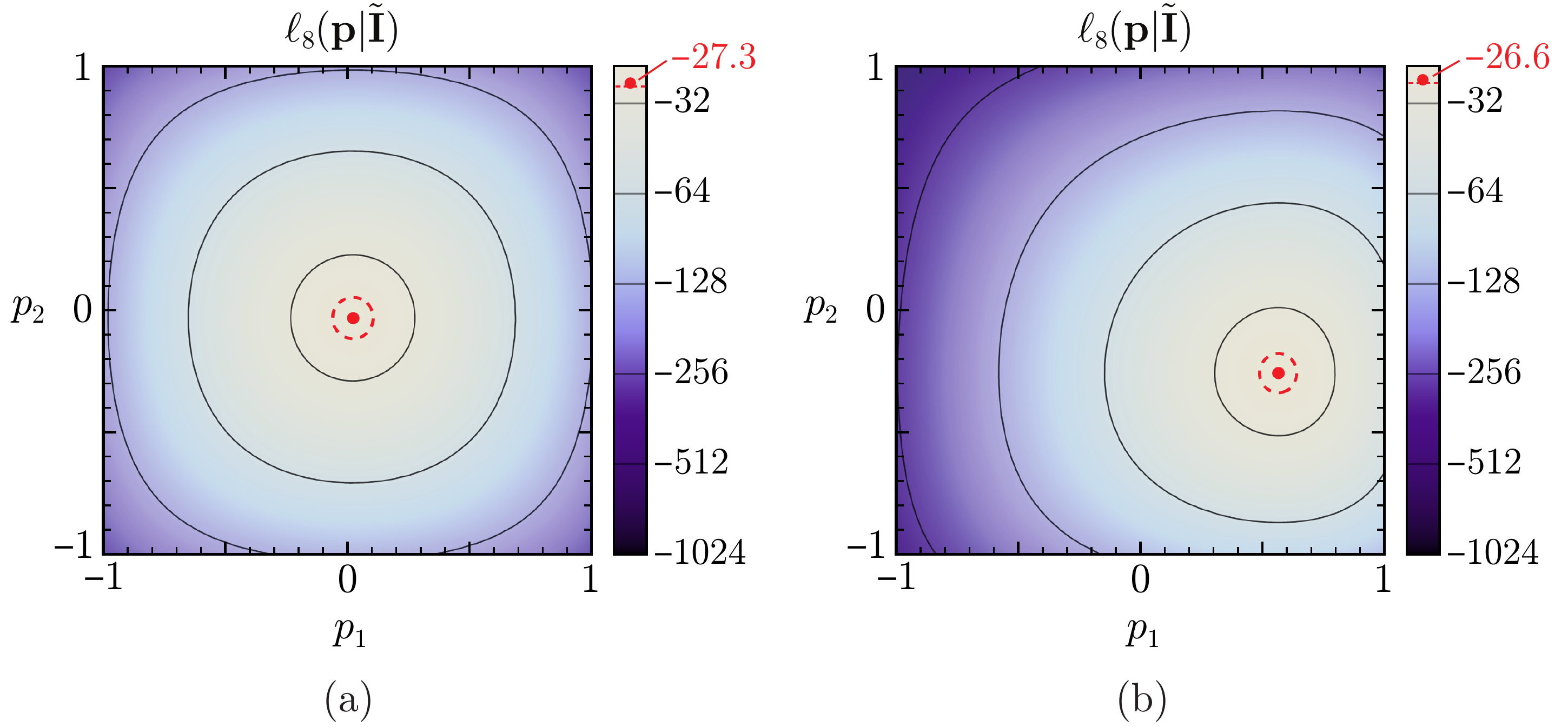}
	\caption{Log-likelihood functions $\ell_8(\bp|\tbI)$ for simulated 1000-photon measurements of $I_8(x;\bp)$ with true parameter values (a) $\bp\nsp=\nsp(0,0)$ and (b) $\bp\nsp=\nsp(0.63,-0.25)$. The plots are shaded on a logarithmic scale with solid contour lines drawn at powers of 2, as indicated in the legend. The peak of each distribution is marked with a red dot. The locations of these maxima (i.e., the MLEs for each measurement) are $\bp=(0.021,-0.029)$ and $\bp=(0.565,-0.256)$, respectively. The dashed contour line indicates where the likelihood $L_8(\bp|\tbI)$ drops to $1/\sqrt{e}$ times its peak value, representing the standard deviation confidence interval for the MLE.}
	\label{fig:MLE_LL2_1000ph_Ex8}
\end{figure}

To recap, the contrast between $I_7(x;\bp)$ and $I_8(x;\bp)$ illustrates a limitation of the MLE approach described in Section \ref{sect:MLE_optics}, as well as one of its key strengths. The shortcoming is that the sole reliance of the parameter estimate on the relative intensity can introduce correlations between parameters that are not present in the absolute (unnormalized) intensity; furthermore, any additional information contained within the overall scale of the intensity is ignored. On the other hand, the advantage of the method is that with good experimental design, the relative intensity can be tailored for optimal sensitivity and minimal coupling between parameters, so that there is no need to analyze the unnormalized intensity. Conveniently, the MLE formalism includes a straightforward error metric (the FIM) that can be used to predict and optimize the sensitivity of the measurement. As stated earlier, the lack of reliance on total intensity has the added benefit of reducing or eliminating errors arising from fluctuations of the source power.

\subsection{Two-parameter off-null measurement}\label{sect:MLE_example9}
The final two examples involve a pair of off-null measurements involving two parameters, starting with the intensity distribution
\begin{equation}
I_9(x;\bp) = 0.125\sp\Pi(x)\bigl[(p_1-x)^2 + (p_2-\cos(\pi x))^2\sp\bigr].
\end{equation}
This is a slightly simplified example of the distribution considered in Ref.~\cite{Vella_2018_fbs_arxiv}, with the contributions from each parameter adding incoherently (i.e., in intensity) rather than coherently (i.e., in electric field). Despite this difference, similar statistical behavior is observed in either case. Notice that the $p_1$ term of $I_9(x;\bp)$ is identical to that of the one-parameter example $I_3(x;p_1)$ considered in Section \ref{sect:MLE_example3}, with \mbox{$c=1$}. The $p_2$ term introduces an additional departure from the null condition, which varies sinusoidally over the sensor. These spatial variations were chosen to allow comparison between $I_9(x;\bp)$ and the earlier two-parameter example $I_6(x;\bp)$, for which the terms with $x$ and $\cos(\pi x)$ dependences were linear in $p_1$ and $p_2$, respectively. The intensity and PMF for $I_9(x;\bp)$ are shown in Fig.~\ref{fig:MLE_IntPlots_Ex9}.
\begin{figure}
	\centering
	\includegraphics[width=\linewidth]{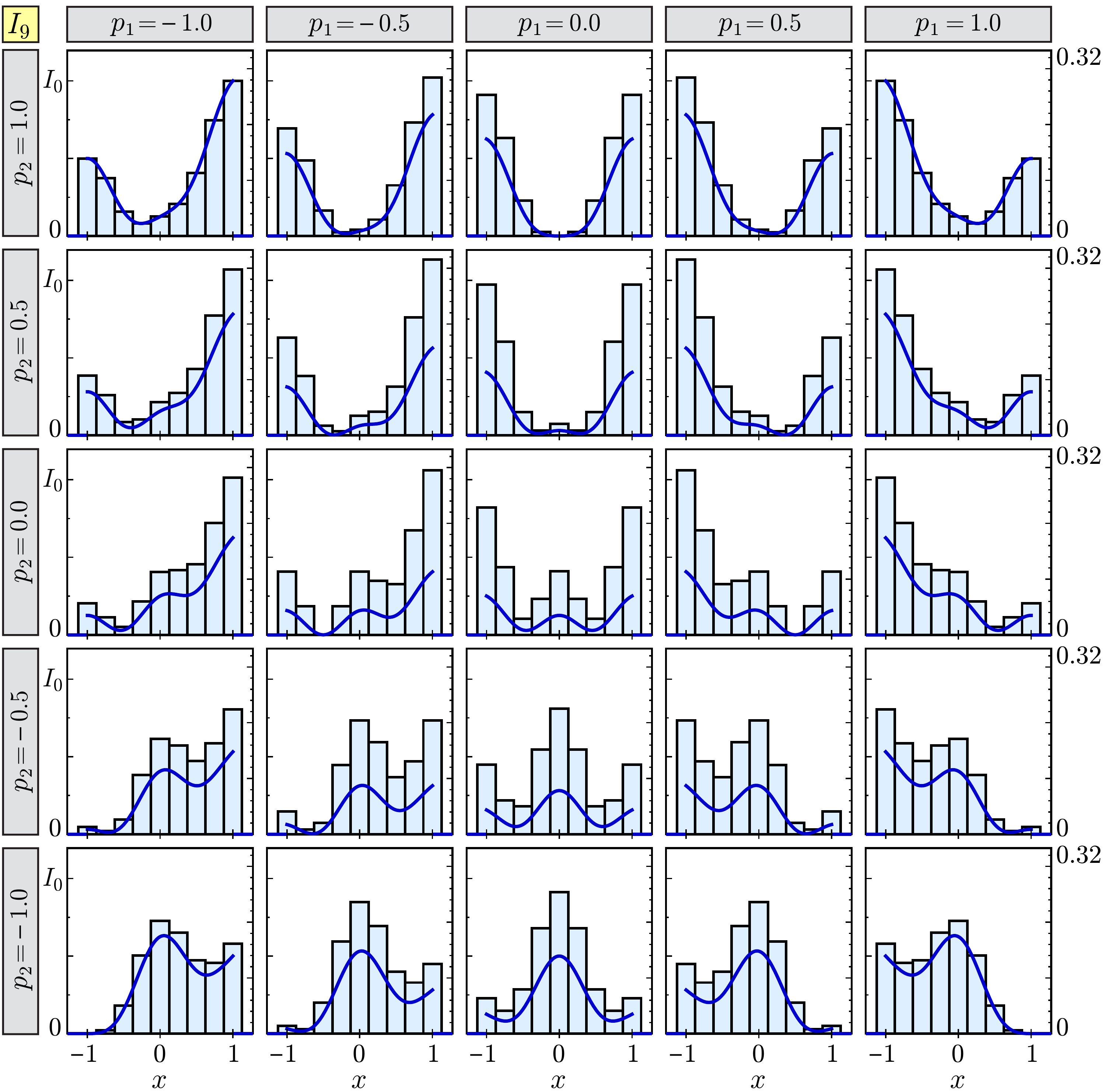}
	\caption{Plots of $I_9(x;\bp)$ (left axes) and $P_9(i|\bp)$ (right axes) for several values of $p_1$ and $p_2$.}
	\label{fig:MLE_IntPlots_Ex9}
\end{figure}
Compared to $I_6(x;\bp)$, observe that the off-null configuration employed in the present example produces more dramatic variations in the shape of the intensity profile with respect to $p_1$ and $p_2$, particularly for parameter values close to zero.

The likelihood functions $L_9(\bp|i)$ for each pixel, which are plotted in Fig.~\ref{fig:MLE_L2_Ex9}, have a far more complex structure than the ones seen in the previous examples. 
\begin{figure}
	\centering
	\includegraphics[width=\linewidth]{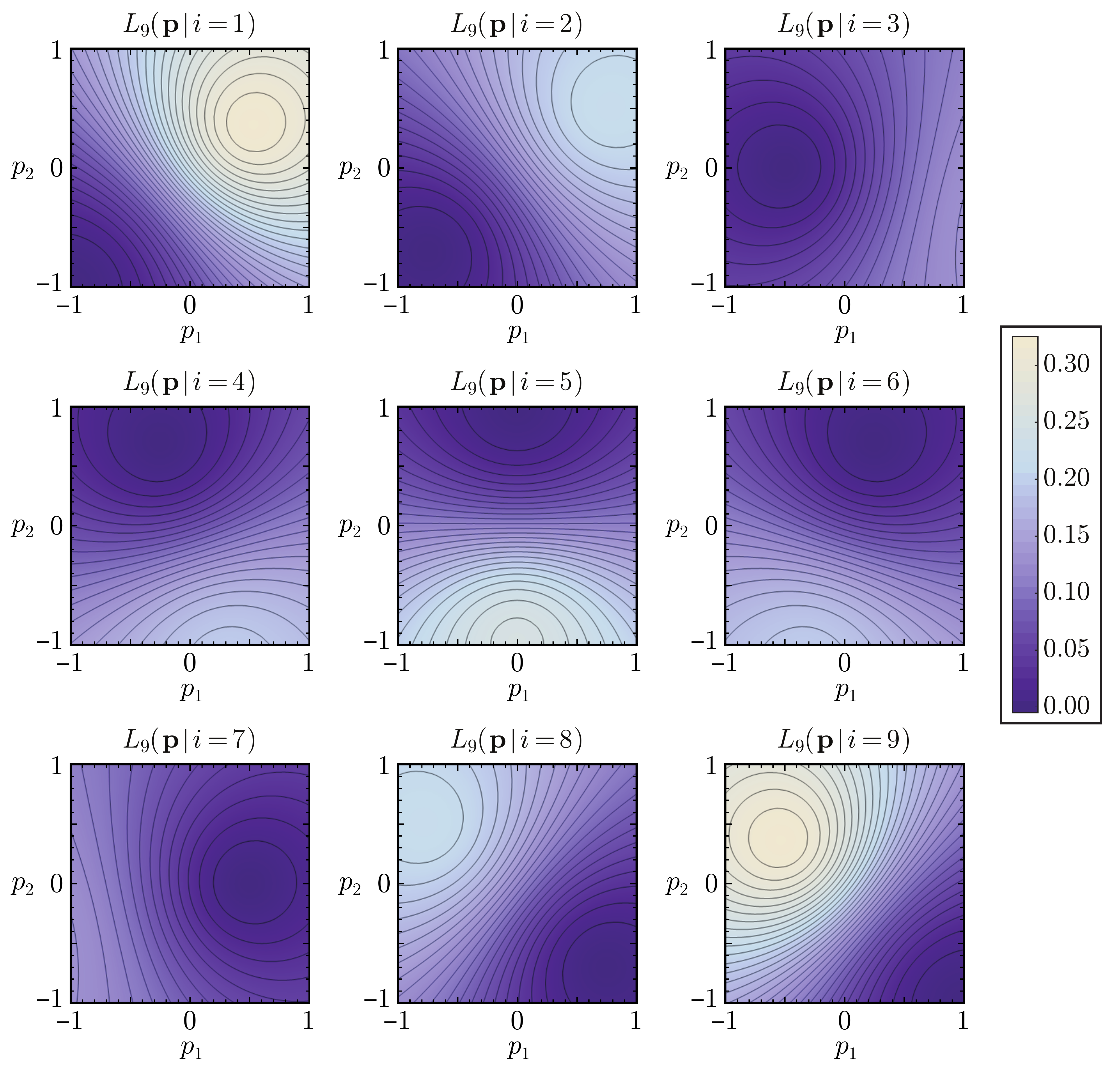}
	\caption{Likelihood functions $L_9(\bp|i)$ associated with each pixel $i$ for a measurement of $I_9(x;\bp)$. Contour lines are shown in increments of $0.01$.}
	\label{fig:MLE_L2_Ex9}
\end{figure}
The contributions of each pixel have similar shapes, consisting of a peaked distribution that rotates clockwise and changes scale as $i$ runs from 1 to 9. The balance between different pixels and the densely spaced contours of constant likelihood suggest that the FIM is likely to be large and diagonal, which would result in a small and diagonal covariance matrix. As indicated by the ellipse map shown in Fig.~\ref{fig:MLE_ellipses_Ex9}, the expected error is indeed quite small, particularly for parameter values near $\bp=(0,0)$, for which the total measured intensity tends to be the lowest. 
\begin{figure}[tb]
	\centering
	\includegraphics[width=.553\linewidth]{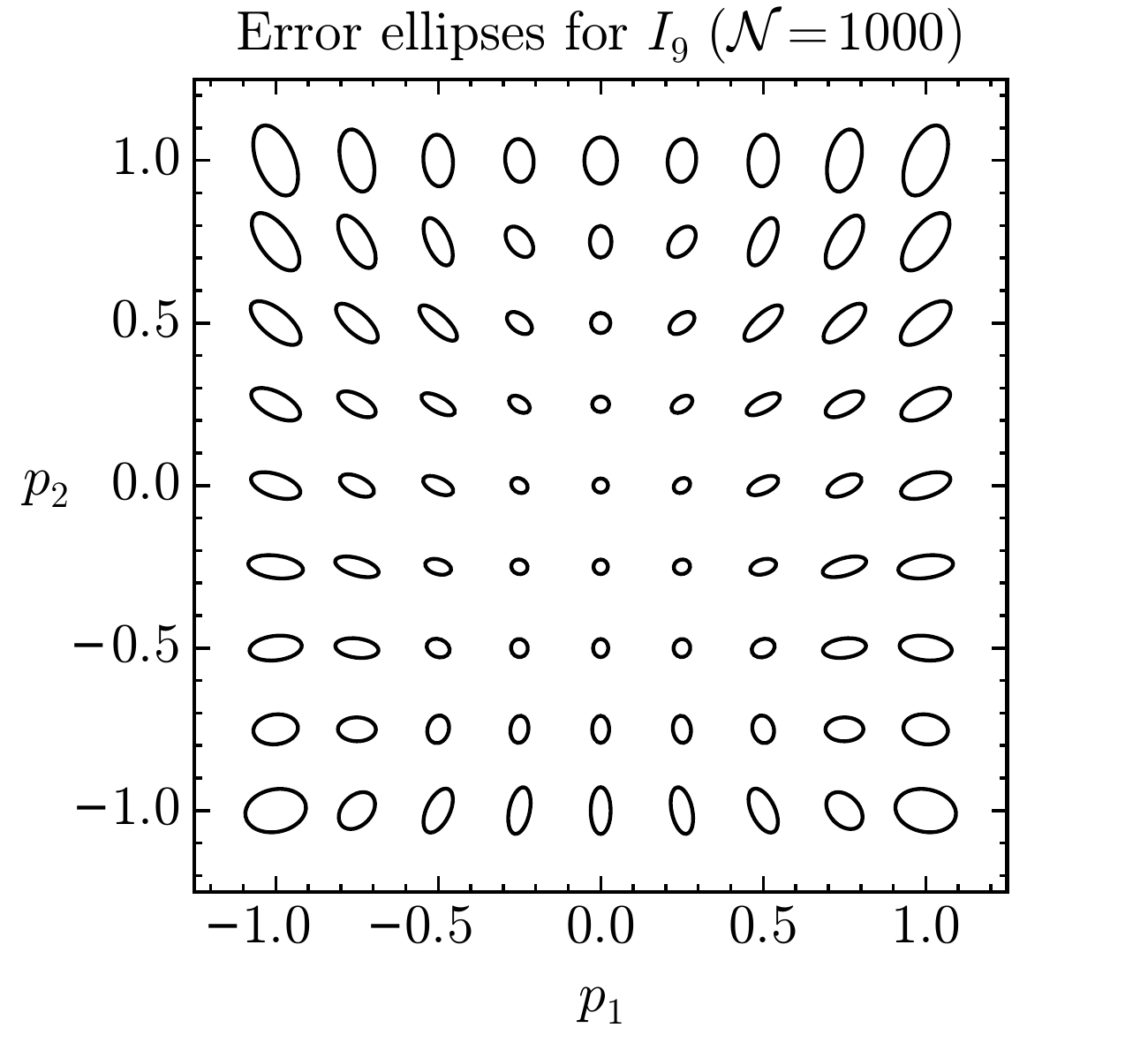}
	\caption{Ellipses representing the expected standard deviation error of a \mbox{1000-photon} measurement of $I_9(x;\bp)$ with true parameter values $p_1$ and $p_2$, sampled over a $9\times 9$ grid in parameter space.}
	\label{fig:MLE_ellipses_Ex9}
\end{figure}
This symmetric ellipse pattern, with the error growing as the departure from null increases, is typical for an off-null measurement. There is a considerable covariance between $p_1$ and $p_2$ near the edge of the parameter range, but in nearly all cases, the error is still smaller (often significantly so) than it would be for a measurement of $I_6(x;\bp)$ (see Fig.~\ref{fig:MLE_ellipses_Ex6} for comparison). 

The log-likelihood functions $\ell_9(\bp|\tbI)$ obtained for two simulated measurements of $I_9(x;\bp)$ with true parameter values $\bp=(0,0)$ and $\bp=(0.63,-0.25)$ can be found in Fig.~\ref{fig:MLE_LL2_1000ph_Ex9}.
\begin{figure}[tb]
	\centering
	\includegraphics[width=\linewidth]{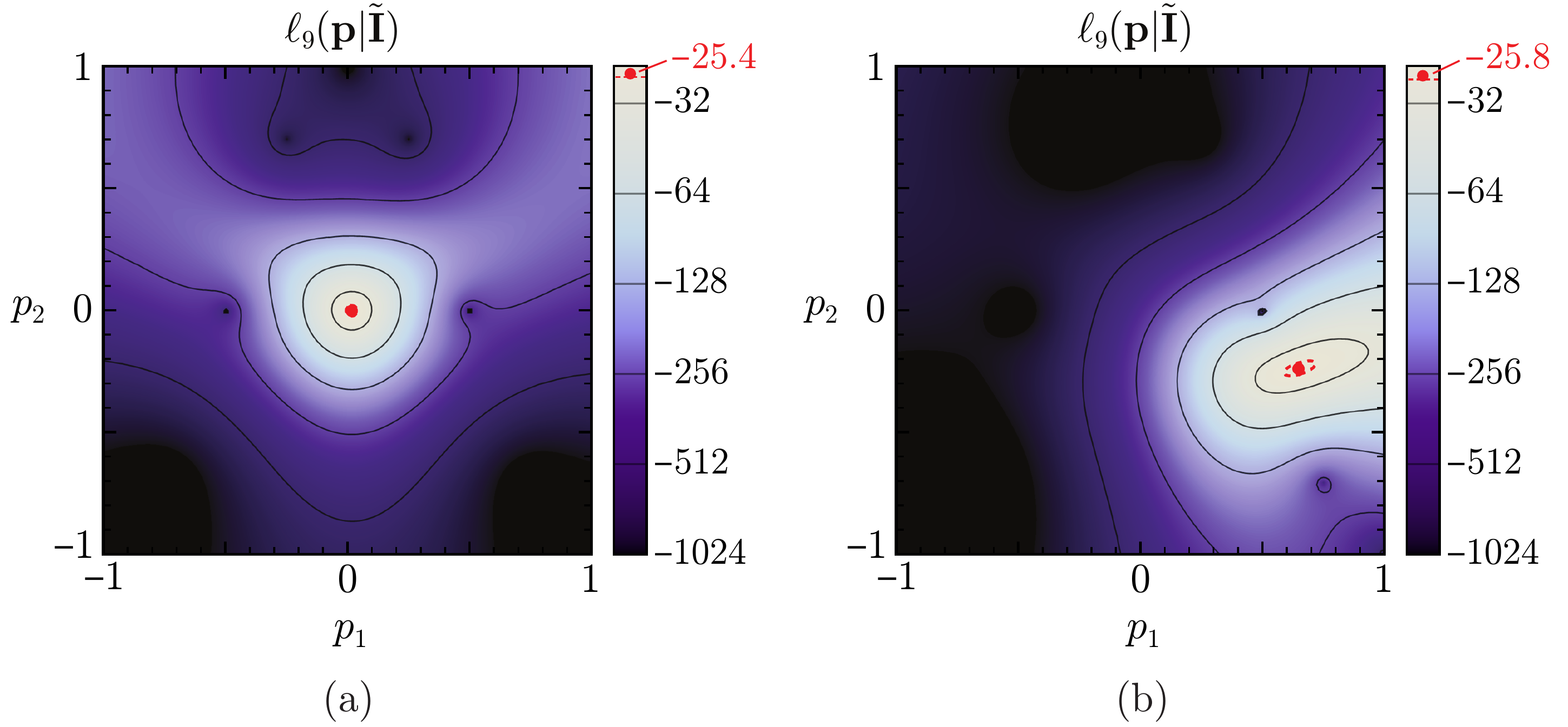}
	\caption{Log-likelihood functions $\ell_9(\bp|\tbI)$ for simulated 1000-photon measurements of $I_9(x;\bp)$ with true parameter values (a) $\bp\nsp=\nsp(0,0)$ and (b) $\bp\nsp=\nsp(0.63,-0.25)$. The plots are shaded on a logarithmic scale with solid contour lines drawn at powers of 2, as indicated in the legend. (Values smaller than $-1024$ are shown in black.) The peak of each distribution is marked with a red dot. The locations of these maxima (i.e., the MLEs for each measurement) are $\bp=(0.016,0.001)$ and $\bp=(0.648,-0.237)$, respectively. The dashed contour line indicates where the likelihood $L_9(\bp|\tbI)$ drops to $1/\sqrt{e}$ times its peak value, representing the standard deviation confidence interval for the MLE. (The dashed contour in plot (a) is too small to be seen.)}
	\label{fig:MLE_LL2_1000ph_Ex9}
\end{figure}
For the $\bp=(0,0)$ case, the likelihood is a sharply peaked distribution, with the location of the peak (the MLE) nearly coinciding with the true value of $\bp$. (The numerical results are provided in the figure caption.) The distribution is considerably wider and less symmetric for the $\bp=(0.63,-0.25)$ case, but the standard deviation uncertainty is still quite small. These results demonstrate the usefulness of an off-null measurement, which enables the simultaneous estimate of multiple parameters with high precision.

\subsection{Two-parameter off-null measurement with smaller departure from null}\label{sect:MLE_example10}
For the final example, consider the intensity distribution
\begin{equation}
I_{10}(x;\bp) = 0.320\sp\Pi(x)\bigl[(p_1-0.25x)^2 + (p_2-0.25\cos(\pi x))^2\sp\bigr].
\end{equation}
Notice that the $x$ dependence of $I_{10}(x;\bp)$ is identical to the previous case except that the departure from null associated with each parameter is four times smaller. As seen in the plots of the intensity profile (Fig.~\ref{fig:MLE_IntPlots_Ex10}) and the likelihood functions for each pixel (Fig.~\ref{fig:MLE_L2_Ex10}), the measurement is very sensitive to variations in $p_1$ and $p_2$ when both parameters are close to zero. 
However, similarly to the $c\ll 1$ case in Section \ref{sect:MLE_example3}, this comes at the expense of greatly reduced sensitivity (i.e., slower variations in likelihood) near the edges of the region of interest.

\begin{figure}
	\centering
	\includegraphics[width=\linewidth]{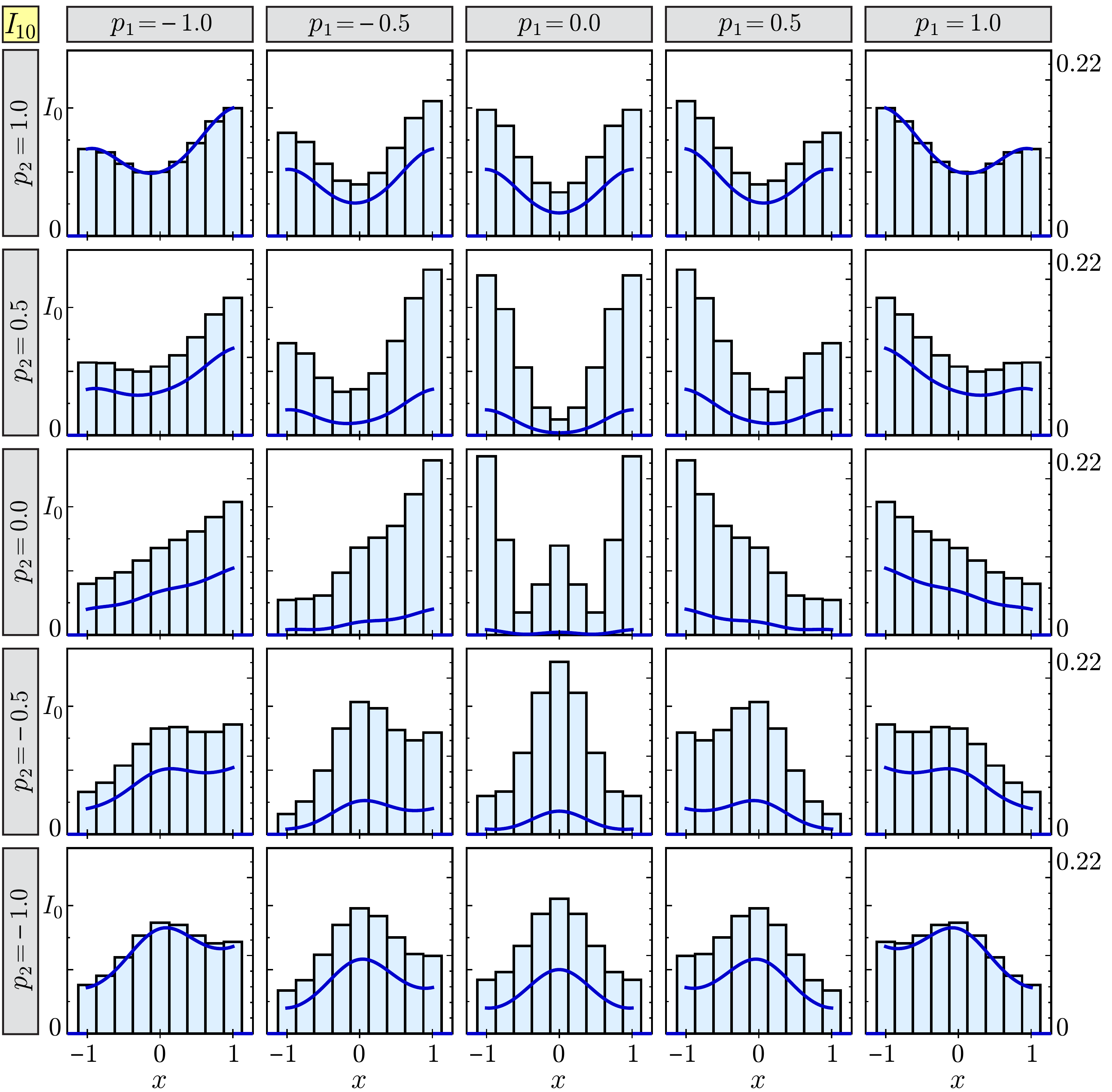}
	\caption{Plots of $I_{10}(x;\bp)$ (left axes) and $P_{10}(i|\bp)$ (right axes) for several values of $p_1$ and $p_2$.}
	\label{fig:MLE_IntPlots_Ex10}
\end{figure}
\begin{figure}
	\centering
	\includegraphics[width=\linewidth]{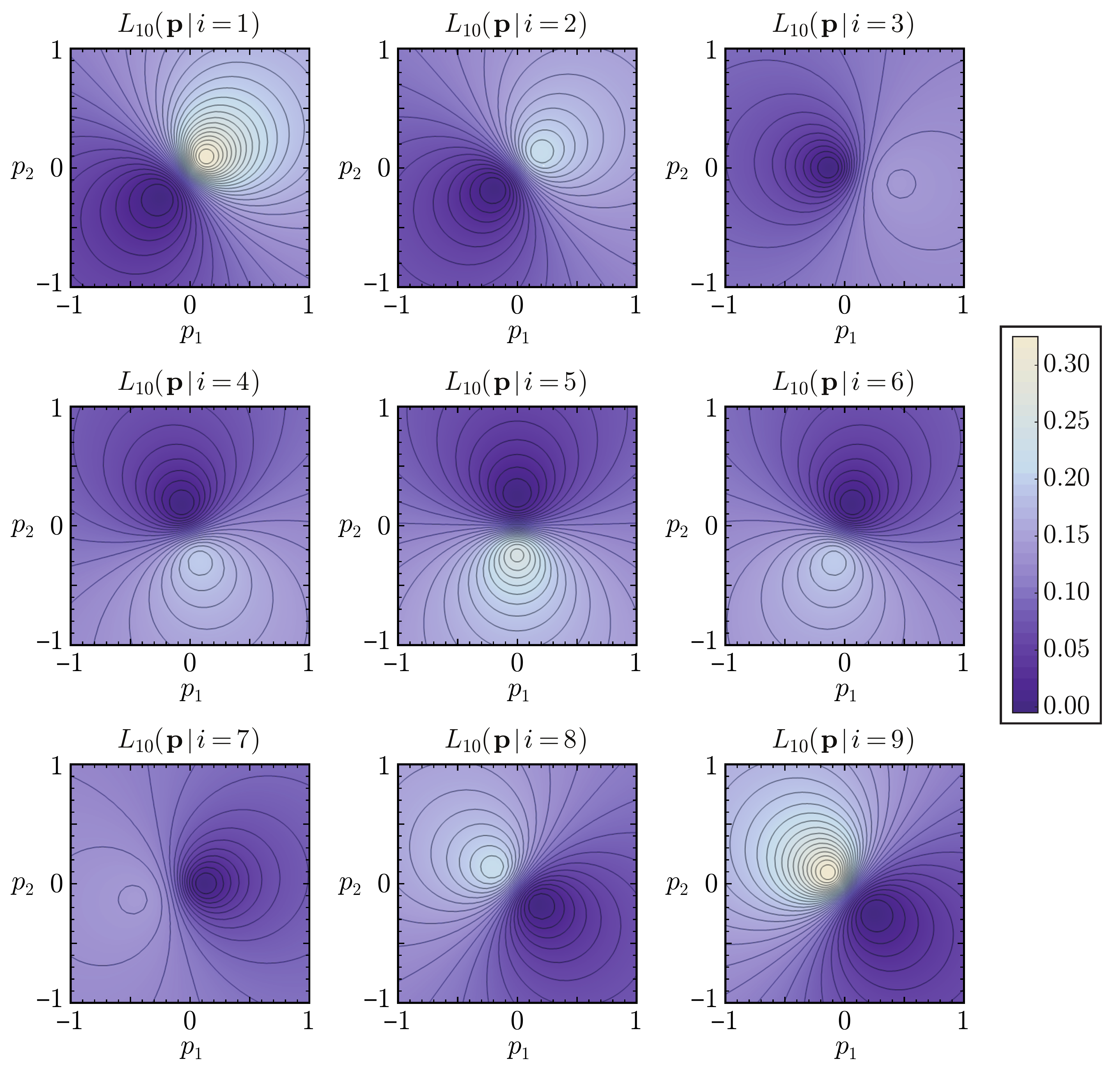}
	\caption{Likelihood functions $L_{10}(\bp|i)$ associated with each pixel $i$ for a measurement of $I_{10}(x;\bp)$. Contour lines are shown in increments of $0.01$.}
	\label{fig:MLE_L2_Ex10}
\end{figure}

The expected error ellipses based on the FIM are plotted for several parameter values in Fig.~\ref{fig:MLE_ellipses_Ex10}. 
\begin{figure}
	\centering
	\includegraphics[width=.553\linewidth]{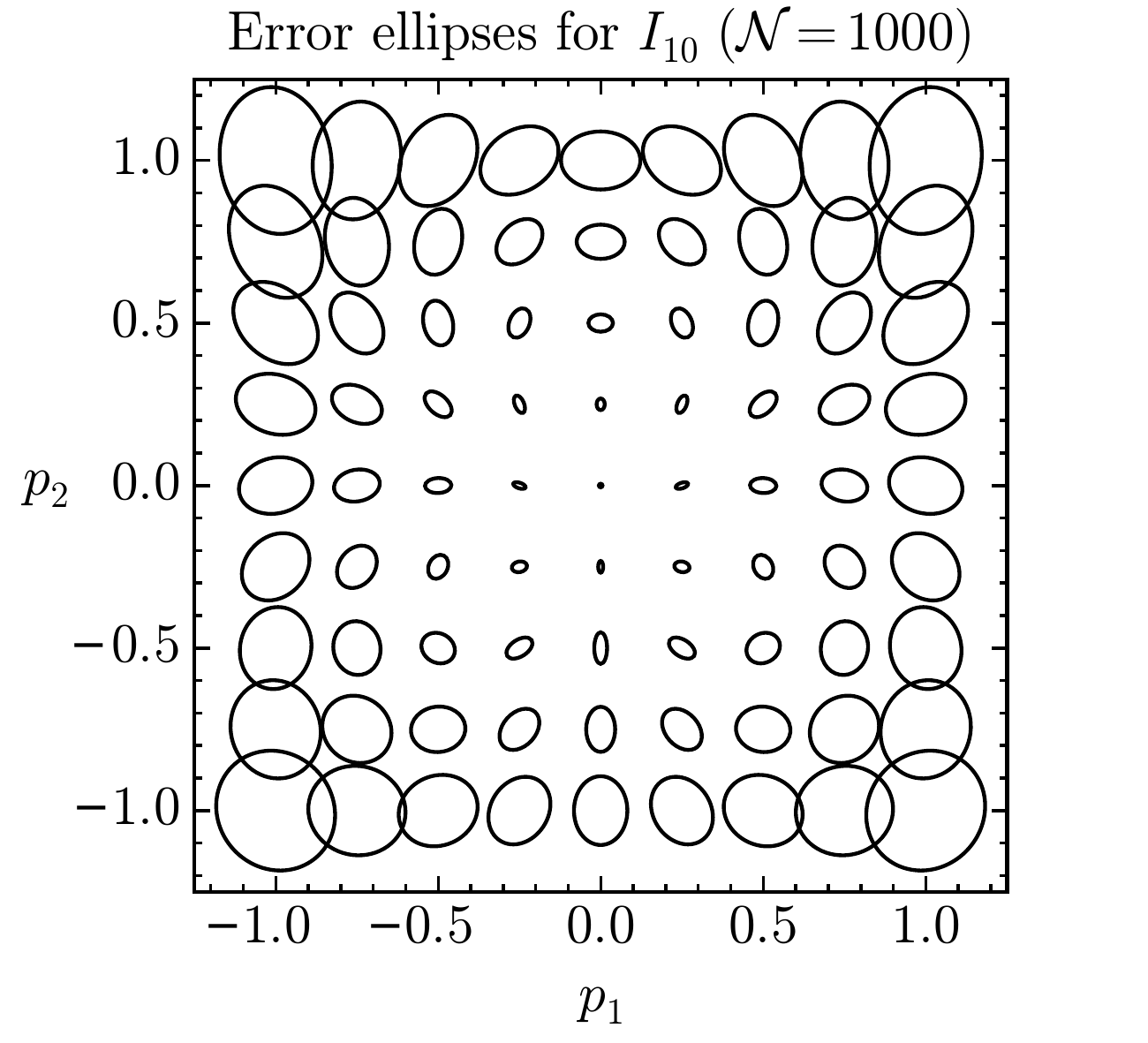}
	\caption{Ellipses representing the expected standard deviation error of a \mbox{1000-photon} measurement of $I_{10}(x;\bp)$ with true parameter values $p_1$ and $p_2$, sampled over a $9\times 9$ grid in parameter space.}
	\label{fig:MLE_ellipses_Ex10}
\end{figure}
The error for a measurement of $I_{10}(x;\bp)$ exhibits the same pattern as that of $I_9(x;\bp)$ (see Fig.~\ref{fig:MLE_ellipses_Ex9}), but with a larger disparity between the magnitudes of the errors near the center and edges of the parameter range. More precisely, for a true parameter value of $\bp=(0,0)$, the expected error is exactly four times smaller for a measurement of $I_{10}$ as it is for a measurement of $I_9$; conversely, the errors near the far corners of the parameter range (where $|p_1|\approx|p_2|\approx 1$) are about two to three times larger for $I_{10}$ than for $I_9$.

Finally, the log-likelihood functions $\ell_{10}(\bp|x)$ for simulated measurements of $I_{10}$ with true parameter values $\bp=(0,0)$ and $\bp=(0.63,-0.25)$ are shown in Fig.~\ref{fig:MLE_LL2_1000ph_Ex10}. 
\begin{figure}[tb]
	\centering
	\includegraphics[width=\linewidth]{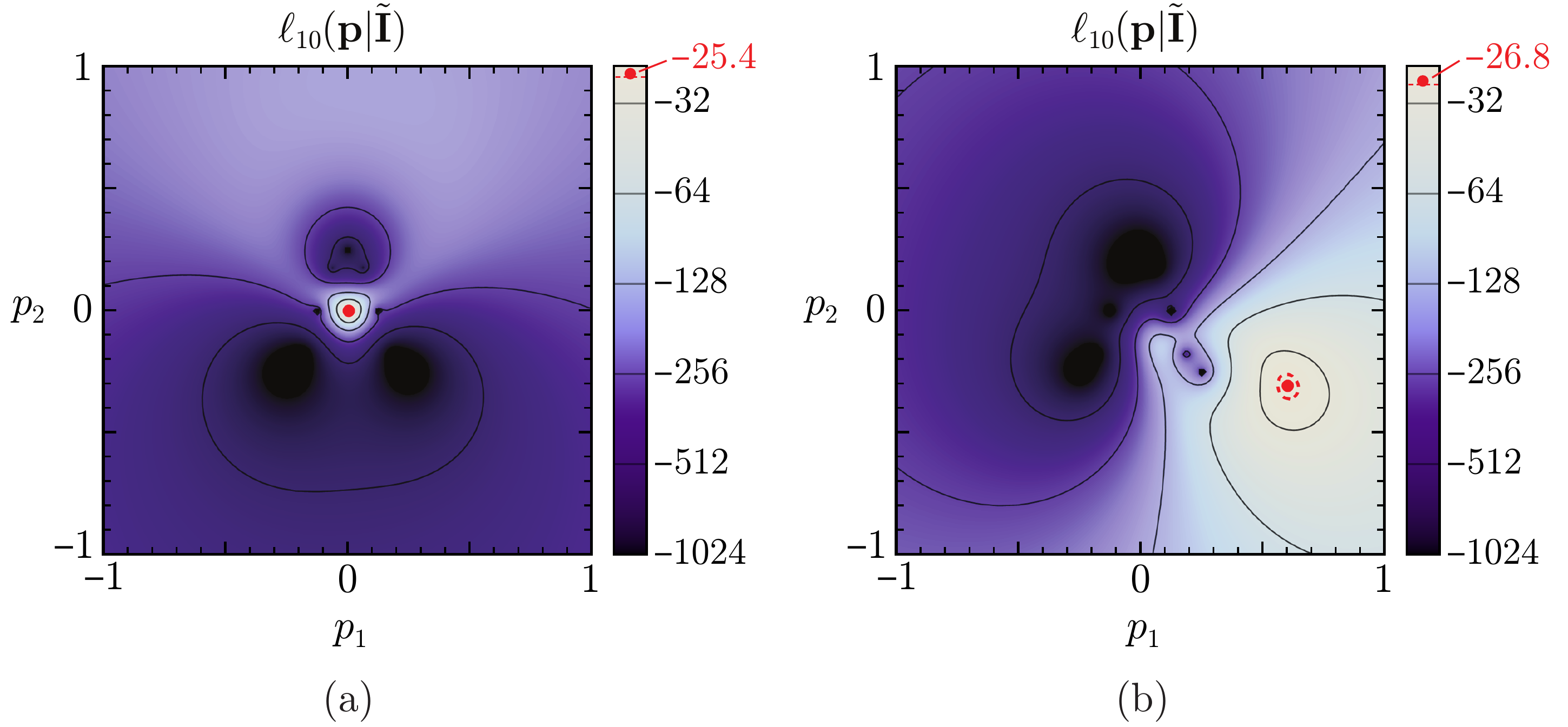}
	\caption{Log-likelihood functions $\ell_{10}(\bp|\tbI)$ for simulated 1000-photon measurements of $I_{10}(x;\bp)$ with true parameter values (a) $\bp\nsp=\nsp(0,0)$ and (b) $\bp\nsp=\nsp(0.63,-0.25)$. The plots are shaded on a logarithmic scale with solid contour lines drawn at powers of 2, as indicated in the legend. (Values smaller than $-1024$ are shown in black.) The peak of each distribution is marked with a red dot. The locations of these maxima (i.e., the MLEs for each measurement) are $\bp=(0.004,-2.6\times 10^{-4})$ and $\bp=(0.602,-0.308)$, respectively. The dashed contour line indicates where the likelihood $L_{10}(\bp|\tbI)$ drops to $1/\sqrt{e}$ times its peak value, representing the standard deviation confidence interval for the MLE. (The dashed contour in plot (a) is too small to be seen.)}
	\label{fig:MLE_LL2_1000ph_Ex10}
\end{figure}
As expected, the likelihood for the $\bp=(0,0)$ case is extremely narrowly distributed about its peak, producing an estimate with error on the order of 0.001. In contrast, the distribution for $\bp=(0.63,-0.25)$ is substantially wider; for parameter values with magnitudes closer to 1, the width of the distribution would continue to grow.

The practical implication of this example is that an off-null measurement can be tailored for high sensitivity over an arbitrarily small range of parameter values. Therefore, it is possible to design an iterative experiment for which the parameter estimate is refined through a series of successive measurements. For example, in the focused beam scatterometry setup described in Ref.~\cite{Vella_2018_fbs_arxiv}, an SLM could be used to produce an arbitrary spatially-varying polarization state, which can be chosen differently for each iteration of the measurement. The experimental details of such an implementation are discussed in Ref.~\cite{Head_2018}. 

As an example of this iterative procedure, suppose that we wish to refine the measurement of $I_9(x;\bp)$ with true parameter values $\bp=(0.63,-0.25)$ obtained in Section \ref{sect:MLE_example9}. The plot of the log-likelihood function $\ell_9(\bp|\tbI)$ for this measurement is shown again in Fig.~\ref{fig:MLE_LL2_1000ph_Ex41-44}(a); the MLE based on this initial measurement is $\bp=(0.648,-0.237)$. To refine the parameter estimate, the experimental configuration could be altered such that the output intensity follows the distribution
\begin{equation}
I_9^{(2)}(x;\bp) = \Pi(x)\bigl[(p_1-0.648-0.5x)^2 + (p_2+0.237-0.5\cos(\pi x))^2\sp\bigr],
\end{equation}
where the constant normalization factor in front of $\Pi(x)$ has been omitted for simplicity.\footnote{In a real experiment, the leading factor (which determines the peak intensity) would typically vary under different experimental configurations. Since the MLE approach ignores any information contained in this scaling factor, it is not important for this discussion.} This distribution is designed so that the departure from null is half as large and centered at the previous MLE. The resulting log-likelihood function $\ell_9^{(2)}(\bp|\tbI)$ for a simulated measurement of 1000 photons, shown in Fig.~\ref{fig:MLE_LL2_1000ph_Ex41-44}(b), is much more narrowly distributed than $\ell_9(\bp|\tbI)$. The MLE based on the refined measurement is found to be $\bp=(0.644,-0.255)$. This process can be applied repeatedly to obtain an estimate with arbitrary precision (barring experimental limitations, as discussed in the next paragraph). The intensity distributions and resulting MLEs for the first four iterations of the process, including the two mentioned above, are listed in Table~\ref{tbl:MLE_iterative_intensities}, and the log-likelihood functions for simulated measurements of $I_9^{(3)}(x;\bp)$ and $I_9^{(4)}(x;\bp)$ are plotted in Fig.~\ref{fig:MLE_LL2_1000ph_Ex41-44}(c,d). 
\begin{figure}
	\centering
	\includegraphics[width=\linewidth]{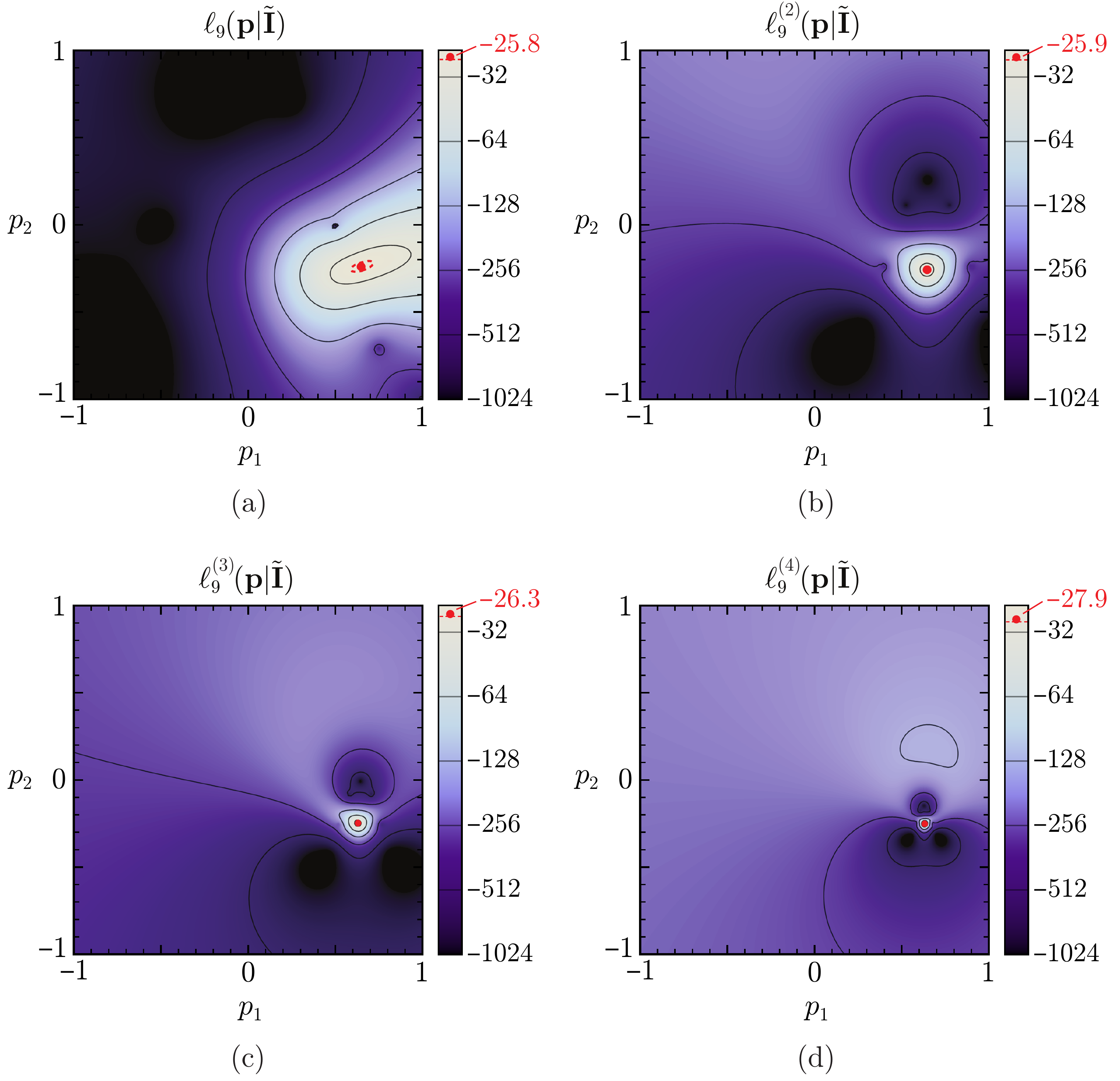}
	\caption{Log-likelihood functions for simulated 1000-photon measurements of intensity distributions (a) $I_9(x;\bp)$, (b) $I_9^{(2)}(x;\bp)$, (c) $I_9^{(3)}(x;\bp)$, and (d) $I_9^{(4)}(x;\bp)$ obtained throughout a four-step iterative measurement with true parameter values $\bp=(0.63,-0.25)$. The peaks of each distribution are indicated with a red dot, and their locations are listed in the rightmost column of Table~\ref{tbl:MLE_iterative_intensities}. The dashed red contour in plot (a) represents the standard deviation confidence interval; the confidence intervals in plots (b-d) are too small to be seen.}
	\label{fig:MLE_LL2_1000ph_Ex41-44}
\end{figure}
As seen in the table, the MLE gets closer to the true value with each iteration, leading to a final estimate of $\bp=(0.631,-0.249)$. As this happens, the likelihood function becomes increasingly compact with an exceptionally sharp peak, which is the reason for the improvement in accuracy. However, note that the calculation of the MLE must be performed carefully in this case since the likelihood function may contain local maxima or regions with very small slopes, which can cause problems with the numerical search procedure. These issues can generally be mitigated by using the previous MLE as the starting point for the search.

\begin{table}
	\small
	\renewcommand{\arraystretch}{.8}
	\begin{center}
		\small
		\begin{tabular}{l@{\hspace{20pt}}l}
			\toprule
			Intensity distribution & MLE for $\bp$\\
			\midrule
			\phantom{a}&\\[-8pt]
			$I_9(x;\bp) \propto \bigl[(p_1-x)^2 + (p_2-\cos(\pi x))^2\sp\bigr]$
			& $(0.648,-0.237)$
			\\[8pt]
			$\mathrlap{I_9^{(2)}(x;\bp) \propto
			\bigl[(p_1-0.648-0.50x)^2 + (p_2+0.237-0.50\cos(\pi x))^2\sp\bigr]}$
			& $(0.644,-0.255)$
			\\[8pt]
			$I_9^{(3)}(x;\bp) \propto \bigl[(p_1-0.644-0.25x)^2 + (p_2+0.255-0.25\cos(\pi x))^2\sp\bigr]$
			& $(0.628,-0.246)$
			\\[8pt]
			$I_9^{(4)}(x;\bp) \propto \bigl[(p_1-0.628-0.10x)^2 + (p_2+0.246-0.10\cos(\pi x))^2\sp\bigr]$
			& $(0.631,-0.249)$
			\\[2pt]
			\bottomrule
		\end{tabular}
		\caption{Intensity distributions used for a simulated four-step iterative measurement with true parameter values $\bp=(0.63,-0.25)$, along with the MLEs obtained from the simulated intensities at each step. The off-null departures for iterations 2 through 4 are each centered at the MLE from the previous iteration. The magnitude of the departure from null decreases with each iteration in order to refine the accuracy of the estimate.}
		\label{tbl:MLE_iterative_intensities}
	\end{center}
\end{table}

\pagebreak[4]

As mentioned above, from a statistical standpoint, this iterative MLE approach can be employed to obtain a parameter estimate with arbitrary precision. That is, for any fixed, reasonably large number of detected photons $\mathcal{N}$, the experiment can be designed to make the Cram\'er-Rao bound arbitrarily small, meaning that there is no fundamental limit to the sensitivity of the measurement. In practice, the accuracy is determined by experimental factors, including but not limited to:
\begin{itemize}
	\item the bit depth and signal-to-noise ratio of the sensor;
	\item the power of the source (which affects the number of photons detected in a given time interval);
	\item the level of precision and temporal stability of the experimental configuration (e.g., SLM control in the application mentioned above);
	\item the validity of the theoretical model and any approximations made;
	\item other sources of random or systematic error (e.g., thermal fluctuations or ghost images).
\end{itemize}
(Note that the second point above can be addressed by optimizing the FIM for emitted photons, as in Section \ref{sect:MLE_example3}.) In any case, the statistical methods discussed in this tutorial are still useful for determining the best nominal design for an experiment, as well as for obtaining parameter estimates from measured data based on a theoretical or empirical model.

\section{Concluding remarks}
This tutorial has summarized the fundamental concepts of maximum likelihood estimation and their application to the measurement of an optical intensity distribution. In this treatment, one or more parameters are estimated from the shape of the intensity profile, without regard for the total measured power. However, the power incident on the detector is still relevant because it determines the uncertainty of the parameter estimate, which scales as the inverse of the square root of the number of detected photons. Depending on the needs of a given application, the methods discussed in this manuscript may be used to optimize the performance of an experiment for minimal estimation error per photon detected by the sensor or per photon emitted by the source. Some sample code for calculating and evaluating the uncertainty of the maximum likelihood estimate in such an experiment can be found in the appendix.

\addcontentsline{toc}{section}{Acknowledgments}
\section*{Acknowledgments}
The author would like to thank Miguel A.~Alonso and Philippe R\'{e}fr\'egier for helpful discussions and suggestions. This work was supported by funding from the National Science Foundation (NSF) (PHY-1507278).

\FloatBarrier




\newpage
\appendix

\vspace{6em}
\section*{\LARGE Appendix}
\section{Mathematica code}\label{app:MLE}
This appendix describes a simple implementation of the MLE approach described in Section \ref{sect:MLE_optics} using the Wolfram Mathematica \cite{Mathematica} programming language. The code includes functions to calculate the PMF, likelihood function, simulated intensity, Fisher information, and MLE for an optical measurement, as well as functions to plot the expected error ellipse(s) for a two-parameter measurement. For simplicity, the code was written for the one-and-two parameter cases explored in Sections \ref{sect:MLE_examples_1param} and \ref{sect:MLE_examples_2param}; as necessary, it could readily be extended for higher-dimensional problems. The code also assumes a one-dimensional spatial coordinate.

Section \ref{sect:MLE_app_overview} below contains a list of the functions defined in this package and the syntax for their use. The function definitions are provided in Section \ref{sect:MLE_app_code}. Finally, a few example calculations are shown in Section \ref{sect:MLE_app_examples}.

\subsection{Syntax and usage}\label{sect:MLE_app_overview}
The functions defined in this package are detailed in Table \ref{tbl:MLE_Mathematica_symbols}.
{\singlespacing
\newpage
\small
\renewcommand{\arraystretch}{1.25}
\begin{longtable}{l@{\hspace{15pt}}p{.08\linewidth}p{.64\linewidth}}
	\caption{Summary of symbols and functions created to perform MLE calculations in Mathematica. When applicable, the relevant equations from the main text are listed in the second column.}
	\label{tbl:MLE_Mathematica_symbols}\\
	\toprule
	\textbf{Symbol} & \textbf{Eq.} & \textbf{Syntax and description}\\
	\midrule
	\endfirsthead\\
	\toprule
	\textbf{Symbol} & \textbf{Eq.} & \textbf{Syntax and description}\\
	\midrule
	\endhead
	
	\midrule
	\multicolumn{3}{r}{\emph{Continued on next page}}\\
	\bottomrule
	\endfoot
	
	\bottomrule
	\endlastfoot
%
%
	\mmaInlineCell[moredefined={xv}]{Code}{xv}
	& (\ref{eq:MLE_examples_u_coords})
	& One-dimensional array of spatial coordinates $x_i$ of each pixel. Can be modified to simulate different pixel arrays.
	\\	
%
%
	\mmaInlineCell[moredefined={Isim}]{Code}{Isim}
	& (\ref{eq:MLE_I_Ex1}),\linebreak (\ref{eq:MLE_I_Ex5})
	& \mmaInlineCell[moredefined={Isim}]{Code}{Isim[}$j$\mmaInlineCell{Code}{][}$p_1$\mmaInlineCell{Code}{][}$x$\mmaInlineCell{Code}{]}
	evaluates the one-parameter intensity distribution $I_j(x;p_1)$ at coordinate $x$ and parameter value $p_1$.
	
	\mmaInlineCell[moredefined={Isim}]{Code}{Isim[}$j$\mmaInlineCell{Code}{][}$\{p_1,p_2\}$\mmaInlineCell{Code}{][}$x$\mmaInlineCell{Code}{]} evaluates the \mbox{two-parameter} intensity distribution $I_j(x;\bp)$ at coordinate $x$ and parameter values $\bp=(p_1,p_2)$.
	In each function below, the argument $j$ identifies which distribution $I_j$ should be used; the corresponding function \mmaInlineCell[moredefined={Isim}]{Code}{Isim} should be defined beforehand. See Section \ref{sect:MLE_app_code} for a few examples.
	\\
%
%
	\mmaInlineCell[moredefined={Iphotonsim}]{Code}{Iphotonsim}
	& N/A
	& \mmaInlineCell[moredefined={Iphotonsim}]{Code}{Iphotonsim[}$j$\mmaInlineCell{Code}{][}$\bp,\mathcal{N}$\mmaInlineCell{Code}{]} randomly generates a simulated measurement of $I_j(x;\bp)$ containing $\mathcal{N}$ photons, such as those shown in Tables \ref{tbl:MLE_ex1_photonsim} and \ref{tbl:MLE_ex2_photonsim}. The output is an array with the same length as \mmaInlineCell[moredefined={xv}]{Code}{xv} containing the number of photons detected at each pixel.\\
%
%
	\mmaInlineCell[moredefined={P}]{Code}{P}
	& (\ref{eq:photon_pmf})
	& \mmaInlineCell[moredefined={P}]{Code}{P[}$j$\mmaInlineCell{Code}{][}$i,\bp$\mmaInlineCell{Code}{]} calculates the PMF $P_j(i|\bp)$, evaluated at pixel $i$ for true parameter value $\bp$, or equivalently the likelihood function $L_j(\bp|i)$ at $\bp$ associated with pixel $i$. The argument $\bp$ should be specified in the form $p_1$ or $\{p_1,p_2\}$ for the one- and two-parameter cases, respectively.\\
%
%
	\mmaInlineCell[moredefined={PInt}]{Code}{PInt}
	& (\ref{eq:P(tbI|p)})
	& \mmaInlineCell[moredefined={PInt}]{Code}{PInt[}$j$\mmaInlineCell{Code}{][}$\tbI,\bp$\mmaInlineCell{Code}{]} calculates the probability $P_j(\tbI|\bp)$ of measuring an intensity distribution $\tbI$ given true parameter value $\bp$, or equivalently the likelihood function $L_j(\bp|\tbI)$. The argument $\tbI$ is an array with the same length as \mmaInlineCell[moredefined={xv}]{Code}{xv}. The argument $\bp$ should be specified in the form $p_1$ or $\{p_1,p_2\}$ for the one- and two-parameter cases, respectively.\\
%
%
	\mmaInlineCell[moredefined={LLIntSum}]{Code}{LLIntSum}
	& (\ref{eq:ell(p|tbI)})
	& \mmaInlineCell[moredefined={LLIntSum}]{Code}{LLIntSum[}$j$\mmaInlineCell{Code}{][}$\bp,\tbI$\mmaInlineCell{Code}{]} calculates the sum appearing in the log-likelihood function $\ell_j(\bp|\tbI)$. The constant term $\ln P_0$ in Eq.~(\ref{eq:ell(p|tbI)}) is ignored to improve computational efficiency when calculating the MLE. The arguments $\bp$ and $\tbI$ are the same as for \mmaInlineCell[moredefined={PInt}]{Code}{PInt} above.\\
%
%
	\mmaInlineCell[moredefined={Fisher1D}]{Code}{Fisher1D}
	& (\ref{eq:MLE_Fisher}),\linebreak(\ref{eq:MLE_Fisher_obs})
	& \mmaInlineCell[moredefined={Fisher1D}]{Code}{Fisher1D[}$j$\mmaInlineCell{Code}{][}$p_1$\mmaInlineCell{Code}{]} calculates the (scalar) expected unit Fisher information $J_j(p_1)$, evaluated at parameter value $p_1$, for the one-parameter intensity distribution $I_j(x;p_1)$.
	
	\mmaInlineCell[moredefined={Fisher1D}]{Code}{Fisher1D[}$j$\mmaInlineCell{Code}{][}$p_1,\tbI$\mmaInlineCell{Code}{]} calculates the (scalar) observed Fisher information $J_j^\text{(obs)}(p_1;\tbI)$ for a measured intensity $\tbI$, which should be specified as an array with the same length as \mmaInlineCell[moredefined={xv}]{Code}{xv}.\\
%
%
	\mmaInlineCell[moredefined={Fisher2D}]{Code}{Fisher2D}
	& (\ref{eq:MLE_Fisher}),\linebreak(\ref{eq:MLE_Fisher_obs})
	& \mmaInlineCell[moredefined={Fisher2D}]{Code}{Fisher2D[}$j$\mmaInlineCell{Code}{][}$\{p_1,p_2\}$\mmaInlineCell{Code}{]} calculates the $2\times 2$ expected unit Fisher information matrix $\mathbb{J}_j(\bp)$, evaluated for parameter values $\bp=(p_1,p_2)$, for the two-parameter intensity distribution $I_j(x;\bp)$.
	
	\mmaInlineCell[moredefined={Fisher2D}]{Code}{Fisher2D[}$j$\mmaInlineCell{Code}{][}$\{p_1,p_2\},\tbI$\mmaInlineCell{Code}{]} calculates the $2\times 2$ observed Fisher information matrix $\mathbb{J}^\text{(obs)}_j(\bp;\tbI)$ for a measured intensity $\tbI$, which should be specified as an array with the same length as \mmaInlineCell[moredefined={xv}]{Code}{xv}.\\
%
%
	\mmaInlineCell[moredefined={MLE1D}]{Code}{MLE1D}
	& N/A
	& \mmaInlineCell[moredefined={MLE1D}]{Code}{MLE1D[}$j$\mmaInlineCell{Code}{][}$\tbI,cons$\mmaInlineCell{Code}{]} finds the maximum likelihood estimate for $p_1$ based on a measurement $\tbI$ of the one-parameter intensity distribution $I_j(x;p_1)$. The optional argument $cons$ may be used to specify a constraint on $p_1$, for example, \mmaInlineCell[moredefined={p1}]{Code}{-1<=p1<=1}.\\
%
%
	\mmaInlineCell[moredefined={MLE2D}]{Code}{MLE2D}
	& N/A
	&\mmaInlineCell[moredefined={MLE2D}]{Code}{MLE2D[}$j$\mmaInlineCell{Code}{][}$\tbI,cons$\mmaInlineCell{Code}{]} finds the maximum likelihood estimate for $\bp=(p_1,p_2)$ based on a measurement $\tbI$ of the one-parameter intensity distribution $I_j(x;p_1)$. The optional argument $cons$ may be used to specify constraints on $p_1$ and/or $p_2$, for example, $\{$\mmaInlineCell[moredefined={p1,p2}]{Code}{-1<=p1<=1, -1<=p2<=1}$\}$.\\
%
%
	\mmaInlineCell[moredefined={ErrorEllipse}]{Code}{ErrorEllipse}
	& N/A
	&\mmaInlineCell[moredefined={ErrorEllipse}]{Code}{ErrorEllipse[}$\mathbb{J},\{p_1,p_2\}$\mmaInlineCell{Code}{]} produces a graphics primitive for the error ellipse associated with the $2\times 2$ Fisher information matrix $\mathbb{J}$. The ellipse is centered at $\bp=(p_1,p_2)$.\\
%
%
	\mmaInlineCell[moredefined={EllipseGrid}]{Code}{EllipseGrid}
	& N/A
	& \mmaInlineCell[moredefined={EllipseGrid}]{Code}{EllipseGrid[}$j$\mmaInlineCell{Code}{][}$\mathcal{N},M$\mmaInlineCell{Code}{]} plots an $M\times M$ grid of error ellipses based on the expected Fisher information matrix for a measurement of $\mathcal{N}$ photons based on the two-parameter intensity distribution $I_j(x;\bp)$. Each ellipse corresponds to a different pair of true parameter values $\bp=(p_1,p_2)$. The grid is sampled over the region $-1\leq p_1,p_2\leq 1$ at $M$ equally spaced points in each dimension.\\
\end{longtable}
}

\subsection{Code}\label{sect:MLE_app_code}
The first step of the calculation is to define an array of spatial coordinates \mmaInlineCell[moredefined={xv}]{Code}{xv} and the intensity distribution \mmaInlineCell[moredefined={Isim}]{Code}{Isim}. For purpose of demonstration, consider the examples discussed in Sections \ref{sect:MLE_examples_1param} and \ref{sect:MLE_examples_2param}. The coordinates of the 9-pixel array given in Eq.~(\ref{eq:MLE_examples_u_coords}) may be defined as follows:

{\mmasize
\begin{mmaCell}[moredefined={xv}]{Input}
xv\,=\,Array[#&,9,\{-1.,1\}]
\end{mmaCell}
\begin{mmaCell}[moredefined={xv}]{Output}
\{-1.,-0.75,-0.5,-0.25,0.,0.25,0.5,0.75,1.\}
\end{mmaCell}
}
\noindent(For users familiar with MATLAB, this command is the Mathematica equivalent of \mmaInlineCell[moredefined={linspace}]{Code}{linspace(-1,1,9)}.) The next step is to define the one- and two-parameter intensity distributions $I_1$ through $I_{10}$, which are summarized in Tables \ref{tbl:MLE_1param_intensity_dist} and \ref{tbl:MLE_2param_intensity_dist} in the main text. Their definitions (with normalization constant $I_0=1$) are below.

{\mmasize
\begin{mmaCell}[moredefined={Isim, getp1, Iphotonsim, xv,rect,UnitBox,BinCounts,RandomChoice},morepattern={p_, p, x_, x, c_, c,d_,d, case_, nphotons_, case, nphotons}]{Input}
(*For one-parameter intensity distributions, use the following function 
 to extract p1 in the event that multiple parameters are supplied*)
getp1[p_]:=First[Flatten[\{p\}]];

(*Hard aperture with radius 1*)
rect[x_]:= UnitBox[x/2];

(*Intensity distributions*)
Isim[1][p_][x_]:=\,rect[x]*(0.5+0.5\,getp1[p]*x);
Isim[2][p_][x_]:=\,rect[x]*(0.9+0.1\,getp1[p]*x);
Isim[\{3,c_\}][p_][x_]:=\,rect[x]/(Abs[c]+1)^2*(getp1[p]-c*x)^2;
Isim[\{4,d_\}][p_][x_]:=\,rect[x]/(Abs[d]+2)^2*(getp1[p]-x-d)^2;
Isim[5][p_][x_]:=0.563\,rect[x]*(2+p[[1]]*x+p[[2]]Sin[Pi*x]);
Isim[6][p_][x_]:=0.25\,rect[x](2+p[[1]]*x+p[[2]]Cos[Pi*x]);
Isim[7][p_][x_]:=0.5\,rect[x]*(1+p[[1]]*x*Boole[x<0]+p[[2]]*x*Boole[x>0]);
Isim[8][p_][x_]:=0.5\,rect[x]*(1+2p[[1]]*(x+.625)*Boole[x<-1/8] +
     2p[[2]]*(x-.625)*Boole[x>1/8]);
Isim[9][p_][x_]:=\,0.125\,rect[x]*((p[[1]]-x)^2+(p[[2]]-Cos[Pi*x])^2);
Isim[10][p_][x_]:=
  0.320\,rect[x]*((p[[1]]-0.25x)^2+(p[[2]]-0.25Cos[Pi*x])^2);
\end{mmaCell}
}

\newpage
\noindent The remaining functions listed in Table \ref{tbl:MLE_Mathematica_symbols} are defined as follows:

{\mmasize
\begin{mmaCell}[
	moredefined={Isim, Iphotonsim, xv,UnitBox,BinCounts,RandomChoice,P,PInt,LLIntSum,prod,Fisher1D,Fisher2D,MLE1D,MLE2D,EllipseGrid, ErrorEllipse,Rotate,Thick},
	morepattern={p_, p, x_, x, c_, c, case_, nphotons_, case, nphotons,case_,i_,case,Imeas_,Imeas,\#,xv_,yv_,\#1,\#2,xv,yv,cons,cons_,J,J_,numpts,numpts_},
	morelocal={pvar,p1var,p2var,Pv,PorI,J11,J12,J22,eig,ellipses,pvals,pp1,pp2},
	morefunctionlocal={i}
	]{Input}
Iphotonsim[case_][p_,\;nphotons_]:=
   BinCounts[
    RandomChoice[(Isim[case][p]/@xv)->Range[Length[xv]],nphotons],
    \{1,Length[xv]+1\}];
    
P[case_][i_,p_]:=\,Isim[case][p][xv[[\mmaPat{i}]]]/Total[Isim[case][p]/@xv];

PInt[case_][Imeas_,p_]:=
   (Factorial[Total[Imeas]]/Apply[Times,(Factorial/@Imeas)])*
    Product[If[Imeas[[i]]==0,1,P[case][i,p]^Imeas[[i]]],\{i,Length[xv]\}];

LLIntSum[case_][p_,Imeas_]:=
   Total[ Imeas * Log[P[case][#,p] &/@ Range[Length[xv]]] ];
   
(*Use the following function to avoid errors from 0*Log[0]*)
prod[xv_,yv_]:=MapThread[If[#1==0,0,#1*#2]&,\{xv,yv\}];

Fisher1D[case_][p_,Imeas_:0]:=
   Module[\{pvar,Pv,PorI\},
    Pv=Table[P[case][i,pvar],\{i,Length[xv]\}];
    PorI=If[Imeas===0,Pv,Imeas];
    Total[prod[PorI,D[Log[Pv],pvar]^2]/.pvar->p]];

Fisher2D[case_][p_,Imeas_:0]:=
   Module[\{Pv,PorI,p1var,p2var,J11,J12,J22\},
    Pv=Table[P[case][i,\{p1var,p2var\}],\{i,Length[xv]\}];
    PorI=If[Imeas===0,Pv,Imeas];
    J11=Total[prod[PorI,D[Log[Pv],p1var]^2]];
    J12=Total[prod[PorI,D[Log[Pv],p1var]*D[Log[Pv],p2var]]];
    J22=Total[prod[PorI,D[Log[Pv],p2var]^2]];
    \{\{J11,J12\},\{J12,J22\}\}/.\{p1var->p[[1]],p2var->p[[2]]\}];	
    
MLE1D[case_][Imeas_,cons_:\{\}]:=
   p1/.Last@Maximize[\{LLIntSum[case][p1,Imeas],cons\},p1];
   
MLE2D[case_][Imeas_,cons_:\{\}]:=
   \{p1,p2\}/.Last@Maximize[\{LLIntSum[case][\{p1,p2\},Imeas],cons\},\{p1,p2\}];
   
ErrorEllipse[J_,p_]:=
   Rotate[Circle[p,\,Sqrt[#[[1]]]],\,ArcTan@@#[[2,1]]] &@
    Eigensystem[Inverse[J]];

EllipseGrid[case_][nphotons_,numpts_]:=
   Module[\{pvals,eig,ellipses\},
    pvals=Array[Identity,numpts,\{-1,1\}];
    ellipses=Table[
       ErrorEllipse[nphotons*Fisher2D[case][\{pp1,pp2\}],\{pp1,pp2\}],
       \{pp1,pvals\},\{pp2,pvals\}];
    Graphics[\{Black,Thick,ellipses\},Frame->True]\,];
\end{mmaCell}
}

\subsection{Examples}\label{sect:MLE_app_examples}
The following examples demonstrate some calculations for the one- and two-parameter intensity distributions $I_1(x;p_1)$ and $I_9(x;\bp)$ from Sections \ref{sect:MLE_example1} and \ref{sect:MLE_example9}, respectively. (Note: numerical results will vary depending on the seed for Mathematica's random number generator.)

{\mmasize
\begin{mmaCell}[moredefined={Imeas1, Iphotonsim, Imeas9}]{Input}
(*Simulate an intensity measurement with 1000 photons, given true 
 parameter values p1=0.63 and p2=-0.25*)
Imeas1=Iphotonsim[1][.63,1000]
Imeas9=Iphotonsim[9][\{.63,-.25\},1000]
\end{mmaCell}
\begin{mmaCell}{Output}
\{45,56,71,98,99,109,161,177,184\}
\end{mmaCell}
\begin{mmaCell}{Output}
\{255,160,124,131,155,88,7,14,66\}
\end{mmaCell}
}

\noindent 

{\mmasize
\begin{mmaCell}[moredefined={MLE1, MLE1D, Imeas1, Imeas9,MLE9, MLE2D, \
		Imeas9}]{Input}
(*Find the maximum likelihood estimates based on each simulated 
 intensity from above. For the one-parameter example, constrain
 p1 to the range over which I1 is valid*)
MLE1=MLE1D[1][Imeas1,\{-1<=p1<=1\}]
MLE9=MLE2D[9][Imeas9]
\end{mmaCell}
\begin{mmaCell}{Output}
0.638811
\end{mmaCell}
\begin{mmaCell}{Output}	
\{0.618961, -0.257141\}
\end{mmaCell}
}

\noindent 

{\mmasize
\begin{mmaCell}[moredefined={J1, Fisher1D, J1obs, Imeas1}]{Input}
(*Calculate the expected and observed Fisher information for the 
 one-parameter measurement of I1. Then invert and take the square root
 to calculate the standard deviation error for each case*)
J1=1000*Fisher1D[1][.63];
J1obs=Fisher1D[1][.63,Imeas1];
Sqrt[1/J1]
Sqrt[1/J1obs]
\end{mmaCell}
\begin{mmaCell}{Output}
0.0406453
\end{mmaCell}
\begin{mmaCell}{Output}
0.0397809
\end{mmaCell}
}

\clearpage
{\mmasize
\begin{mmaCell}[moredefined={J9, Fisher2D, J9obs, Imeas9}]{Input}
(*Calculate the expected and observed Fisher information matrices for 
 the two-parameter measurement of I9. Then invert the matrices to
 obtain the covariance matrices for each case*)
J9=1000*Fisher2D[9][\{.63,-.25\}];
J9obs=Fisher2D[9][\{.63,-.25\},Imeas9];	
Inverse[J9]
Inverse[J9obs]
\end{mmaCell}
\begin{mmaCell}{Output}
\{\{0.00299494, 0.000763432\}, \{0.000763432, 0.000823367\}\}
\end{mmaCell}
\begin{mmaCell}{Output}
\{\{0.00283021, 0.000765918\}, \{0.000765918, 0.000839408\}\}
\end{mmaCell}
}

{\mmasize
\begin{mmaCell}[moredefined={ErrorEllipse,Thick,J9, MLE9}]{Input}
(*Plot the error ellipse for the measurement of I9, centered at the MLE*)
Graphics[\{Thick,ErrorEllipse[J9,MLE9]\},Frame->True]
\end{mmaCell}
\begin{mmaCell}{Output}
 
\end{mmaCell}
}
\vspace{-1em}
\begin{figure}[!h]
	\centering
	\includegraphics[width=.7\linewidth]{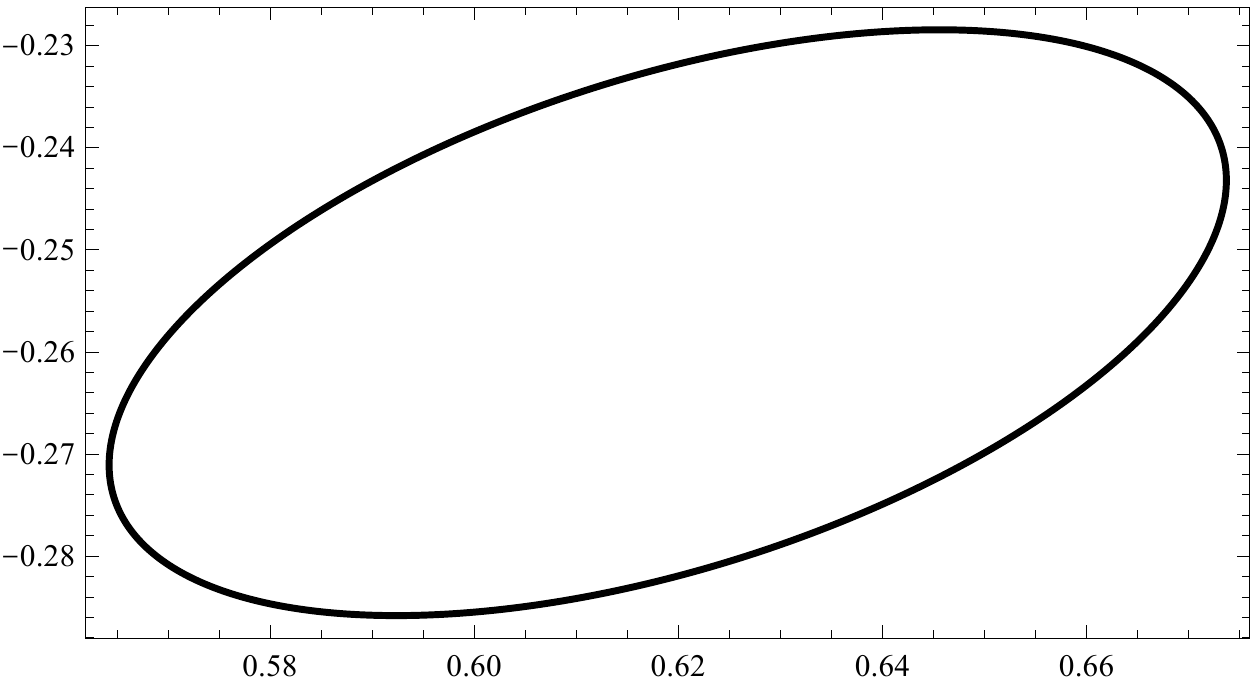}
	\label{fig:MLE_app_ellipse}
\end{figure}

{\mmasize
\begin{mmaCell}[moredefined={EllipseGrid}]{Input}
(*The following command reproduces Fig. \refellipse (sans formatting), which 
 contains a grid of error ellipses representing the expected error for 
 a measurement of I9.*)
EllipseGrid[9][1000,9]
\end{mmaCell}
}

\end{document}